# Rapid Parameter Estimation from Photoluminescence Decays of Halide Perovskite Thin Films

*Robin Heumann, Toby Rudolph, Gaosheng Huang, Thomas Kirchartz*, Chris Dreessen**

R. Heumann, T. Rudolph, G. Huang, T. Kirchartz, C. Dreessen
IMD-3 Photovoltaics, Forschungszentrum Jülich, 52425 Jülich, Germany

T. Kirchartz
Faculty of Electrical Engineering and Information Technology, RWTH Aachen, 52062 Aachen, Germany

E-mail: t.kirchartz@fz-juelich.de, c.dreessen@fz-juelich.de

**Abstract**

Extracting material parameters from experimental data is often challenging if no invertible analytical equation can be used to link the data with the quantities of interest. If the link between experiment and material parameters is given mathematically by a set of non-linear differential equations, these must be solved repeatedly during the traditional fitting procedure, resulting in long optimization times and limited insight into parameter uncertainty. Here, we present a parameter estimation workflow specifically aimed at transient photoluminescence measurements performed on lead-halide perovskite films. This workflow is accelerated using artificial neural networks for rapid comparison between experiment and simulation. An advantage of the method is the ability to rapidly scan multidimensional material parameter spaces and identify correlations between parameters that provide insights into the physics of non-radiative recombination in halide perovskites. Finally, we compare steady-state and transient photoluminescence and show how uncertainty in parameters such as the defect density can be reduced by including steady-state data in the parameter estimation workflow.

## 1. Introduction

Solution-processable solar cells present specific challenges in understanding their electronic properties and their relationship with device performance[1]. As both chemistry and processing affect the eventual functionality of a photovoltaic device[2,3], the large variability in chemistry and processing, and often even the lack of reproducibility in largely manual processes, imply a large number of unknown electronic parameters[4] for any given sample. In crystalline silicon, correlations between e.g. mobility and temperature and doping concentration are well understood and described by empirical relations[5]. Furthermore, parameters are tabulated and reliable[6]. In most solution processable solar cell materials, such relations do not exist and every given sample has a multitude of unknown parameters that would have to be determined by fitting a model to experiments that provide a significant amount of discriminatory power[7,8]. In most of these situations, the mathematical relations that link the material or device parameters of interest (charge-carrier lifetimes, mobilities, band offsets, doping densities, etc…) to observables are not analytical equations but rather systems of non-linear differential equations[9]. In some highly idealized situations[10,11], these equations can be linearized and analytical approximations become accessible. However, in a multitude of practical situations, only numerical solutions are possible[4,12,13]. The downside of the numerical approaches is that they are often slow, they are not invertible (one cannot solve the equation for the parameter of interest), and they often replace the physical experiment (understanding the experimental data) with a mathematical experiment (understanding the outcome of a system of differential equations). Often the latter is not significantly easier to grasp for the human mind than the former which may lead to a model that fits the data but provides little intuitive understanding.

A possible solution to some of these challenges has emerged over the last years, often discussed under the term Bayesian inference or Bayesian parameter estimation[14–19]. The reference to Bayes implies an understanding of likelihood that is related to the confidence in a result rather than the frequency of an occurrence (e.g. the frequency of one result when throwing a die). The likelihood in Bayesian inference can be understood as the likelihood of a model correctly describing an experiment and thereby the confidence of the researcher in the parameters of the model. A significant advantage of such a confidence-based approach is the ability to quantify the usefulness of additional experimental data[20,21]. Thus, within the framework of Bayesian inference, it is possible to quantify the increase in confidence by the decrease of the Shannon entropy[14] which is a single number reflecting the width of a probability distribution – i.e. a broad, uninformative distribution has high Shannon entropy whereas a narrow, well-constrained distribution has low Shannon entropy. One of the key challenges that is common to both Bayesian inference and basic fitting approaches is the need to ideally compare experiments with simulations representing a large number of parameter combinations. As performing simulations in the context of solar cell and material characterization usually implies solving several coupled differential equations, finding good agreements between experiment and simulation in a multidimensional space (> 5 dimensions) can be quite time consuming[16,19,22]. This can be considered a bottleneck for the design of novel, highly efficient solar cells[23]. However, machine learning and high performance computing[24] can help mitigate this problem. Here, the strategy can be broadly categorized into a purely data-driven approach whereby a model itself is inferred from the experimental data on the one hand[25] and on the other hand a physics-informed model approach whereby the model is known and the inferred model parameters are physically interpretable. Regarding the latter approach, an important conceptual improvement was implemented by Ren et al.[15] and later Das et al.[26] and involved the use of a neural network as a surrogate model for the numerical simulation. This idea makes use of the ability of neural networks to interpolate multidimensional functions. Thus, if we train the neural network with numerical simulations, we can then replace the simulations with an appropriately trained neural network for the purpose of fitting or Bayesian inference as long as the neural network is not forced to extrapolate. Thereby, the process of comparing experiment with theory is accelerated by several orders of magnitude[15,20,21,26].

Whereas Ren et al.[15] used the concept of neural-network-based surrogate models for GaAs solar cells, the present study applies the general idea to the specific case of transient photoluminescence data obtained on lead-halide perovskite thin films. We train neural networks with $>10^4$ simulations based on solving zero-dimensional rate equation models that typically contain two traps as a minimum requirement to reproduce typical experimental datasets. Note that the model used here does not consider transport of charge carriers and is meant for application on films. However, extensions of the general idea to multilayer stacks are at least conceptually straightforward. Within the framework of Shockley-Read-Hall (SRH) recombination[27,28], each singly charged trap is described by four continuous parameters (trap depth, density, electron and hole capture coefficients). In addition, there is technically a fifth, categorical parameter in the SRH model, namely the charge state of the trap which can be acceptor-like (0/-) or donor-like (+/0). If we make an assumption on the charge state (e.g. acceptor-like) and in combination with the one parameter needed for radiative recombination, we then have $4m+1$ parameters, where $m$ is the number of individual species of traps. For two traps, this leads to a minimum of 9 parameters for the description of a decay. Note that any reduction of the number of parameters per trap, in particular, the popular omission[16,19,22,29–31] of detrapping (thermal emission of a trapped carrier back into its excited band) from the SRH model would lead to a model that would be entirely unsuitable to describe the actual – highly non-exponential – charge-carrier kinetics of most halide perovskite films[32,33].

Such a high-dimensional description of the recombination dynamics makes the fitting process notoriously difficult. To date, the applied Bayesian inference schemes[16,19,22,31] are usually based on Markov-Chain Monte-Carlo (MCMC) sampling, which generates sequences of parameter samples (referred to as chains) to obtain posterior distributions for the material parameters. MCMC suffers from two major drawbacks, namely the rapidly growing computational cost with more parameters, often requiring several hours of computation time on a conventional CPU. Secondly, the chains may get stuck in certain regions of the parameter space or move rather slowly between strongly correlated parameter combinations resulting in insufficiently sampled posteriors yielding unreliable uncertainty estimates when the posterior is highly degenerate[34] (which is the case for transient photoluminescence, where multiple parameter sets can reproduce the same decay). To mitigate inefficient parameter exploration by the chains, many previous works[16,19,22,31] have simplified the SRH model by neglecting detrapping of electrons (holes) from the traps back into the conduction (valence) band, reducing the parameter correlations but thereby discarding physics that often quite accurately describes the experimentally observed decay.

In this paper we apply an approach that retains all the SRH terms (including detrapping) and transform the experimental tr-PL data into a differential decay time versus Fermi-level splitting space. In this representation, the type (shallow or deep) of the dominant trap becomes immediately apparent, already before the data is fitted. We also replace the computationally expensive MCMC algorithm with a rapid genetic optimization algorithm coupled to the neural-network-based surrogate model whose starting values are defined by suitable analytical approximations to the shape of the tr-PL decay. To better understand the physical meaning of the results, we then do a sliced grid search about the optimum parameter combination to study how parameters are correlated with each other. We still obtain approximations for the posterior probability distributions for the relevant material parameters within tens of seconds instead of hours utilizing a computer with a 24 GB graphical processing unit. We also quantify the error introduced by choosing a zero-dimensional model that neglects carrier diffusion (leading to an initial PL drop at early times[35,36]) by studying the effects of diffusion on a suit of synthetic transient photoluminescence decays. We incorporate this uncertainty in the Bayesian inference scheme and use pairwise conditional probability plots to uncover parameter correlations that are otherwise obscured in the marginalized posterior.

The paper is organized as follows. First, we define the differential decay time and Fermi-level splitting and the recombination model used for the description of transient photoluminescence and we study the effect of the trap depth on the tr-PL decay of synthetic data in section 2. In section 3, we present the machine-learning assisted parameter estimation workflow. Hereby, it is necessary to discuss the theory of Bayesian inference (section 4). We first apply the workflow to a known synthetic dataset for a single shallow trap in section 5.1 and obtain a fit as well as conditional posterior probability densities. In section 5.2, we apply our scheme to the experimental tr-PL decay obtained from a $\mathrm{Cs_{0.05}FA_{0.73}MA_{0.22}\,Pb\,I_{2.56}Br_{0.44}}$ triple-cation perovskite thin film post-treated with n-octylammonium iodide (referred to as an 85:15 film) and describe the decay by two traps. In this context, a comparison of the obtained best fit and another tr-PL fit with an enforced lower trap density is made for transient and steady-state photoluminescence data. We finish the manuscript with concluding remarks and a description of the scientific methods used in the research for this paper.

## 2. Theory

### 2.1. Differential Decay Time

Transient photoluminescence (tr-PL) is a frequently-used technique to study recombination and sometimes charge extraction in semiconductor films, layer stacks and complete devices[13,37–39]. The high luminescence quantum efficiency[40] of halide perovskites has contributed to the appeal of the method also in the context of halide perovskite research. The underlying idea is to excite a semiconductor with a short laser pulse, generate electron-hole pairs, and study their recombination as a function of time after photogeneration. The observable is radiative recombination of the electron-hole pairs, which is directly connected to the product of electron and hole concentrations, and the subsequent emission of photons. As recombination in most semiconductors is controlled by non-radiative processes, the tr-PL experiment usually probes the kinetics of non-radiative recombination. As radiative recombination can also contribute to the decay, there is sometimes a confusion arising from mixing up the observable (photon emission by radiative recombination) and the parameters that can be inferred from the decay (recombination parameters, radiative and non-radiative whatever is dominant).

To obtain a mathematical representation of the tr-PL decay, we need to understand the initial value problem defined by the differential equations for the decay of the electron and hole concentrations. Thus, let us briefly study a simple situation, where we have an equal density of electrons and holes, radiative recombination and SRH recombination[27,28] via a deep trap. Then we could write down the time-resolved change in electron concentration $n$ as

$$\frac{\mathrm{d}n}{\mathrm{d}t} = -k_{\mathrm{rad}}(n^2 - n_{\mathrm{i}}^2) - \frac{n - n_0}{\tau_{\mathrm{SRH}}}, \quad (1)$$

where $t$ is the time after photoexcitation, $\tau_{\mathrm{SRH}}$ is the SRH lifetime for the deep trap, $k_{\mathrm{rad}}$ is the radiative recombination coefficient and $n_{\mathrm{i}}^2$ is the intrinsic carrier concentration. In this simple situation, we would still be able to solve the problem analytically. There are now two options how to proceed. Option (1) would be to solve **Equation 1** for the explicit solution of $n(t)$, then use that the PL intensity $\phi \propto n^2$. Then we can fit the PL intensity $\phi(t)$ with the solution for $n^2(t)$ and adjust the parameters $k_{\mathrm{rad}}$ and $\tau_{\mathrm{SRH}}$ such that they reproduce the experimental data.

Option (2) involves finding the implicit solution for Equation 1, which is found by determining a differential decay time via[41]

$$\tau_{\mathrm{diff}} = \frac{-n}{\mathrm{d}n/\mathrm{d}t} = \frac{1}{k_{\mathrm{rad}}n + \frac{1}{\tau_{\mathrm{SRH}}}}, \tag{2}$$

where we assumed $n \gg n_{\mathrm{i}}$. We note that we can directly determine the implicit solution from Equation 1 without performing any sophisticated mathematics. Furthermore, the implicit solution is much simpler than the explicit solution which can be found e.g. in ref. [38]. **Equation 2** has the significant advantage that in those situations, where a deep defect dominates recombination, the equation simplifies to $\tau_{\mathrm{diff}} = \tau_{\mathrm{SRH}}$. Thus, the SRH lifetime can in certain situations be directly derived from the differential decay time[13].

Thus, we established[13,41] an approach for tr-PL data analysis over the last years that is based on the determination of the differential decay time from the tr-PL decay. For an intrinsic semiconductor, the most logical definition is

$$\tau_{\mathrm{diff}} = \frac{-2\phi}{\mathrm{d}\phi/\mathrm{d}t}, \tag{3}$$

whereby the factor 2 originates from the assumption that $\phi \propto n^2$ [13]. Thereby, if $n(t)$ was controlled by Equation 1, **Equation 3** would also lead to $\tau_{\mathrm{diff}} = \tau_{\mathrm{SRH}}$.

The differential decay time can be plotted either vs. time or vs. an assay of carrier density. As we will show in section 2.3, there are situations, where plotting the decay time vs. time leads to an information loss[33]. This is primarily the case, where quadratic or other higher order recombination terms (radiative recombination or shallow traps) dominate. In this situation, it is decisive to not lose the information about the carrier density. Thus, we generally plot the differential decay time $\tau_{\mathrm{diff}}$ vs. either the harmonic mean $\sqrt{np}$ of the carrier densities or the Fermi-level splitting $\Delta E_{\mathrm{F}}$ as both are readily available from the PL axis of the original PL vs. time plot (PL is proportional to $\exp(\Delta E_{\mathrm{F}}/k_{\mathrm{B}}T)$[42]). Note that to go from $\sqrt{np}$ to $\Delta E_{\mathrm{F}}$ we need an approximate value of the intrinsic carrier concentration $n_{\mathrm{i}}$, which can be obtained from the knowledge of the effective mass of halide perovskites[43–46] (see Methods Numerical Simulation). In this context, the latter approach to plot the data vs. $\Delta E_{\mathrm{F}}$ has the advantage that it can be more easily compared to the bias conditions encountered during a current voltage curve as shown in Figure 3 of ref. [47] or to bias voltages set during small signal measurements (see refs. [10,41]).

## 2.2. Rate Equation Model

Whereas Equation 1 and 2 are useful to illustrate the general idea of tr-PL and the determination and meaning of the differential decay time, they are insufficiently complex to properly explain typical experimental data obtained on halide perovskite thin films. Thus, a slightly more complicated approach is needed. The next level of complication beyond the purely analytical approach of Equation 1 is the solution of a zero-dimensional rate equation model that disregards any spatial gradients that are relevant at early times[47,48]. In such a rate equation model, we use two defect states with time dependent concentrations $n_{\mathrm{t1}}$ and $n_{\mathrm{t2}}$ as well as free electrons and holes (with concentrations $n$ and $p$). Furthermore, we consider the charge-neutrality equation for which we must make an assumption related to the possible charge states of the defects. For the sake of simplicity, we assume here that both defects are acceptor like, that is negative if filled with an electron and neutral if empty. As lead-halide perovskites behave like undoped or lowly doped semiconductors in many assays of doping[49], we therefore assume that we primarily have acceptor-like defects above or around midgap but not close to the valence band (as this would lead to considerable doping already in the dark). In this case, the charge neutrality equation can be formulated as $n + n_{\mathrm{t1}} + n_{\mathrm{t2}} = p$. The rate equations are then given by

$$\begin{aligned}\frac{\mathrm{d}n}{\mathrm{d}t} &= -R_{\mathrm{rad}} - R_{\mathrm{n}}^{\mathrm{t1}} - R_{\mathrm{n}}^{\mathrm{t2}} \\ &= -k_{\mathrm{rad}}(np - n_{\mathrm{i}}^2) - \beta_{\mathrm{n}}^{\mathrm{t1}} n(N_{\mathrm{t1}} - n_{\mathrm{t1}}) + e_{\mathrm{n}}^{\mathrm{t1}} n_{\mathrm{t1}} - \beta_{\mathrm{n}}^{\mathrm{t2}} n(N_{\mathrm{t2}} - n_{\mathrm{t2}}) + e_{\mathrm{n}}^{\mathrm{t2}} n_{\mathrm{t2}}, \qquad (4)\end{aligned}$$

$$\begin{aligned}\frac{\mathrm{d}p}{\mathrm{d}t} &= -R_{\mathrm{rad}} - R_{\mathrm{p}}^{\mathrm{t1}} - R_{\mathrm{p}}^{\mathrm{t2}} \\ &= -k_{\mathrm{rad}}(np - n_{\mathrm{i}}^2) - \beta_{\mathrm{p}}^{\mathrm{t1}} p n_{\mathrm{t1}} + e_{\mathrm{p}}^{\mathrm{t1}}(N_{\mathrm{t1}} - n_{\mathrm{t1}}) - \beta_{\mathrm{p}}^{\mathrm{t2}} p n_{\mathrm{t2}} + e_{\mathrm{p}}^{\mathrm{t2}}(N_{\mathrm{t2}} - n_{\mathrm{t2}}), \qquad (5)\end{aligned}$$

$$\frac{\mathrm{d}n_{\mathrm{t1}}}{\mathrm{d}t} = R_{\mathrm{n}}^{\mathrm{t1}} - R_{\mathrm{p}}^{\mathrm{t1}} = \beta_{\mathrm{n}}^{\mathrm{t1}}(N_{\mathrm{t1}} - n_{\mathrm{t1}}) - e_{\mathrm{n}}^{\mathrm{t1}} n_{\mathrm{t1}} - \beta_{\mathrm{p}}^{\mathrm{t1}} p n_{\mathrm{t1}} + e_{\mathrm{p}}^{\mathrm{t1}}(N_{\mathrm{t1}} - n_{\mathrm{t1}}), \qquad (6)$$

$$\frac{\mathrm{d}n_{\mathrm{t2}}}{\mathrm{d}t} = R_{\mathrm{n}}^{\mathrm{t2}} - R_{\mathrm{p}}^{\mathrm{t2}} = \beta_{\mathrm{n}}^{\mathrm{t2}}(N_{\mathrm{t2}} - n_{\mathrm{t2}}) - e_{\mathrm{n}}^{\mathrm{t2}} n_{\mathrm{t2}} - \beta_{\mathrm{p}}^{\mathrm{t2}} p n_{\mathrm{t2}} + e_{\mathrm{p}}^{\mathrm{t2}}(N_{\mathrm{t2}} - n_{\mathrm{t2}}), \qquad (7)$$

where all instances of $\beta$ represent capture- and $e$ emission coefficients. The superscripts t1 and t2 denote the trap number and the subscripts n and p refer to the affected carrier types (electrons and holes, respectively). $N_{\mathrm{t1,2}}$ represents the two trap densities which are constants within the logic of the model (see **Figure 1**), i.e. we assume that our samples remain reasonably stable during the measurement. **Equation 5** could be replaced by the charge neutrality condition ($p = n + n_{\mathrm{t1}} + n_{\mathrm{t2}}$) but is still included above for clarity. Note that if the initial values obey the charge-neutrality condition, the above system of differential equations will keep obeying charge neutrality even if it is not explicitly enforced by replacing Equation 5 by $p = n + n_{\mathrm{t1}} + n_{\mathrm{t2}}$. This system of differential equations must be solved given some initial values for the vector $(n, p, n_{\mathrm{t1}}, n_{\mathrm{t2}})$. The initial values are determined as discussed in the methods section.

Within the SRH formalism, it is important to note that the emission coefficients $e_{\mathrm{n,p}}^{\mathrm{t1,2}}$ cannot vary freely but depend on the capture coefficients $\beta_{\mathrm{n,p}}^{\mathrm{t1,2}}$ due to the principle of detailed balance[50], whereby their ratio is indicative of the trap depth $E_{\mathrm{t}}$ (see equation 2.9 in ref. [27]). We can write

$$E_{\mathrm{C}} - E_{\mathrm{t}} = k_{\mathrm{B}} T \ln\left(\frac{\beta_{\mathrm{n}} N_{\mathrm{C}}}{e_{\mathrm{n}}}\right), \tag{8}$$

and

$$E_{\mathrm{t}} - E_{\mathrm{V}} = k_{\mathrm{B}} T \ln\left(\frac{\beta_{\mathrm{p}} N_{\mathrm{V}}}{e_{\mathrm{p}}}\right), \tag{9}$$

where we omitted the superscript indices, as these relations are independently valid for every trap state.

The pairs $E_{\mathrm{C}}, N_{\mathrm{C}}$ ($E_{\mathrm{V}}, N_{\mathrm{V}}$) represent the energy level and effective density of states at the conduction (valence) band edge. We can understand e.g. **Equation 8** intuitively by saying that a high trap depth (high energy difference on the left-hand side of Equation 8) corresponds to a low emission coefficient $e_{\mathrm{n}}$ on the right-hand side of the equation. In practice, we rearrange **Equation** 8 and **9** such that we obtain an expression for the emission coefficients $e_{\mathrm{n}} = \beta_{\mathrm{n}} N_{\mathrm{C}} \exp\left(\frac{E_{\mathrm{t}} - E_{\mathrm{C}}}{k_{\mathrm{B}} T}\right) := \beta_{\mathrm{n}} n_1$ and $e_{\mathrm{p}} = \beta_{\mathrm{p}} N_{\mathrm{V}} \exp\left(\frac{E_{\mathrm{V}} - E_{\mathrm{t}}}{k_{\mathrm{B}} T}\right) := \beta_{\mathrm{p}} p_1$ where $n_1$ ($p_1$) can be regarded as the electron (hole) concentration if the electron (hole) Fermi-level were at the trap energy level $E_{\mathrm{t}}$. The rate equation model (defined by **Equations 4** to 9, the charge neutrality condition and the initial values) is then solved numerically using a Python script that can be found via a data repository. For our Bayesian inference workflow, we deal with carrier lifetimes instead of capture coefficients and we use $\tau_{\mathrm{SRH,n}} = (\beta_{\mathrm{n}} N_t)^{-1}$ as the SRH electron lifetime and $\tau_{\mathrm{SRH,p}} = \left(\beta_{\mathrm{p}} N_t\right)^{-1}$ as the SRH hole lifetime.

We emphasize that the reader should be cautious to use any variant of an SRH-like model that does not introduce the trap depth as an explicit parameter. These variants are unfortunately frequently found in the halide perovskite literature[16,22,29,30]. The absence of the trap depth means that detailed balance between capture and emission is not enforced and the key quantity that can be most easily used to compare the fit results with e.g. density functional theory simulations[51,52] is missing or hidden in the model. Furthermore, the knowledge of the trap depth and its relative position to the equilibrium Fermi level is essential to identify the initial conditions (e.g. trap filling level before the laser pulse hits the sample).

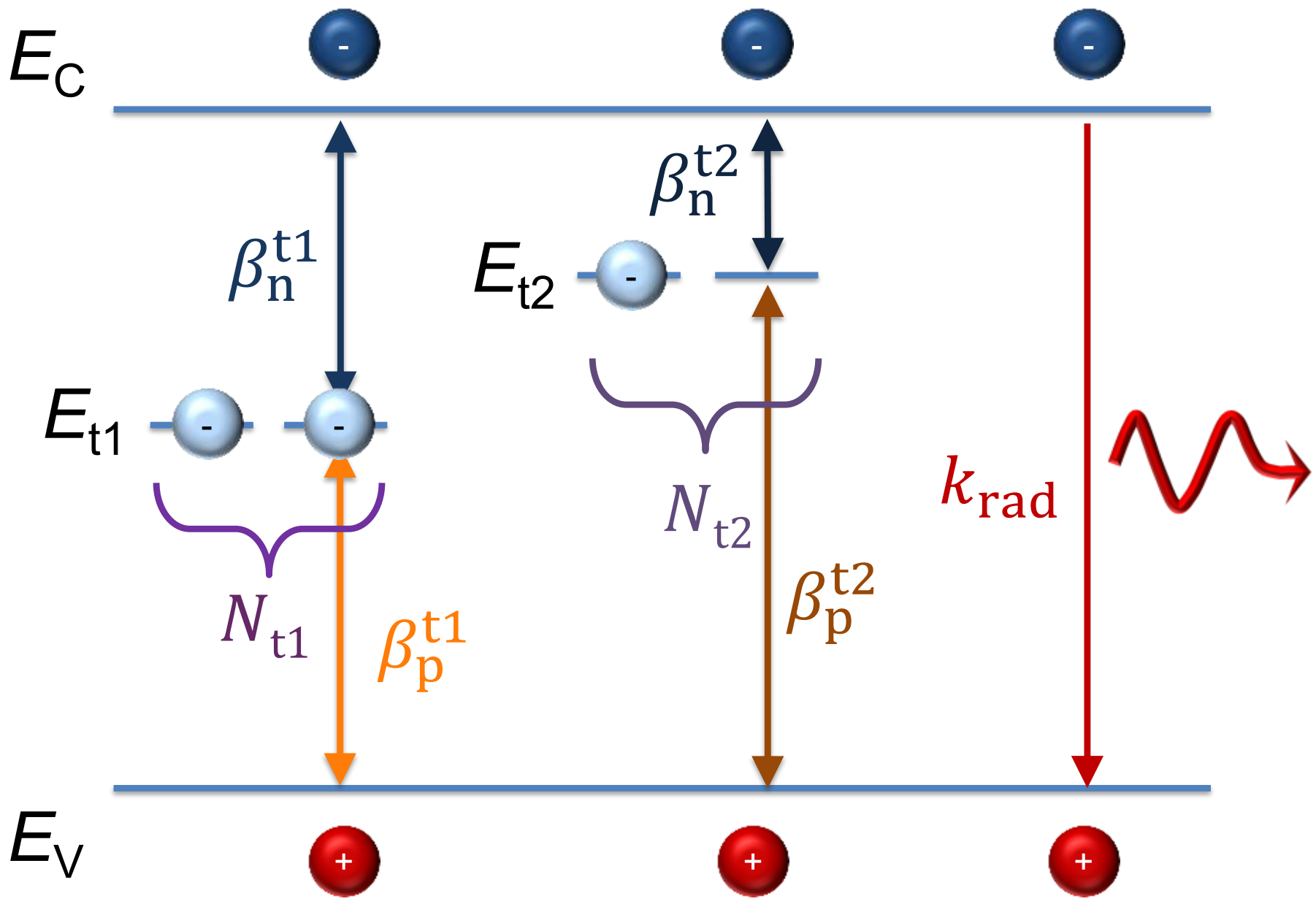

**Figure 1.** Illustration of the rate equation model used in the study. Both traps have a density $N_t$ and they interact with both bands via the capture coefficients $\beta_n^t$ and $\beta_p^t$. This includes trapping and detrapping, whereby the two rates are connected to each other via detailed balance (see Equation 8 and 9). The numbers 1 and 2 in the superscript of the parameters denote the trap number.

### 2.3 Impact of Trap Depth on tr-PL Decays

To illustrate the results we obtain from the rate equation model, we present simulations of tr-PL decays including one trap level and radiative recombination as a function of the trap depth in **Figure 2**. Figure 2a shows the normalized transient photoluminescence as a function of time (on a linear time axis), whereby the color of the lines encodes the trap depth. The dark yellow curve represents midgap, while the shades towards green and blue represent increasingly shallower acceptor-like traps close to the conduction band. The shallower the trap becomes, the more the initially exponential decay (see yellow curve) transitions into a decay that becomes increasingly slower. The shape of the decay of the shallow traps is non-exponential. When studying Figure 2b, we note that the decays in the darker shades of green and blue (shallower traps) transition from an initially exponential decay at early times towards a power law (straight line in a double-logarithmic plot) at longer times. The dashed lines in Figures 2a and b indicate a specific power law with the slope defined by $\phi \propto t^{-2}$ (i.e. $\mathrm{d}\ln(\phi)/\mathrm{d}\ln(t) = -2$), which the simulated curves eventually assume at longer times. The origin of this phenomenon has been discussed recently in refs. [32,33,53]. The general idea is that the shallower a trap becomes, the more its occupation will be controlled by the Fermi-level describing the occupation of the nearest band. So, if the trap was close to the conduction band, then the ratio of the electron density in the trap and the free electron density would saturate at longer times (or in a steady-state experiment) meaning that the capture and emission rates that control the exchange between trap and nearest band approach each other. This leads to the situation that SRH recombination via the shallow trap becomes dominated by the direct recombination of the trapped electrons with the free holes and therefore similar to radiative recombination in their statistics. This is a direct consequence of the complete SRH formalism if detrapping is properly included and is particularly relevant to materials that have a low doping density. Then both radiative band-to-band recombination as well as recombination via shallow traps will eventually follow a differential equation of the form $\mathrm{d}n/\mathrm{d}t = -kn^2$, where $k$ is a recombination coefficient that includes all rates that are approximately quadratic in free carrier density. The general solution of this differential equation (assuming an initial density $n(0)$) is $n(t) = n(0)(1 + n(0)kt)^{-1}$, implying that these bimolecular mechanisms eventually lead to a situation, where $\phi \propto n^2 \propto t^{-2}$ will hold at long times, where $n(0)kt \gg 1$. Thus, we understand that recombination via shallow defects can lead to power law decays at longer times, whereby the condition of longer times is relevant for two distinct reasons. First, the condition $n(0)kt \gg 1$ must hold, because otherwise bimolecular recombination does not lead to a perfect power law. Second, only at longer times, the occupation of the traps will have reached a situation, where the occupation is approximately controlled by the same Fermi level as that of the nearest band. Figures 2c and d show the differential decay times calculated from the tr-PL decays plotted as a function of (c) a logarithmic time axis and (d) the quasi-Fermi-level splitting. The Fermi-level splitting is essentially a parameter that can be entirely mapped to the original y-axis of the data (the PL axis) as we know that each variation $\phi_1/\phi_2$ in PL will lead to a variation in Fermi-level splitting that scales with $\Delta E_F = kT\ln(\phi_1/\phi_2)$. Thus, any order of magnitude in PL intensity ($\phi_1/\phi_2 = 10$) will lead to a change in Fermi-level splitting of $\Delta E_F = kT\ln(10) \approx 60$ meV at room temperature.

**Table 1.** Trap parameter used for the time-resolved photoluminescence simulation for different trap depths in Figure 2.

| | **Parameter** | **Value** |
|---|---|---|
| **Basic** | Bandgap $E_\mathrm{g}$ (eV) | 1.625 |
| | Radiative recombination coefficient $k_\mathrm{rad}$ (cm$^3$/s) | 3×10$^{-11}$ |
| | Intrinsic carrier concentration $n_\mathrm{i}$ (1/cm$^3$) | 4.96×10$^4$ |
| | Initial carrier concentration $n(0)$ (1/cm$^3$) | 10$^{18}$ |
| **Defect 1** | Defect level $E_\mathrm{t}$ (eV) | 0.8 – 1.6 |
| | $n_1$ (1/cm$^3$) | 3×10$^4$ – 8×10$^{17}$ |
| | Defect density $N_\mathrm{t}$ (1/cm$^3$) | 10$^{18}$ |
| | Electron SRH lifetime $\tau_\mathrm{SRH,n}$ (s) | 10$^{-7}$ |
| | Hole SRH lifetime $\tau_\mathrm{SRH,p}$ (s) | 10$^{-7}$ |

When studying the decay times as a function of time, we note that all curves start with the same continuously increasing decay time, governed by radiative recombination. This increase slows down as the decay time approaches a constant value of 200 ns, which is the value of $\tau_\mathrm{SRH}$ ($= \tau_\mathrm{SRH,n} + \tau_\mathrm{SRH,p}$) chosen in the simulation (see also **Table 1)**. At intermediate times, the trajectories become dependent on the trap depth and exhibit a jump to higher decay times (abrupt change of slope visible in both Figures 2a and b). This transition occurs earlier for shallow traps and progressively later for deeper traps. Following this trap-depth-dependent regime, all curves ultimately converge onto a common trajectory. This trajectory, highlighted by the grey curve, corresponds to the situation $\tau_\mathrm{diff} = t$ that was discussed in detail in ref. [33]. Briefly, the origin of the trajectory is applying Equation 3 to $\phi \propto t^{-2}$. Thus, independent of the proportionality constant connecting the PL intensity $\phi$ with the $t^{-2}$ power law decay, the decay time will eventually (at long times) always follow the delay time (the point in time after the laser pulse, where the derivative $\frac{-2\phi}{\mathrm{d}\phi/\mathrm{d}t}$ was taken to calculate $\tau_\mathrm{diff}$). Thus, it will be independent of recombination coefficient $k$ that indicates the effective strength of all bimolecular contributions to the recombination rate. In theory, this will happen in any intrinsic semiconductor for all acceptor-like traps when the electron Fermi level will drop below the trap level which can be influenced by the ratio of electron-to-hole concentrations. In practice, it will be only visible for all traps that are shallow enough such that this happens during the visible transient. Note that for a donor-like trap, the condition would be that the hole Fermi level rises above the trap level. If this point is not reached, before the signal reaches the noise floor, the transition to the power law will be invisible in the experimental data.

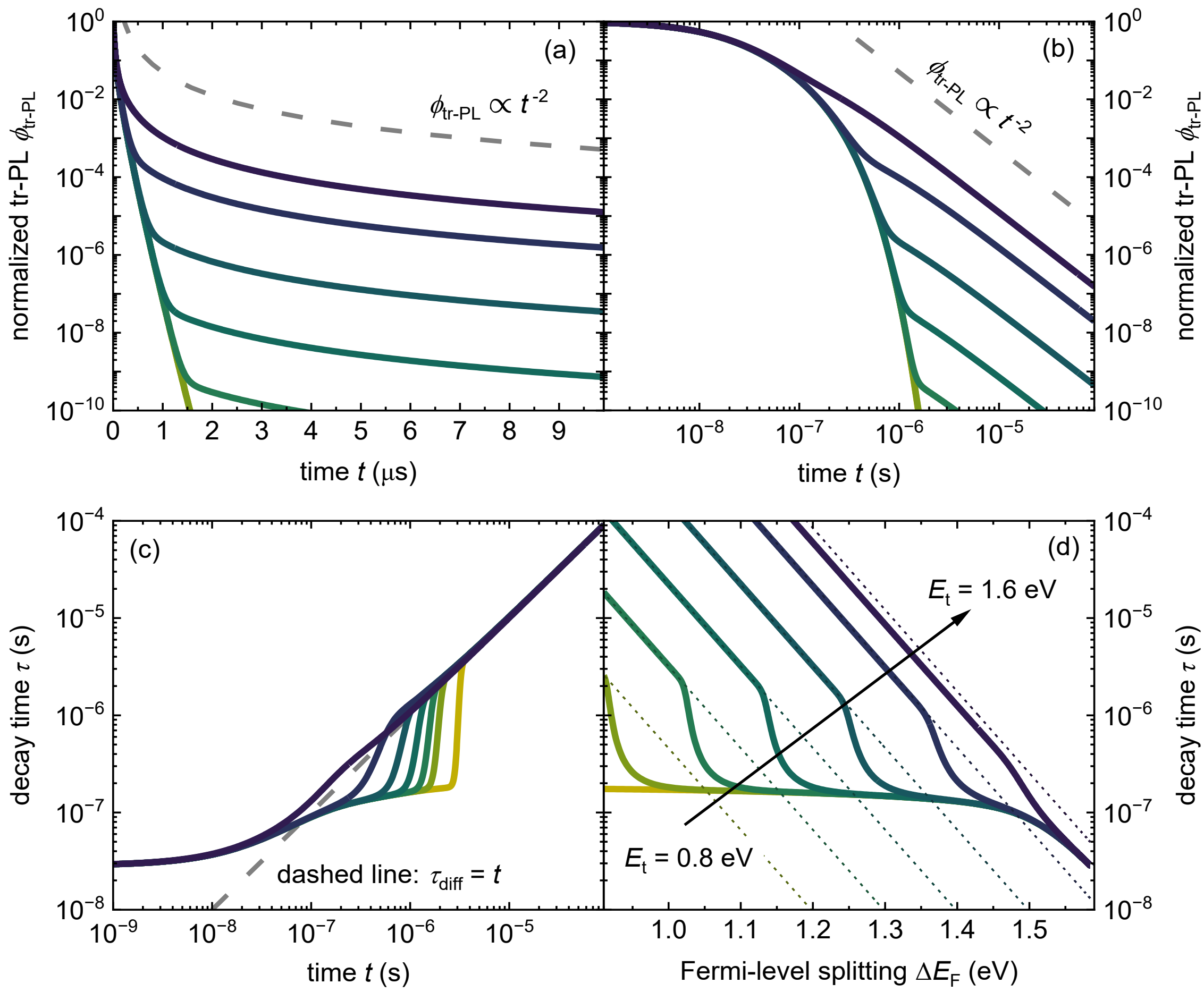

**Figure 2.** Time-resolved photoluminescence simulation for different trap depths. (a) Simulated tr-PL decay on a semi log scale. (b) Same simulation on a double log scale. (c) differential decay time $\tau_{\mathrm{diff}}$ in a double log scale. (d) Decay time $\tau_{\mathrm{diff}}$ vs Fermi-level-splitting $\Delta E_{\mathrm{F}}$ whereby the dotted lines represent the analytical solutions.

To avoid the problem caused by the $\tau_{\mathrm{diff}} = t$ trajectory being void of any information on material properties (i.e. being independent of $k$), it is therefore necessary to keep the information on either carrier density, photoluminescence intensity or Fermi-level splitting. In general, it would be rather difficult to know for certain, whether traps are preferentially capturing electrons or holes, which would lead to $n \neq p$ during the transient. Thus, the PL-axis will only provide information on $np$ or the Fermi-level splitting. Here, we note that a measurement of the fluence (energy per laser pulse per area) will provide information about the initial electron and hole densities, while the transition from $np$ to Fermi-level splitting requires an estimate of the effective density of states, which results from measurements or from calculated values of the effective masses[54,55]. The Fermi-level splitting can be easily compared to voltages[41] and to the band gap and is therefore slightly more intuitive than the $np$ product (in the somewhat non-intuitive unit of cm[-6]). When studying the decay times vs. Fermi-level splitting, we note that the midgap curve (yellow) again saturates to a constant value (lower left part of the figure), while all other curves eventually transition to a continuously varying decay time that changes as $\tau_{\mathrm{diff}} \propto \exp(-\Delta E_{\mathrm{F}}/2kT)$ (see dashed lines). The decay times follow an analytical approximation that is given by

$$\tau_{\mathrm{diff}} \approx \tau_{\mathrm{SRH,p}} \sqrt{\frac{n_1 N_T}{np}} \frac{1}{1 + k_{\mathrm{rad}} \tau_{\mathrm{SRH,p}} n_1} = \tau_{\mathrm{SRH,p}} \frac{\sqrt{n_1 N_T}}{n_{\mathrm{i}}} \frac{1}{1 + k_{\mathrm{rad}} \tau_{\mathrm{SRH,p}} n_1} \exp\left(-\frac{\Delta E_{\mathrm{F}}}{2kT}\right) \quad (10)$$

where $n_1 = N_{\mathrm{C}} \exp\left(-\frac{E_{\mathrm{C}} - E_{\mathrm{t}}}{kT}\right)$ and $\tau_{\mathrm{SRH,p}} = \left(\beta_{\mathrm{p}} N_t\right)^{-1}$. **Equation 10** assumes that the density of (acceptor-like) trap states is significant enough to contribute to the charge-neutrality equation in such a way that $n_{\mathrm{t}} \approx p$. The equivalent equation for lower trap densities is

$$\tau_{\mathrm{diff}} \approx \tau_{\mathrm{SRH,p}} \frac{n_1}{n_{\mathrm{i}}} \frac{1}{1 + k_{\mathrm{rad}} \tau_{\mathrm{SRH,p}} n_1} \quad (11)$$

and derivations for both situations can be found in refs.[32,33] (see also **Figure 3**).

To summarize the insights, we obtained from studying decays for different trap depths, we note that decays can have different characteristic shapes that lead to either exponential or power-law behaviour. For a single trap, these often feature a sharp transition in slope that may or may not be inside the experimentally measurable range for the carrier densities or the Fermi-level splitting. The transition in slope happens when the Fermi-level drops below the trap level and is therefore fairly sharp for the case of an individual trap with a sharp well-defined trap energy. In experiment, we regularly see transitions between exponential and power law decays but one rarely sees changes in slope as abrupt as those seen in Figure 2. Thus, we conclude that the successful model-based description of the tr-PL of realistic halide perovskite films will require either two or more traps or a single trap with a distribution of energy levels. In the following, we will choose the former option and discuss data that is described with two trap levels.

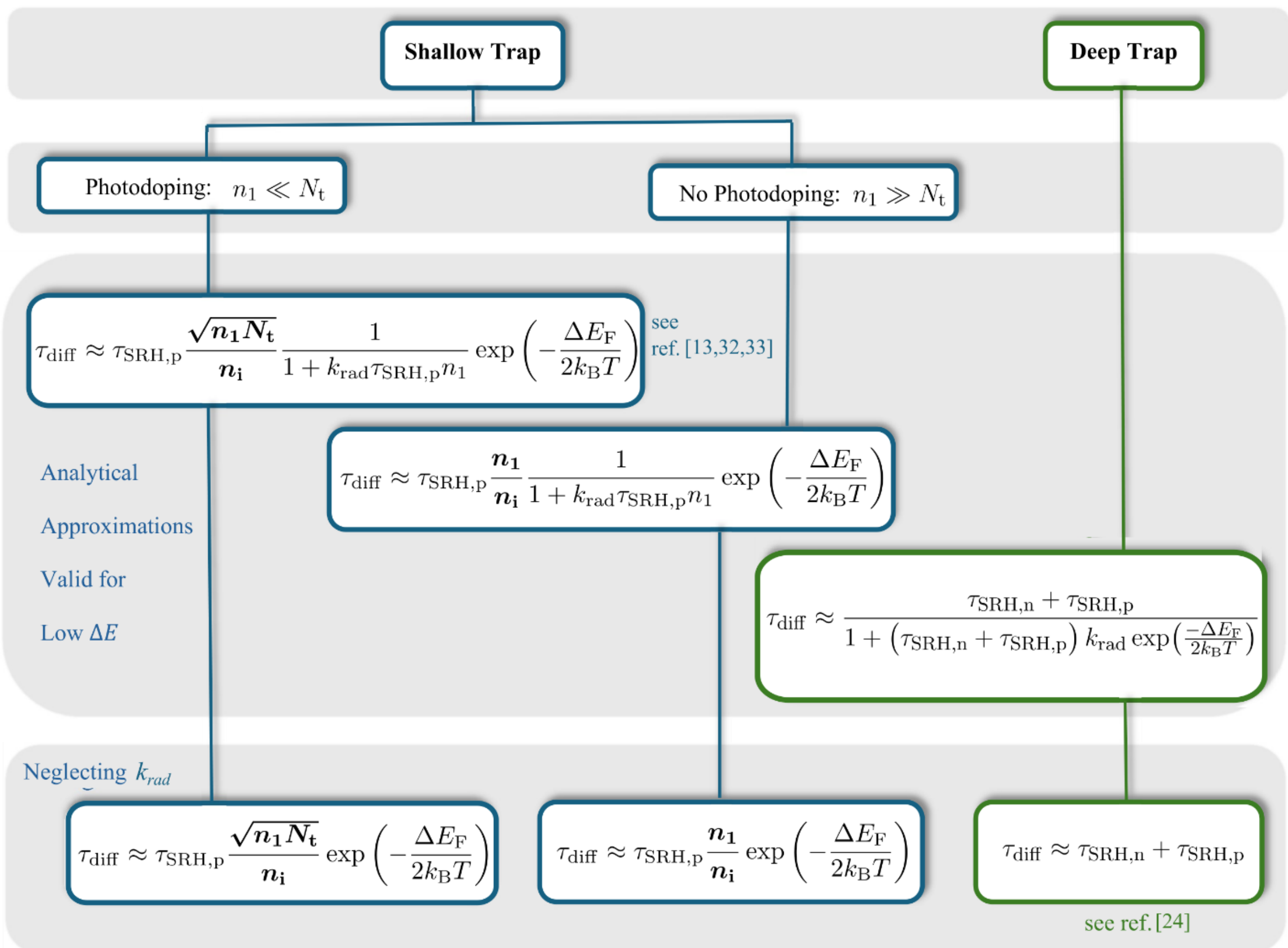

**Figure 3.** Summary of the analytical equations for the start point estimation for the case of an extremely shallow acceptor-like defect close to the conduction band and a deep defect at midgap for which charge neutrality is ignored. For the shallow trap, the analytical approximations for the differential decay time distinguish between situations for a high defect density (photodoping) and a low defect density (no photodoping). In case of a high defect density, the assumption is that $n_t = p$, i.e. the free electron density is small relative to the trapped electron density and therefore does not contribute to the charge-neutrality condition. In case of a low defect density, the charge-neutrality condition reads $n = p$, i.e. $n_t$ is small relative to $n$. Additionally, the equations can be further simplified when radiative recombination is neglected.

## 3. Parameter Estimation Workflow

To extract physically meaningful parameters from transient PL (tr-PL) data, we follow a workflow that links experimental preprocessing with surrogate-model–based optimization and that is visualized in **Figure 4**. Starting from the raw PL transients, we first subtract the background signal and apply a spline fit to suppress high-frequency noise without distorting the decay. From these pre-processed data sets, we calculate the differential decay time via Equation 3, which serves as a sensitive indicator of the dominant recombination pathway and reduces ambiguities associated with absolute signal scaling (see Figure 4a).

Based on the shape of this differential decay time, we classify each transient into two characteristic types (see Figure 4b), reflecting whether the decay is primarily governed by deep or shallow traps. For this purpose, we test whether the decay time is eventually saturating to a constant value (as seen in Figure 2c and d for the yellow curve) or whether the decay time is continuously changing without a clear indication of saturation (as seen in Figure 2c and d for the curves in different shades of green and blue). For a deep trap, the plateau region corresponds to the regime where $\tau_{\mathrm{diff}} \approx \tau_{\mathrm{SRH,n}} + \tau_{\mathrm{SRH,p}}$ holds. For an acceptor-like trap close to the conduction band, the differential lifetime follows a distinct power law with a clearly defined slope of $1/2k_{\mathrm{B}}T$ in the exponential (see shallow trap expressions in Figure 3). We use this slope as a classification criterion and compare it to the experimentally extracted slope whenever the differential decay time changes with Fermi-level splitting. This classification helps effectively confine the relevant parameter search space for the optimizer and ensures that the subsequent optimization remains within physically meaningful regimes and keeps the optimization tractable. For each class of decays, analytical approximations to simplified rate-equation models are used to derive reasonable starting values for the trap parameters (see equations in Figure 3 and Section 5.2.1).

To enable rapid parameter estimation, we replace the full numerical solution of the coupled rate equations with a neural-network-based surrogate model that serves essentially as a sophisticated look-up table and interpolation engine for the previously calculated solutions of the rate equation model (see Figure 4c). Here, the number of free parameters is restricted to those describing trap-assisted recombination and band to band recombination. Specifically, we treat the trap depths, trap densities, and carrier lifetimes for one or two independent traps as free parameters as defined in the theoretical framework in Section 2.2. Auger recombination is neglected in the present analysis, while the bandgap, temperature, and effective density of states are fixed to experimentally determined or literature values. The neural network is trained on a dataset of more than $6.5 \times 10^4$ ($2.6 \times 10^5$) simulated PL transients for the 1 (2) defect case spanning a wide range of trap densities, SRH lifetimes, trap energies and radiative recombination coefficients. During training, the neural network learns to map parameter sets to PL decay curves with high fidelity while avoiding extrapolation beyond the range of the simulated dataset (see Methods Bayesian Inference Workflow and Section S1 in the SI).

Starting from the analytical initial guesses, we then couple the surrogate model to a CMA-ES optimizer[26], which efficiently determines the parameter set that best reproduces the experimental decay (see Figure 4d). The CMA-ES code is chosen for its ability to find optima quickly in situations where the cost (i.e. time) for one function evaluation is particularly fast.

Finally, we perform a local grid search around the optimum to generate corner plots that reveal correlations between parameters within the multidimensional parameter space (see Figure 4e) from which a conditional posterior for each material parameter can be constructed.

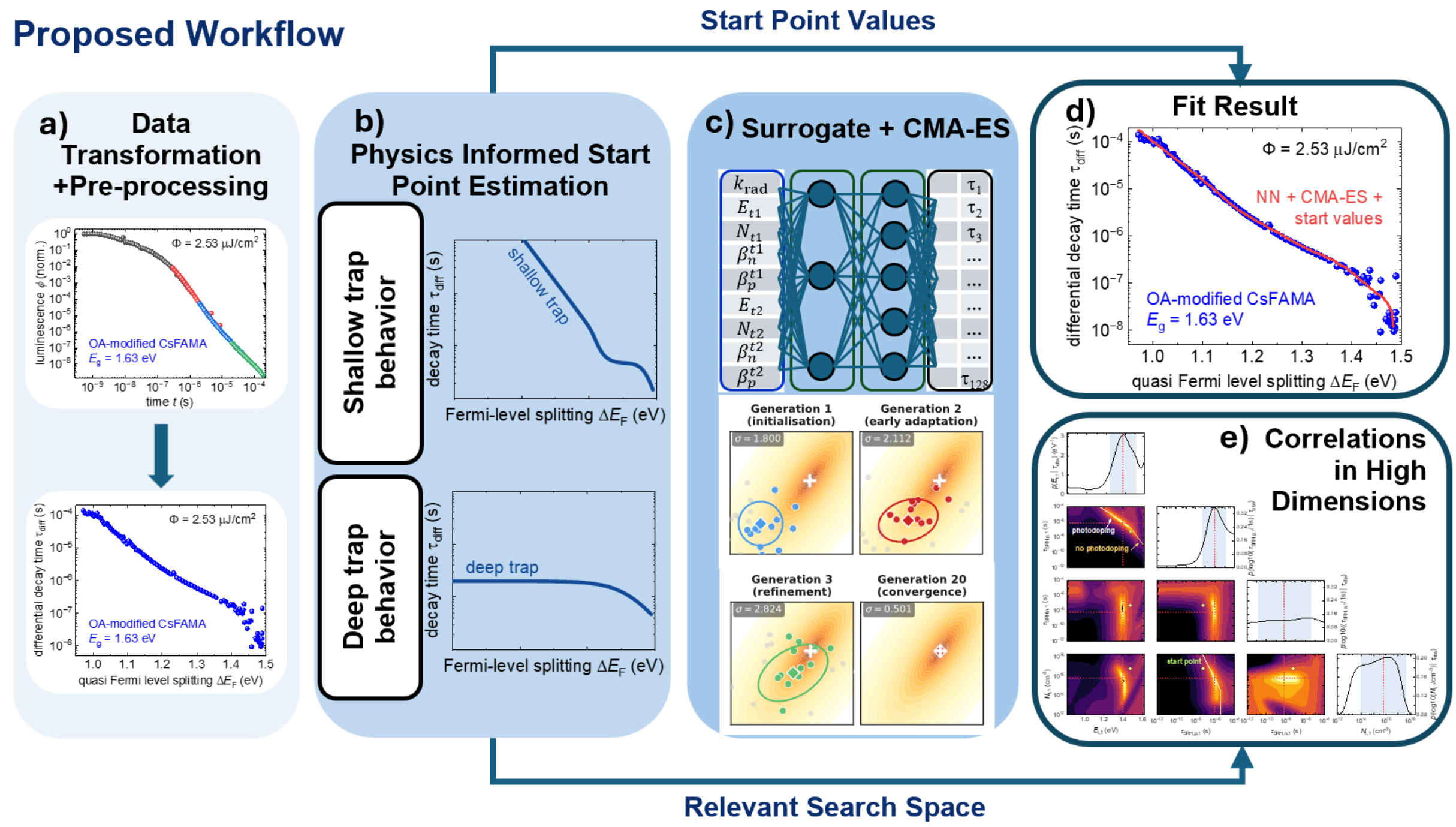

**Figure 4.** Proposed workflow for the tr-PL data analysis. (a) The initial step is to determine the differential decay time from the tr-PL raw data and (b) subsequently group the decays into two types that are dominated either by deep or by shallow traps. We then use analytical approximations to determine starting values for the parameters of the traps and (c) then apply the neural network combined with a CMA-ES algorithm to fit the data (d). (e) Around the best fit, we prepare sliced corner plots that show the correlations between the parameters in a multidimensional parameter space.

## 4. Bayesian Inference for Transient Photoluminescence

Traditional least square fitting approaches can yield reasonable fitting parameters for an assumed model. For the case of recombination dynamics in perovskites, material parameters such as the lifetime $\tau_{\mathrm{SRH}}$ are highly correlated to other material parameters such as $N_{\mathrm{t}}$ ($\tau_{\mathrm{SRH}} = (N_{\mathrm{t}}\beta)^{-1}$). Hence, many material parameter combinations could lead to equally good fits for the same experimental data. This motivates why we should not only be interested in the material parameter values obtained but also how uncertain we are about these acquired parameters or in other words how well our experiment constrains them. This question may be answered by utilizing an adapted Bayesian inference approach.

Bayesian inference is part of a statistical school of thought in which beliefs about a hypothesis $H$ are updated based on new data $D$. The Bayes' theorem[56]

$$P(H|D) = \frac{P(D|H)P(H)}{P(D)} \quad (12)$$

relates the posterior $P(H|D)$ to the likelihood $P(D|H)$, the prior $P(H)$, and the marginal likelihood $P(D)$. The posterior is the probability that our hypothesis $H$ is true given the observed data $D$ and is the quantity that we are interested in but usually cannot evaluate directly. However, the theorem allows us to connect this quantity to quantities we can evaluate. The likelihood $P(H|D)$ quantifies how likely our data is given we assume our hypothesis $H$ to be true. The prior $P(H)$ encodes all our knowledge and assumptions about the hypothesis before we even see any evidence. This could be literature values, physical constraints or the outcome of previous experiments. The marginal likelihood $P(D)$ is a normalizing factor that ensures that the posterior integrates to 1 and is hard to compute as it is calculated by averaging over all possible hypotheses $P(D) = \sum_H P(D|H)P(H)$.

For our case, the data $D$ corresponds to our tr-PL dataset. Each of the $N$ data points consist of a measured Fermi-level splitting $\Delta E_{\mathrm{F},i}$ and a measured differential decay time $\tau_{\mathrm{diff},i}$, i.e. $D = \{(\Delta E_{\mathrm{F},1}, \tau_{\mathrm{diff},1}), (\Delta E_{\mathrm{F},2}, \tau_{\mathrm{diff},2}), \ldots, (\Delta E_{\mathrm{F},N}, \tau_{\mathrm{diff},N})\}$. The hypothesis $H$ is a set of model parameters $\Theta = \{k_{\mathrm{rad}}, E_{\mathrm{t},1}, N_{\mathrm{t},1}, \tau_{\mathrm{SRH},n1}, \tau_{\mathrm{SRH},p1}, \ldots\}$ that we want to fit to our data and consists of the radiative recombination coefficient and four more parameters for each trap. Because these parameters are continuous variables, the probabilities in Bayes' theorem are replaced by probability density functions and the sum in the calculation of the marginal likelihood becomes an integral over the parameter space. Thus, the posterior becomes a probability distribution over the parameters given the observed data.

The likelihood connects the physical model to our measurement and is expressed as a normal distribution for which we have assumed additive Gaussian measurement noise[14]

$$P(D|\Theta) = \frac{1}{\sigma\sqrt{2\pi}} \exp\left( -\frac{\frac{1}{N}\sum_{i=1}^{N}\left(\log_{10}(\tau_{\mathrm{diff},i}/1\mathrm{s}) - \log_{10}(\tau^*_{\mathrm{diff},i}/1s)\right)^2}{2\sigma^2} \right). \quad (13)$$

**Equation 13** also incorporates a sum of least-squares in the exponent, where the $\tau^*_{\mathrm{diff},i}$ indicates the simulated dataset based on the hypothesis parameters $\Theta$, scaled by the standard deviation $\sigma$ assuming a uniform standard deviation across data points. We took the logarithm of the differential decay times here to avoid neglecting the contributions from fast differential decay times as the differential decay times can vary over orders of magnitude. It follows that maximizing the likelihood corresponds to minimizing the least-squares error function. The uncertainty $\sigma$ not only incorporates the uncertainty of the measurement apparatus or random noise but, in practice, also absorbs numerical errors of the forward model (e.g., optimizer or surrogate-model accuracy). It can also include the error introduced by the choice of the model

$$\sigma = \sqrt{\sigma_{\mathrm{model}}^2 + \sigma_{\mathrm{noise}}^2 + \sigma_{\mathrm{optimizer}}^2 + \sigma_{\mathrm{nn}}^2 + \cdots}. \tag{14}$$

Careful consideration should be given to the $\sigma$ value because if it is too small the model enforces a perfect fit and can lead to overfitting. If the uncertainty value is too large the model ignores the data entirely. Generally, because uncertainties are squared in **Equation 14**, the largest uncertainty component dominates the total uncertainty.

## 5. Results

### 5.1. Parameter Estimation on Synthetic Data

To validate the performance and generality of the proposed parameter-estimation workflow, we first apply it to synthetically generated tr-PL data. For the validation, we first analyze the neural network error. The neural network was trained with synthetic data which were produced by numerically solving the full rate-equation model (see Methods and -Section S1 in SI). The neural network performance was evaluated on a parameter set that is *not* included in the neural-network training dataset and is referred to as the test set. This check therefore evaluates the ability of the surrogate model to interpolate within the multidimensional parameter space rather than relying on memorized solutions (see also Methods and -Section S1 in SI for neural network training details).

The $R^2$ value quantifies the quality of the ability of the surrogate model to interpolate by comparing the neural network predictions $\tau_{\mathrm{diff},i}^*$ on the test set to a baseline predictor that only predicts the mean $\overline{\tau_{\mathrm{diff}}}$ of the observed data sets $\tau_{\mathrm{diff},i}$

$$R^2 = 1 - \frac{\sum_i \left(\log_{10}(\tau_{\mathrm{diff},i}/1\mathrm{s}) - \log_{10}(\tau_{\mathrm{diff},i}^*/1\mathrm{s})\right)^2}{\sum_i \left(\log_{10}(\tau_{\mathrm{diff},i}/1\mathrm{s}) - \log_{10}(\overline{\tau_{\mathrm{diff}}}/1\mathrm{s})\right)^2}. \tag{15}$$

The logarithms in **Equation 15** mitigate the problem of a biased weighting towards greater differential decay time values as the decay time typically spans orders of magnitude. In case the surrogate model perfectly replicates the observed test data, $R^2 = 1$ holds whereas $R^2 = 0$ means that the surrogate model only predicts the mean. Our surrogate model has a $R^2$value of above 0.998 and is therefore an excellent interpolator on unseen data that lies within the training parameter bounds demonstrating that the surrogate model can accurately reproduce decay dynamics it has never encountered before.

To showcase the validity of our machine learning workflow and the high fidelity of the surrogate model, we now focus on fitting the tr-PL decay of one shallow trap for an artificial dataset not contained in the training set. For this dataset, the experimentally inaccessible region (decay times within the first nanosecond after excitation) of the decay was disregarded. The top panel of **Figure 5** (Figure 5a) shows the best fit obtained from the neural-network–based optimization compared to the synthetic transient. The agreement between both curves is excellent, with a root mean square error

$$\epsilon_{\mathrm{rms}} = \sqrt{\frac{1}{N}\sum_{j}^{N}\left(\log_{10}(\tau_{\mathrm{diff},j}/1\mathrm{s}) - \log_{10}(\tau^{*}_{\mathrm{diff},j}/1\mathrm{s})\right)^{2}} \tag{16}$$

on the order of $3\times10^{-2}$. The small error highlights that the trained network captures the essential structure of the underlying physical model and can reliably replace computationally expensive numerical simulations.

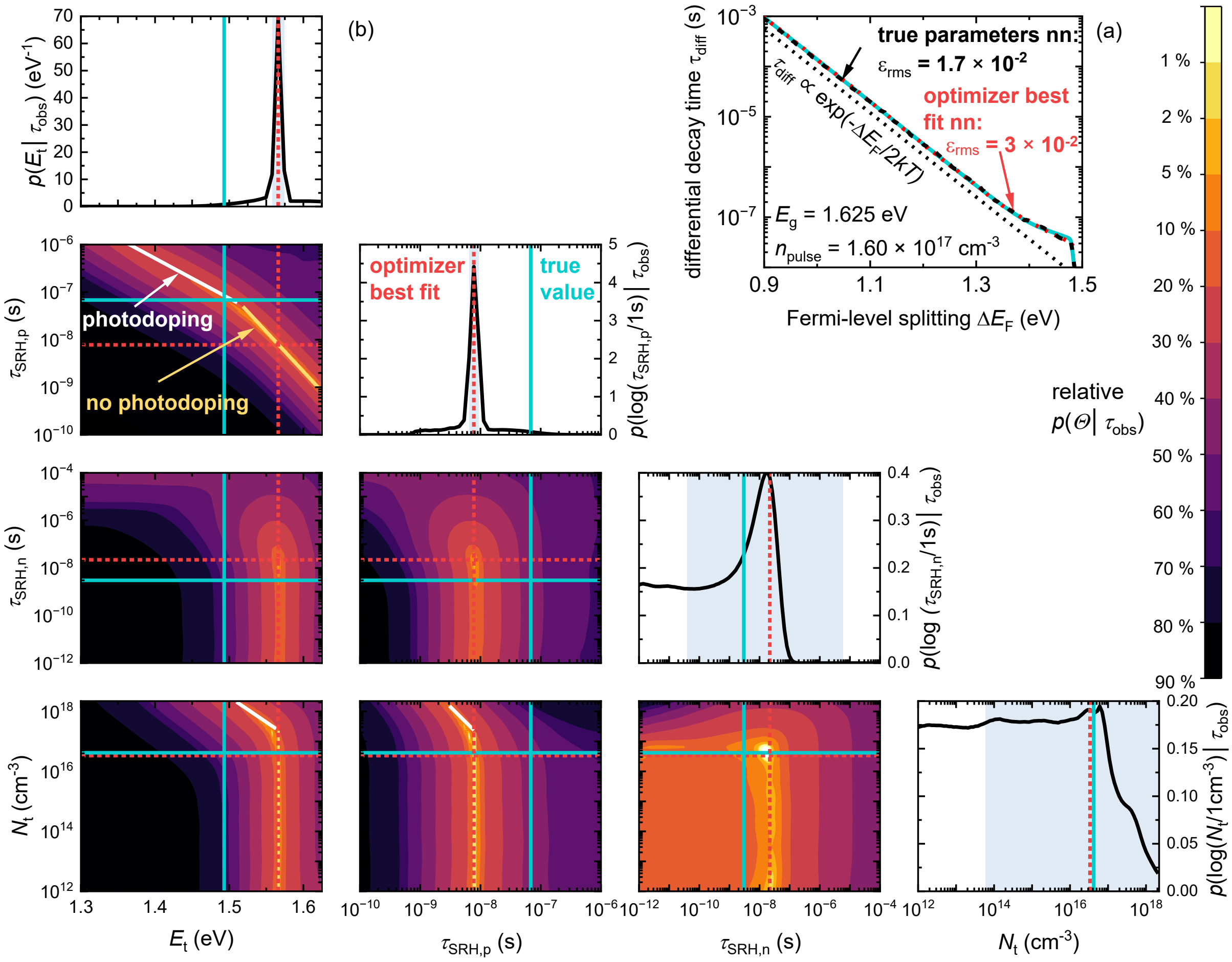

**Figure 5.** Best fit and parameter uncertainty estimation on synthetic tr-PL data. The chosen dataset corresponds to one shallow trap and was not included in the training dataset for the neural-network surrogate model. The bandgap $E_g$ is 1.625 eV and the excited carrier concentration $n_{pulse}$ is $1.6\times10^{17}$ cm$^{-3}$. Top right (a): Best fit of the neural-network (nn) surrogate

model to the synthetic decay and neural-network output for the true parameters with their root mean square errors ($\epsilon_{rms}$). Bottom (b): Sliced corner plot around the optimizer best fit value showing posterior probability density functions (diagonal panels) and relative joint posterior probability densities (off-diagonal panels) for all trap parameters. For each axis 100 points were used yielding $10^4$ points per panel. The dotted red lines indicate the optimizer best fit value and the solid cyan lines show the true parameter values of the artificial dataset. In the panels ($E_t$, $\tau_{SRH,p}$), ($E_t$, $N_t$) and ($\tau_{SRH,p}$, $N_t$) the solid white line originates from **Equation. 10** for high trap densities (photodoping) and the orange solid line for low trap densities from **Equation. 11** (no photodoping, see ref. [26]) having neglected the radiative recombination. In the diagonal panels the blue shaded regime around the optimizer best fit is the approximate $1\sigma$ (68 %) conditional credible interval. The uncertainty for the likelihood is $\sigma = 3.3\times10^{-2}$. The colour scheme ranks the grid points by their posterior probability density relative to that at the best fit point; the X % band comprises the top X % of grid points, with the thresholds determined once from the pooled values of all panels so that the levels are common to the whole figure. Since the panel normalisation and the constant prior cancel in this ratio, it is identical to the likelihood ratio with respect to the best fit and is therefore the same quantity in every panel (see discussion in methods section). Note that X % refers to the fraction of grid points and not to enclosed probability: an X % band does not contain X % of the posterior mass, unlike the $\sigma$ contours, which are defined by enclosed mass (see also $k_{rad}$ corner plot panels in Section S3 of the SI and the methods section for more information on the corner plots).

The lower panels of Figure 5 (Figure 5b) display the corresponding sliced corner plot obtained from a local grid search around the optimum parameter set (see Methods for more details). Synthetic datasets lack any kind of noise and we are applying the correct model. Therefore, the uncertainty in the likelihood equation (see Equation 13) equals the root mean square error (see **Equation 16**) at the optimizer best fit and captures the uncertainty introduced by the optimizer. In principle, we also add uncertainty to the system by using the neural network as a surrogate model. We estimate this uncertainty from the root means square error from the transient simulation at the ground truth which is about $2 \times 10^{-2}$(see Figure 5) and conclude via Equation 14 that the surrogate uncertainty has to be included to yield a total uncertainty of $\sigma = \sqrt{\sigma_{\text{optimizer}} + \sigma_{\text{nn}}} = 3.3 \times 10^{-2}$ (the subscript nn stands for neural network).

The diagonal entries of Figure 5b show the one-dimensional conditional posterior probability density functions for each individual parameter while the off-diagonal panels illustrate the conditional joint posterior probability densities relative to the optimizer best fit posterior probability density. This analysis enables the identification of parameter correlations, degeneracies, and the presence of multiple local optima, providing important insights into the robustness and identifiability of the extracted parameters. In this way, the method not only yields best fit values but also quantifies the confidence and limitations associated with the inferred trap parameters.

From the high probability density regions in panels ($E_{\mathrm{t}}$, $\tau_{\mathrm{SRH,p}}$), ($\tau_{\mathrm{SRH,p}}$, $N_{\mathrm{t}}$) and ($E_{\mathrm{t}}$, $N_{\mathrm{t}}$), we can recover the analytic approximations for the coupled rate equations for an acceptor-like shallow trap when neglecting the influence of the radiative recombination $k_{\mathrm{rad}}$ and considering photodoping and non-photodoping (see equations in Figure 3). For this, the pre-factor $\tau_0$ was obtained from a fit in a Fermi-level splitting region where $\tau_0 = \tau_0 \exp(-\frac{\Delta E_F}{2k_{\mathrm{B}}T})$ holds and then equated with the analytic pre-factor $\tau_0 \approx \tau_{\mathrm{SRH,p}} \frac{\sqrt{n_1 N_{\mathrm{t}}}}{n_{\mathrm{i}}}$ (photodoping) or $\tau_0 \approx \tau_{\mathrm{SRH,p}} \frac{n_1}{n_{\mathrm{i}}}$ (no photodoping). Here, the change from the photodoping to the non-photodoping formula occurs when $N_t \geq n_1$ switches to $N_t < n_1$, respectively, and is visible as a change in slope in panel ($E_{\mathrm{t}}$, $\tau_{\mathrm{SRH,p}}$). For panels ($\tau_{\mathrm{SRH,p}}$, $N_{\mathrm{t}}$) and ($E_{\mathrm{t}}$, $N_{\mathrm{t}}$), only the photodoping formula can be used as $N_{\mathrm{t}}$ does not appear in the no photodoping formula. Therefore, the probability densities do not depend on $N_{\mathrm{t}}$ for trap densities below $n_1$. Once $N_{\mathrm{t}}$ rises above the crossing point $N_{\mathrm{t}}^{\mathrm{cross}} = n_1^{\mathrm{cross}}$, $E_{\mathrm{t}}$ and $\tau_{\mathrm{SRH,p}}$ must change according to Equation 10 to stay in the high probability region. The crossing point $N_{\mathrm{t}}^{\mathrm{cross}}$ is determined by rearranging the formula for the non-photodoping case and we arrive at a constant $N_{\mathrm{t}}^{\mathrm{cross}} = n_1^{\mathrm{cross}} = \frac{\tau_0 n_{\mathrm{i}}}{\tau_{\mathrm{SRH,p}}^*}$ for the ($E_{\mathrm{t}}$, $N_{\mathrm{t}}$) panel and a constant $\tau_{\mathrm{SRH,p}}^{\mathrm{cross}} = \frac{\tau_0 n_{\mathrm{i}}}{n_1^*}$ for the ($\tau_{\mathrm{SRH,p}}$, $N_{\mathrm{t}}$) panel where $\tau_{\mathrm{SRH,p}}^*$ and $n_1^*$ originate from the best fit parameters.

Panels ($E_{\mathrm{t}}, \tau_{\mathrm{SRH,n}}$) and ($\tau_{\mathrm{SRH,p}}, \tau_{\mathrm{SRH,n}}$) show that there exists a confined area for the SRH electron lifetime at around $10^{-8}$ s with high probability density. For lower probability densities ( 20 %), it appears that $\tau_{\mathrm{SRH,n}}$ can vary over several magnitudes once a suitable energy level or SRH hole lifetime are found without changing the probability densities. In comparison to the other off-diagonal panels, the probability density landscape in panel ($\tau_{\mathrm{SRH,n}}$, $N_{\mathrm{t}}$) is rather broad and the trap density seems to matter little once it is below $n_1$ and a suitable $\tau_{\mathrm{SRH,n}}$ has been discovered (see Table 2 for values). This is also reflected in the conditional probability density distributions. The energy level and the SRH hole lifetime posterior probability density functions are very narrow, resembling Dirac-Delta distributions. In stark contrast, the SRH electrons lifetime's posterior exhibits a shoulder and the trap density's posterior shows a much broader posterior probability density function with a plateau that spans many orders of magnitude compared to the other posteriors. This indicates that the defect density seems especially poorly constrained by the synthetic transient photoluminescence decay as does the SRH electron lifetime. Therefore, $N_\mathrm{t}$ and $\tau_{\mathrm{SRH,n}}$ are subject to a higher variance than the other parameters even though the optimizer was able to find a defect density that lies very close to the ground truth (see also Table2). Such features in the probability density landscape are valuable, as they reveal which parameters are robustly identifiable and which show intrinsic ambiguity even under idealized synthetic conditions.

Here, we remind the reader that these posterior probability densities originate from certain slices around the optimizer best fit parameter combination which only captures the pairwise relationships between two parameters while keeping the other remaining parameters fixed at their best fit values. Therefore, the posterior distributions shown in the corner plots in this paper are conditional posteriors obtained by marginalizing within these conditional slices (see also Methods for more details). Here, we chose to use a sliced grid search approach in order to avoid a tractability problem when conducting a full grid search. Nevertheless, our sliced approach still reveals parameter correlations (see Figure 5). Our sliced approach constructs slices perpendicular to the parameter axes of the coordinate system that is given by the entries of the parameter vector $\boldsymbol{\Theta}$ which causes the true parameter values to seemingly appear in low probability regions in some of the off-diagonal panels. Here we note, that in principle one could also construct slices along a vector with a different orientation to the coordinate axes.

From the obtained posteriors, we find that the obtained best fit parameters significantly differ to the true material parameter combination except for the defect density. For the energy level and the SRH hole lifetime, the true parameter values even lie outside of the $1\sigma$ credibility interval (see **Table 2**). The credibility intervals are only local approximations around the best fit parameters and can only be used to compare how confined the obtained best fit parameters are due to the decay relative to each other. The discrepancy between the best fit and true parameters does not imply a failure of the fitting procedure. Solving the rate equations of our model directly at the best fit parameters reproduces the ground truth transient over the experimentally accessible window (disregarding decay times within the first ns after excitation which in any real experiment are obscured by the instrument response function and noise) with a root mean square error of about $1.4\times10^{-2}$. Therefore, the two parameter sets lead to physically near-identical photoluminescence decays (see Figure S8 in SI). Thus, whether or not the optimizer terminated at a local or at the global minimum in the highly dimensional error landscape is therefore without practical consequence, since the two solutions cannot be distinguished by the 'experimental' data they are fitted to.

**Table 2.** Trap parameters obtained from the machine learning assisted parameter estimation workflow for a known synthetic dataset depicted in Figure 4. The bandgap $E_g$ is 1.625 eV, the initially excited carrier concentration by the laser pulse $n_{pulse}$ is $1.6\times10^{17}$and the intrinsic carrier concentration $n_i$ is $4.96\times10^4$. The lower and upper confidence bounds are the conditional smallest symmetric contiguous interval around the optimizer best fit and are only approximations of the $1\sigma\,\chi^2$ credibility interval within which the true parameter value lies with 68% probability.

| | Parameter | True Value | Best Fit Value | Lower Credibility Bound | Upper Credibility Bound |
|---|---|---|---|---|---|
| Bi-molecular | Radiative recombination coefficient $k_{\mathrm{rad}}$ (cm$^3$/s) | $5.9\times10^{-11}$ | $1.0\times10^{-10}$ | $5.3\times10^{-12}$ | $6.7\times10^{-10}$ |
| Defect | Defect level $E_{\mathrm{t}}$ (eV) | 1.493 | 1.566 | 1.56 | 1.57 |
| | $n_1$ (1/cm$^3$) | $1.4\times10^{16}$ | $2.2\times10^{17}$ | $1.6\times10^{17}$ | $3.1\times10^{17}$ |
| | Defect density $N_{\mathrm{t}}$ (1/cm$^3$) | $4.2\times10^{16}$ | $3.4\times10^{16}$ | $5.9\times10^{13}$ | $1.9\times10^{19}$ |
| | Electron SRH lifetime $\tau_{\mathrm{SRH,n}}$ | 3 ns | 22 ns | 80 ps | 5.8 μs |
| | Hole SRH lifetime $\tau_{\mathrm{SRH,p}}$ | 68 ns | 7.8 ns | 6.5 ns | 6.5 ns |

We can also rule out the surrogate as the cause of the deviation in the obtained material parameters. The neural network replicates the rate equation solution at the best fit and the true parameters with a root mean square error of $\epsilon_{\mathrm{rms}} = 1\times10^{-2}$ and $\epsilon_{\mathrm{rms}} = 1.7\times10^{-2}$, respectively, and these error are comparable to the errors of the rate equation solutions for these two parameters sets within the experimentally accessible region. Therefore, and because the best fit looks satisfactory upon visual inspection we can say that the error of the fit is fully accounted for by a genuine degeneracy of the model and the two parameter sets are indistinguishable.

These results together with the observed parameter correlations reveal a pronounced practical non-identifiability of the material parameters which is already observable for a single trap in the tr-PL analysis. There seem to exist many parameter combinations that equally well describe tr-PL decays within experimental resolution and lie in a degenerate region where these solutions have similar errors (probability densities) in the high dimensional error (probability density) landscape. The identifiability problem shown here is of real relevance as the evolution of the charrier densities versus Fermi-level splitting (see Figure S15, SI) proofs that the best fit parameters lead to a decay with no photodoping ($n = p$) even though the true parameters lead to a decay with photodoping ($n \neq p$) and the whole inference results discussed in this chapter are also of practical relevance as the chosen ground truth parameters have been obtained from a one trap fit to the 85:15 film's experimental transient.

In the remaining panels including the radiative recombination coefficient (see Section S3, Figure S7 in SI) the conditional posteriors of $k_{\mathrm{rad}}$ are also rather broad. Even though the corner plots here only depict approximations to the probability density functions and their derived credibility intervals, we see real physical relationships between the parameters and how the analysis of transient photoluminescence confines some parameters more than others.

### 5.2. Parameter Estimation for Experimental Data

In section **Error! Reference source not found.**, we established the framework for fitting synthetic data. Next, we analyse experimental data of a $\mathrm{Cs_{0.05}(FA_{0.77}MA_{0.23})_{0.95}Pb(I_{0.85}Br_{0.15})_3}$ triple-cation perovskite thin film post-treated with n-octylammonium iodide (OAI), referred to as an 85:15 film according to its iodide to bromide ratio. The data was previously published and described in ref. [32] and was obtained using a high dynamic range measurement performed with a gated intensified CCD camera as described in ref. [32]. The existence of a continuously changing decay time over >0.5 eV in Fermi-level splitting enabled by a high dynamic range in PL intensity (around 9 orders of magnitude) helps to increase the confidence in the inferred parameters.

#### *5.2.1. Deriving Physics-Informed Start Values*

To improve convergence and reduce computation time, the algorithm is initialized using analytically derived starting values, followed by a brief pre-fitting stage that restricts the search to physically plausible regions of parameter space.

First, we need to establish whether the observed tr-PL is shallow trap or deep trap dominated. This classification is done by analyzing the slope of the decay at low $\Delta E_{\mathrm{F}}$. In case the transient photoluminescence follows a power-law, the decay is shallow trap dominated[53] which means that for a certain Fermi-level splitting regime $\tau_{\mathrm{diff}} = \tau_0 \exp\left(-\frac{\Delta E_{\mathrm{F}}}{2k_{\mathrm{B}}T}\right)$ holds. We can extract the pre-factor $\tau_0$ by fitting a straight line to the decay. Equation 10 or 11 translate this pre-factor to the corresponding material parameters. For deep defects, we obtain the Shockley-Read-Hall lifetime $\tau_{\mathrm{SRH}}$ from the $\tau_{\mathrm{diff}}$ plateau value at low Fermi-level splitting and relate it to the electron and hole SRH lifetime by using $\tau_{\mathrm{diff}} = \tau_{\mathrm{SRH,n}} + \tau_{\mathrm{SRH,p}}$.

For both cases, the defect density remains somewhat arbitrary. Steady-state photoluminescence measurements (see discussion in Chapter 5.3) give further insight into whether high or low defect densities should be used. By further assuming equal electron and hole SRH lifetimes $\tau_{\mathrm{SRH,n}} = \tau_{\mathrm{SRH,p}}$ and assuming the radiative recombination coefficient $k_{\mathrm{rad}}$ to equal $10^{11}$ $\mathrm{cm^3/s}$, we then arrive at a complete material parameter set for both cases.

For the 85:15 film studied in this paper, the tr-PL decay cannot be explained by one trap and follows the distinct power-law with a slope of $1/2k_{\mathrm{B}}T$ for low $\Delta E_{\mathrm{F}}$. This indicates the presence of shallow traps with energy levels close to the conduction band and we assume two shallow traps with different energy levels. Motivated by the steady-state PL measurement (see Figure S11), we further assume that the shallow traps significantly affect the charge-neutrality ($N_t \gg n_1$) leading to photodoping and we set $N_{\mathrm{t}} = 10\ n_1$.

We also study the tr-PL decay of a $\mathrm{Cs_{0.05}(FA_{0.77}MA_{0.23})_{0.95}Pb(I_{0.77}Br_{0.23})_3}$ + 3% $\mathrm{MAPbCl_3}$ film on glass which has a different iodide to bromide ratio and is referred to as a 77:23 film. At low $\Delta E_{\mathrm{F}}$ the differential decay time does not follow our shallow trap power-law nor does it fully saturate. The behaviour of the decay cannot simply be explained by one type of trap. Due to the presence of a regime where $\tau_{\mathrm{diff}} \propto \exp\left(-\frac{\Delta E_{\mathrm{F}}}{2k_{\mathrm{B}}T}\right)$ at around $\Delta E_{\mathrm{F}} \approx 1.4$ eV (see **Figure 6**), we assume the existence of a shallow trap additional to a deep trap at midgap. We also assume that the shallow trap affects the charge-neutrality and use $N_{\mathrm{t,shallow}} = 10\ n_{1,\mathrm{shallow}}$. Moreover, we assume that the $\tau_{\mathrm{diff}}$ point at the highest Fermi-level splitting is significantly influenced by the trapping of the excited electrons into both traps. It follows that we can estimate $\tau_{\mathrm{SRH,n}}$ for the deep and shallow trap from this differential decay time.

Figure 6 highlights the pre-fitting regimes for both samples and shows that the physics-informed start point simulations are in good agreement with the experiments across a large Fermi-level splitting range. For the 85:15 film at high Fermi-level splittings, the approximation overestimates the strength of the decay and leads to smaller decay times (see Figure 6a). When the decay for the 85:15 film is decomposed into the decay curves of two separate shallow traps, it becomes apparent that both shallow traps seem to follow the same decay for $\Delta E_{\mathrm{F}} < 1.37$ eV which is offset in parallel to the simulation with two traps. For the 77:23 film, the deep trap and the shallow trap primarily govern the recombination dynamics in their respective pre-fitting regimes. For both cases, the single trap simulations show that the trap with the stronger decay (smaller differential lifetime) at a certain Fermi-level splitting dominates the recombination dynamics at this Fermi-level splitting. The complete start values for both samples can be found in **Table 3**.

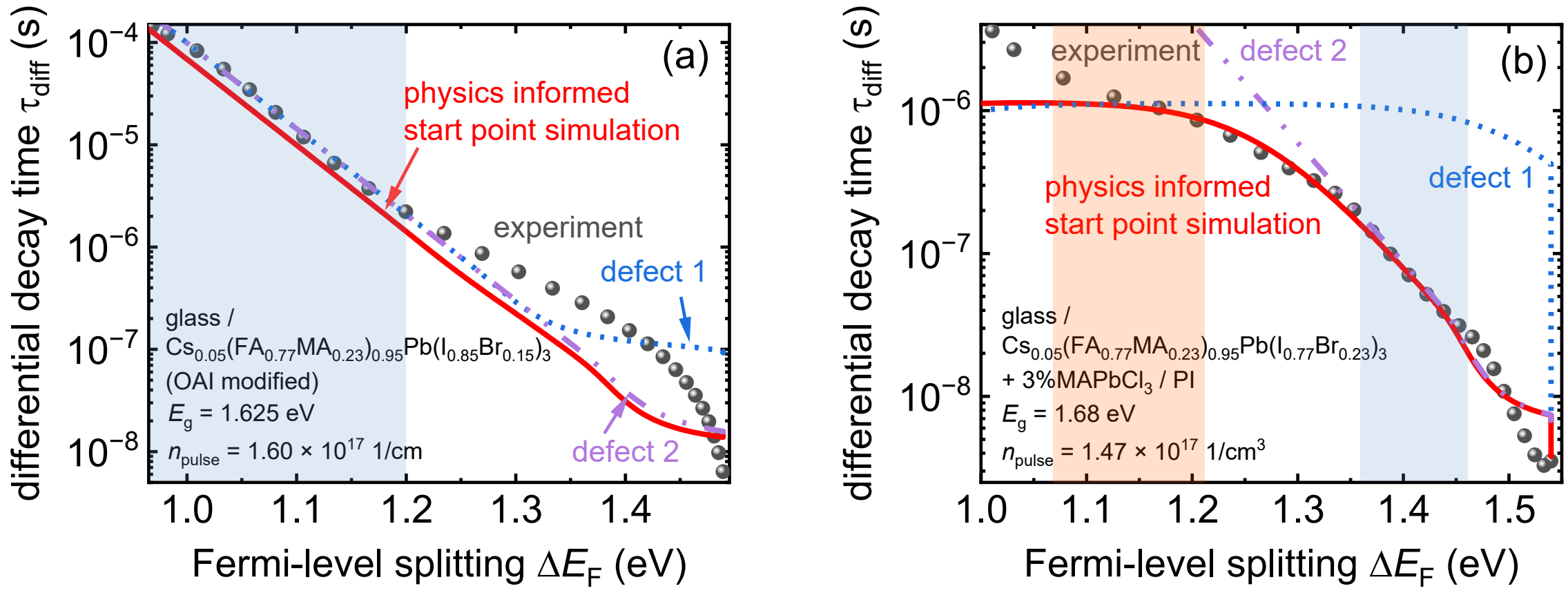

**Figure 6.** Comparison of spline fitted experimental data (black points) to its physics-informed start point simulation using two traps (red line) for two different perovskite films. The decay of a $\mathrm{Cs_{0.05}(FA_{0.77}MA_{0.23})_{0.95}Pb(I_{0.85}Br_{0.15})_3}$ film is shown in panel (a) and in panel (b) the decay for a $\mathrm{Cs_{0.05}(FA_{0.77}MA_{0.23})_{0.95}Pb(I_{0.77}Br_{0.23})_3}$ film. The dashed purple and blue lines represent the decay contributions for each individual trap. The light blue color regime represents the pre-fitting regime for which the shallow trap Equation (10) was applied and for the light orange regime the deep trap formula was used (see equations in Figure 3).

**Table 3.** Trap parameters obtained from analytic start point approximations for the 85:15 and the 77:23 film.

| | Parameter | 85:15 Film | 77:23 Film |
|---|---|---|---|
| Basic | Bandgap (eV) | 1.625 | 1.68 |
| | Film thickness (nm) | 450 | 1000 |
| | Intrinsic carrier concentration (1/cm$^3$) | 4.96×10$^4$ | 1.71×10$^4$ |
| | Radiative recombination coefficient $k_{\mathrm{rad}}$ (cm$^3$/s) | 1×10$^{-11}$ | 1×10$^{-11}$ |
| | Excited carrier concentration (1/cm$^3$) | 1.6×10$^{17}$ | 1.5×10$^{17}$ |
| Defect 1 | Defect level $E_{\mathrm{t1}}$ (eV) | 1.475 | 0.84 |
| | $n_{1,1}$ (1/cm$^3$) | 6.7×10$^{16}$ | 5×10$^5$ |
| | Defect density $N_{t1}$ (1/cm$^3$) | 6.6×10$^{16}$ | 1.0×10$^{12}$ |
| | Electron SRH lifetime $\tau_{SRH,n1}$ | 56 ns | 4 ns |
| | Hole SRH lifetime $\tau_{SRH,p1}$ | 56 ns | 1.1 µs |
| Defect 2 | Defect level $E_{t2}$ (eV) | 1.525 | 1.59 |
| | $n_{1,2}$ (1/cm$^3$) | 4.6×10$^{16}$ | 5.7×10$^{17}$ |
| | Defect density $N_{t2}$ (1/cm$^3$) | 4.6×10$^{17}$ | 6.8×10$^{17}$ |
| | Electron SRH lifetime $\tau_{SRH,n2}$ | 8.1 ns | 3.8 ns |
| | Hole SRH lifetime $\tau_{SRH,p2}$ | 8.1ns | 3.8 ns |

*5.2.2. Transient Photoluminescence Fit*

Having shown that the physics-informed start point estimation technique can produce high-fidelity simulations for both a shallow and shallow-deep trap case, we now focus on the next steps of our interpretation approach (as outlined in section 3) and restrict ourselves to the parameter estimation of the 85:15 film.

**Figure 7** confirms that the neural network fitting approach is capable of achieving a high-fidelity fit with a root mean square error of $\epsilon_{\mathrm{rms}} = 3.0 \times 10^{-2}$ for our 85:15 film tr-PL data. The best fit parameters (see Table 3) were reinserted into the rate equations, and the resulting fit matches the experiment not only in the Fermi-level splitting vs. differential decay time but also in the back-calculated time vs. photoluminescence space or time vs. differential decay time plots. Note, due to the availability of additional steady-state PL data for the 85:15 film, the fitting routine also considered the steady-state error (see also Figure S11 in the SI for the steady-state fit) during the optimization but we only discuss the results of the transient here.

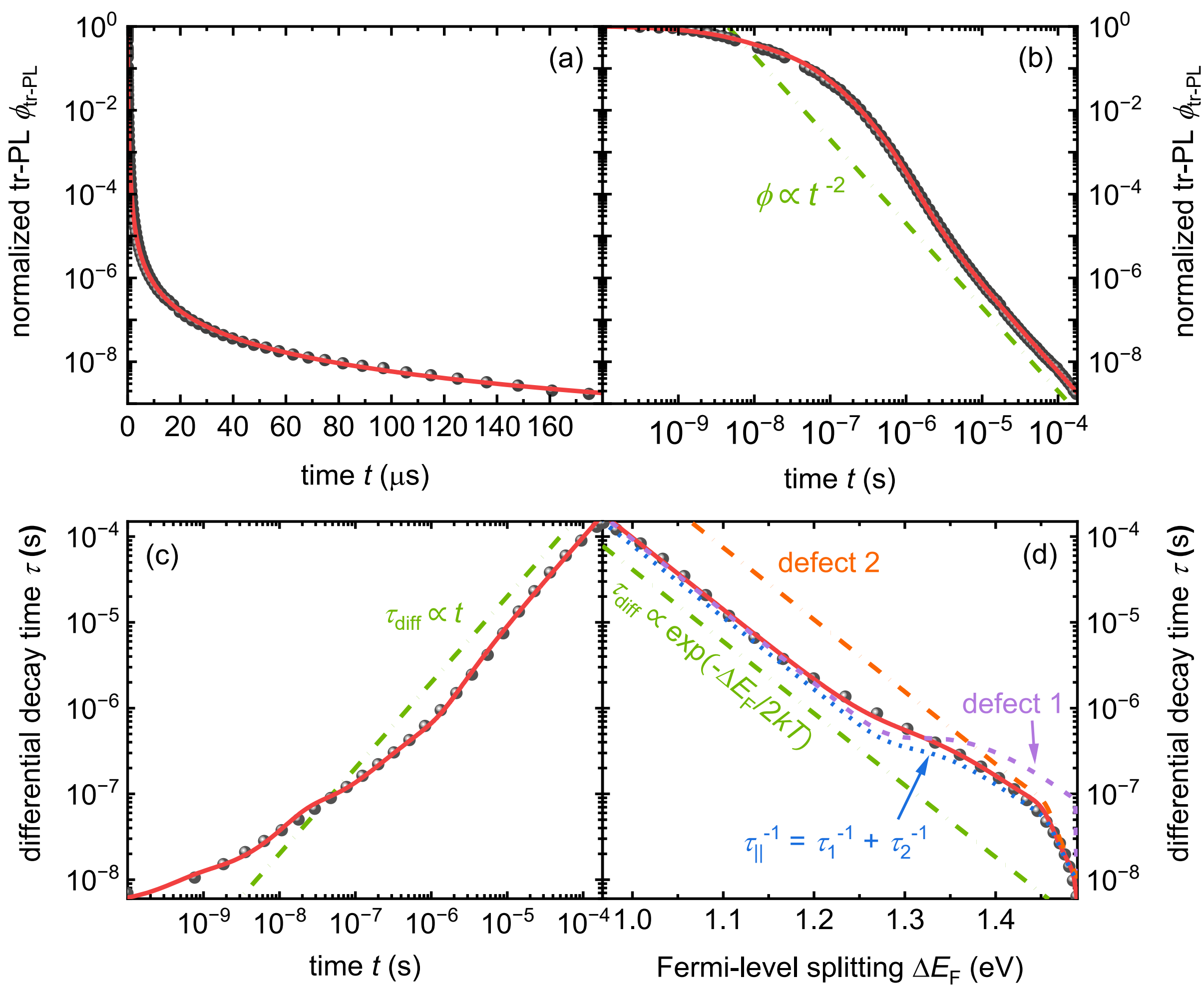

**Figure 7.** Transient photoluminescence data (black) and two trap fit (red line) for the 85:15 modified film. The normalized PL of the raw data is shown in panels a and b semi-logarithmically and double logarithmically, respectively. The differential decay time vs time (quasi-Fermi-level splitting) is shown in panel c (d). The fitting was carried out in the ($\Delta E_F$, $\tau_{diff}$) space with the help of the neural network surrogate and the obtained parameter set reinserted into the rate equations and plotted here. The dotted purple and orange lines denote the individual contributions of the two-trap fit to the decay and the blue dotted line is the parallel connection of the decays of these two traps when viewed independently.

We also analyzed the influence of the individual traps on the total decay and we find that the parallel circuit of these individual traps cannot fully explain the observed data (see Figure 7 dashed blue vs solid red curve). The reason for this is the photodoping nature of both fitted trap states. Defect 2 is a shallow trap close to the conduction band influencing charge carrier continuity by having $N_{t2} \approx 5.6\, n_{1,2}$. Defect 1 has a significantly lower trap density but due to the deeper energy level $N_{t1} > n_{1,1}$ is still true. This is reflected in the fact that $n \neq$ $p$ throughout the decay (see **Figure 8a**). From the simulated carrier concentrations and recombination rates (see Figure 8) we find that defect 1 becomes completely filled with electrons right after the initial excitation whereas defect 2 takes longer for this process and does not become completely filled. The net capture rate of defect 2 $R_{\mathrm{n}}^{\mathrm{t2}}$ changes its sign at around $\Delta E_{\mathrm{F}} \approx 1.4$ eV. The sign change means that the emission rate $e_{\mathrm{n}}^{\mathrm{t2}} n_{\mathrm{t2}}$ exceeds the capture rate $\beta_{\mathrm{n}}^{\mathrm{t1}} n(N_{\mathrm{t1}} - n_{\mathrm{t1}})$ and the trapped electrons are effectively released back into the conduction band. At approximately the same Fermi-level splitting defect 1 begins to release its captured electrons into the valence band (see Figure 8a). At around 1.3 eV defect 2's trapped electron concentration $n_{\mathrm{t2}}$ becomes smaller than defect 1's $n_{\mathrm{t1}}$ and we see that the net hole recombination of defect 2 stays smaller than its net capture rate. Here, more electrons are sent back into the conduction band than recombine back into the valence band via defect 2. The situation is vice versa for the second trap. So, defect 2 influences the recombination of defect 1 indirectly by emitting electrons back into the conduction band which immediately fall into defect 1 and then the trapped electrons recombine with the free holes in the valence band (≙ hole trapping) predominantly via defect 1 (its $R_{\mathrm{p}}^{\mathrm{t1}}$ is significantly larger than any other rate in Figure 8b). Hence, the discrepancy of the simple parallel circuit to the observed recombination dynamics.

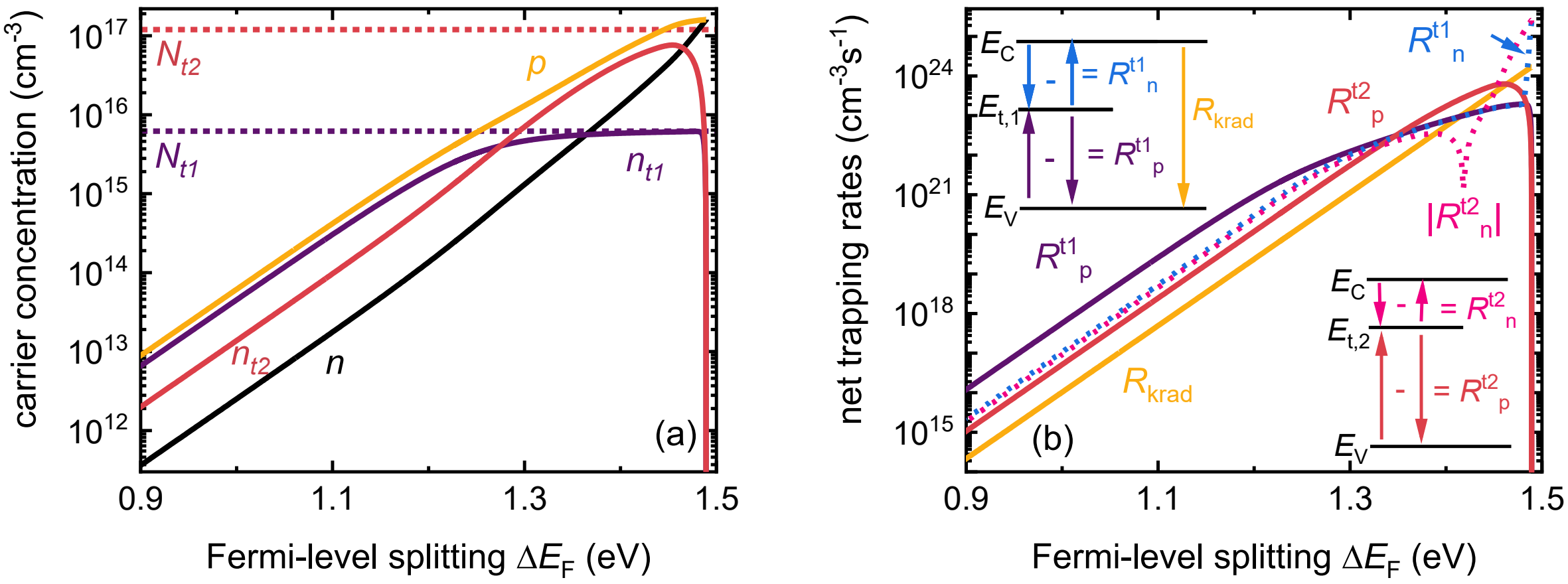

**Figure 8.** a) Carrier concentrations vs Fermi-level splitting for electrons in the conduction band $n$ and holes in the valence band $p$ as well as trapped electrons in acceptor-like defect 1 $n_{\mathrm{t1}}$ and defect 2 $n_{\mathrm{t2}}$. The dotted lines mark the value of the fitted defect densities for each trap. b) Net tapping rates plotted against Fermi-level splitting. The insets serve as a visual definition of the rates for defect 1 and the shallower defect 2. For the mathematical definitions see Equation 4 to 7. Note that the absolute value of $R_{\mathrm{n}}^{\mathrm{t2}}$ is plotted since the rate is negative for $\Delta E_{\mathrm{F}} < 1.4$ eV.

**Table 4.** Trap parameters obtained from the machine learning assisted parameter estimation workflow for the 85:15 film. The bandgap $E_{\mathrm{g}}$ is 1.625 eV, the initially excited carrier

concentration by the laser pulse $n_{pulse}$ is $1.6\times10^{17}$and the intrinsic carrier concentration $n_i$ is $4.96\times10^4$. The lower and upper credibility bounds are the conditional smallest symmetric contiguous interval around the best fit and are only approximations of the $1\sigma$ $\chi^2$ credibility interval within which the true parameter value lies with 68% probability.

| | Parameter | 85:15 Film Best Fit | Lower Credibility Bound | Upper Credibility Bound |
|---|---|---|---|---|
| | Radiative recombination coefficient $k_{\mathrm{rad}}$ ($cm^3/s$) | $6.6\times10^{-11}$ | $4.5\times10^{-12}$ | $9.9\times10^{-10}$ |
| Defect 1 | Defect level $E_{\mathrm{t1}}$ (eV) | 1.39 | 1.25 | 1.55 |
| | Defect density $N_{\mathrm{t1}}$ ($1/cm^3$) | $6.2\times10^{15}$ | $8.6\times10^{13}$ | $4.5\times10^{17}$ |
| | Electron SRH lifetime $\tau_{\mathrm{SRH,n1}}$ | 6.9 ns | 12 ps | 3.9 μs |
| | Hole SRH lifetime $\tau_{\mathrm{SRH,p1}}$ | 0.8 μs | 48 ns | 13 μs |
| Defect 2 | Defect level $E_{\mathrm{t2}}$ (eV) | 1.51 | 1.39 | 1.62 |
| | Defect density $N_{\mathrm{t2}}$ ($1/cm^3$) | $1.2\times10^{17}$ | $1.2\times10^{14}$ | $1.2\times10^{20}$ |
| | Electron SRH lifetime $\tau_{\mathrm{SRH,n2}}$ | 6.3 ns | 30 ps | 1.4 μs |
| | Hole SRH lifetime $\tau_{\mathrm{SRH,p2}}$ | 0.1μs | 2.8 ns | 7 μs |

*5.2.2. Sliced Transient Photoluminescence Corner Plot*

In the following, we focus only on the analysis of our Bayesian inference approach for the transient photoluminescence data. The inference results that consider both the steady-state and the transient photoluminescence data goes beyond the scope of the main text and is provided in the Supporting Information for the interested reader (see Section S3.2.2, Figures S11-S14).

The sliced tr-PL corner plot around the optimum best fit value reveals relatively broad posterior probability density functions for both defects (see diagonal panels in Figure 9). The posteriors for the trap energy levels have the highest probability densities followed by the SRH hole lifetimes, defect densities and the SRH electron lifetimes in receding order for each defect. This is also the case for the 1 trap synthetic dataset in section 5.1 but due to greater uncertainty $\sigma$ introduced from neglecting diffusion the posterior probability density distributions are much broader for the experimental dataset (compare Equation 13). Here, the model uncertainty outweighs the uncertainty introduced from measurement noise and the neural network uncertainty (see Section S2 in SI) yielding a model uncertainty of $\sigma = \sigma_{\mathrm{model}} =$ $4.9 \times 10^{-1}$.The relatively large uncertainty $\sigma$ leads to large credibility bounds (see Table 4) which can span orders of magnitude especially for the defect density. Here, the large credibility bounds highlight that the information contained in one tr-PL decay cannot uniquely identify all nine material parameters independently. This problem does not arise from the contained noise as we have observed this practical identifiability problem of the material parameters already for one trap and in the absence of any measurement noise (see Section 5.1). Therefore, the non-identifiability is an intrinsic property of the differential equations of the model[57–60] within the experimentally observable regime.

It is noteworthy that the posterior for the defect density of defect 1 shows the broadest plateau of all material parameters ranging roughly from $N_{\mathrm{t1}} \approx 10^{14}$ cm$^{-3}$ to $N_{\mathrm{t1}} \approx 10^{17}$ cm$^{-3}$. The defect density of defect 2 exhibits another significant plateau at even lower defect densities $N_{\mathrm{t2}} \approx 10^{13}$ cm$^{-3}$. The remaining pairwise panels of the corner plot also indicate additional regions of high probability density at these lower defect densities (see Figure S9-10 in SI). This is investigated further in section 5.3 for the simpler situation of one shallow trap. The analytic approximations for shallow traps (see equations in Figure 3) are situated close to regions of high probability density (see off-diagonal panels in Figure 9) but deviate more from the observed behavior than for the 1 trap case in Section 5.1. The SRH electron lifetime of defect 1 seems to be the parameter that remains the most uncertain, reflected in its shallow probability density function and also by the joint probability density function plots (see also Figure S9). A similar behavior is observed for the radiative recombination coefficient (see Figure S9-10) but its posterior probability density function has a more well-defined peak. Thus, it seems that once suitable material parameters are found $\tau_{\mathrm{SRH,n1}}$ and $k_{\mathrm{rad}}$ can vary over at least one order of magnitude while remaining in a high probability density region (top 1%). The highest probability density region (top 1%) is more constricted in the remaining joint probability density plots.

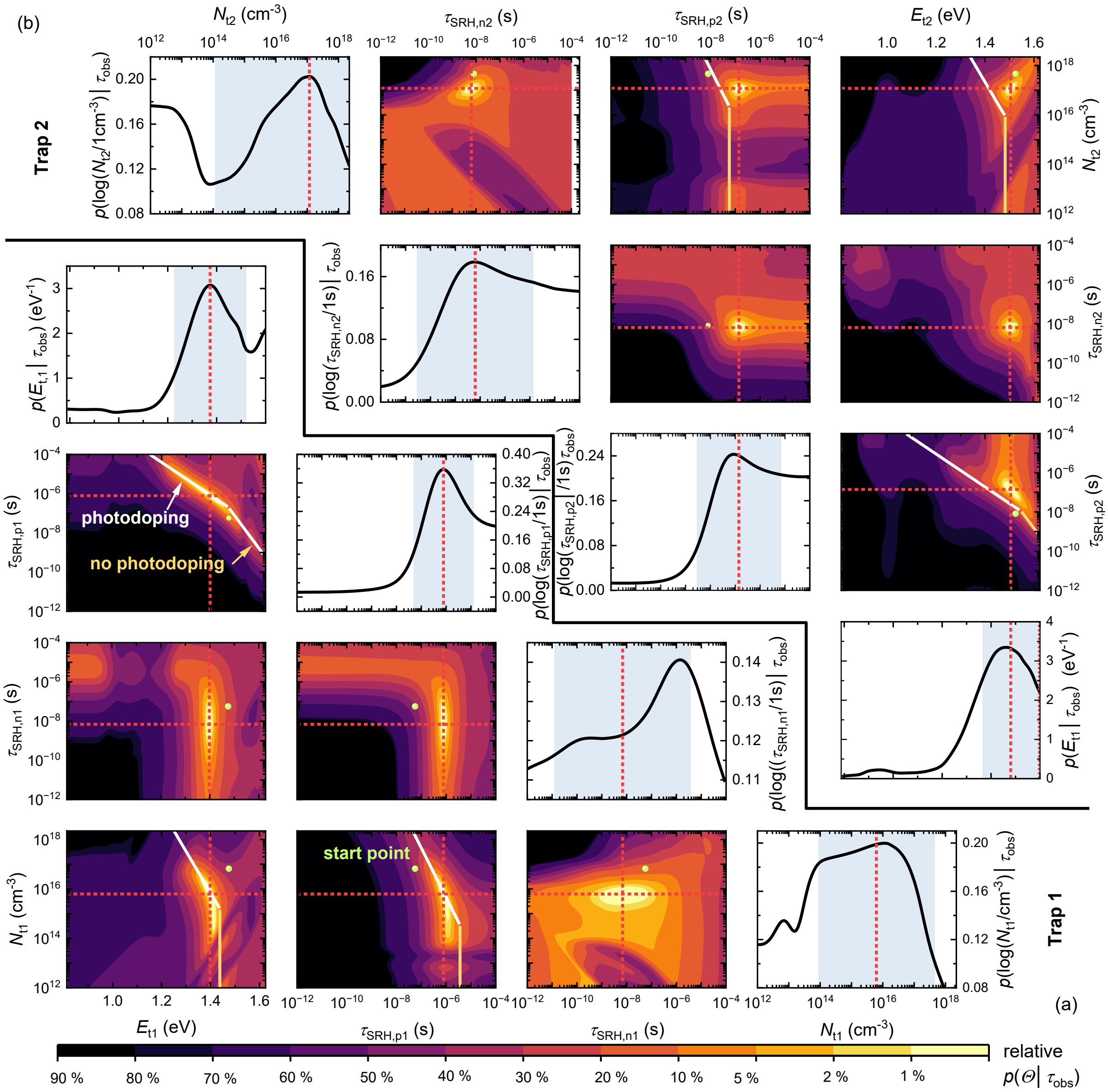

**Figure 9.** Sliced corner plot around the optimizer best fit values for the tr-Pl decay of the experimental 85:15 film data (see Figure 6). Panel (a) focuses on the defect parameters associated with defect 1 and panel (b) with defect 2. The diagonal panels show the one-dimensional conditional marginal posterior probability density functions for each free parameter, while the off-diagonal panels display the conditional joint probability density functions, revealing correlations and degeneracies in multidimensional parameter space. The dotted red lines denoted the optimizer best fit values. The black point in the off-diagonal panels marks the physics informed start point used for the fitting routine. In panels ($E_t$, $\tau_{SRH,p}$), ($E_t$, $N_t$) and ($\tau_{SRH,p}$, $N_t$,) the solid white line originates from Equation 10 for high trap densities (photodoping) and the orange solid line for low trap densities from Equation 11 (no photodoping, see ref. [26]). The color scheme was constructed as stated in Figure 5 (see also $k_{rad}$ corner plot panels in the SI). The diagonal panels show the approximate $1\sigma$ (68 %) conditional

credibility intervals around the best fit (blue shaded regime). The uncertainty for the likelihood is $\sigma = \sigma_{model} = 4.9\times10^{-1}$.

### 5.3. Comparison with Steady-State Photoluminescence

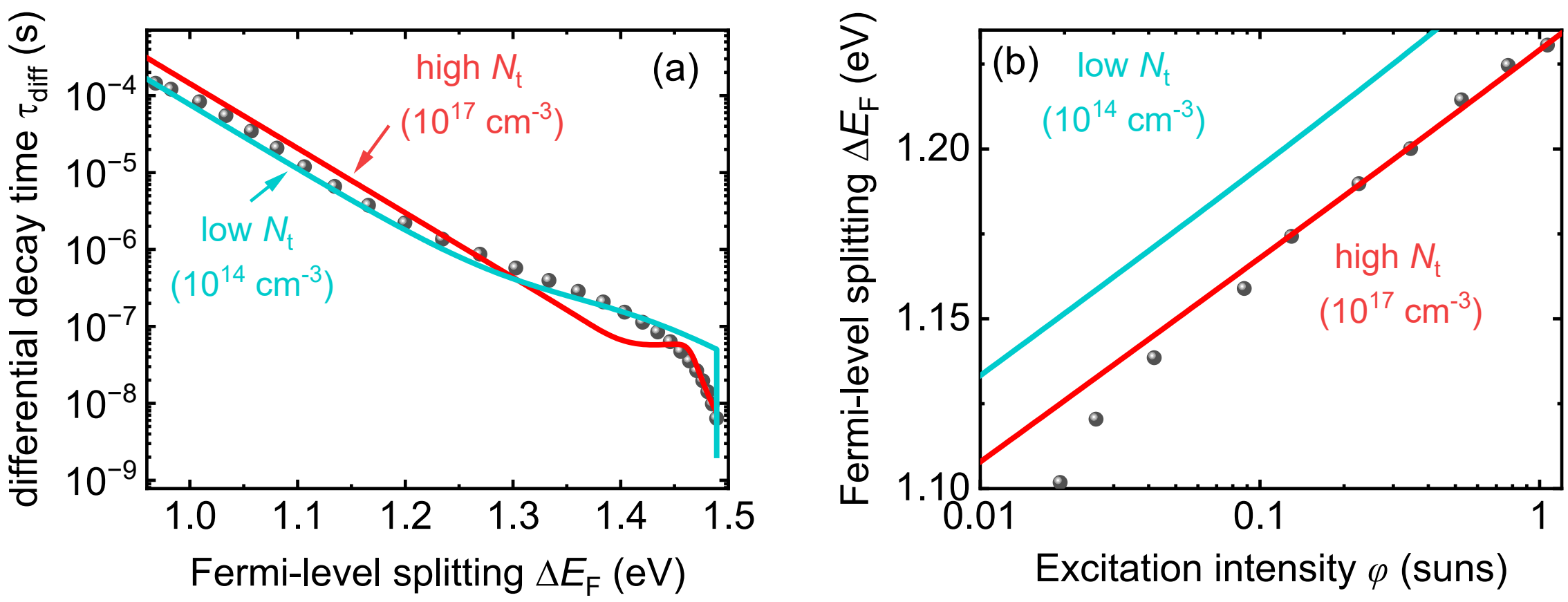

**Figure 10.** Comparison of an artificially enforced high trap density fit (red solid line) to an artificially enforced low trap density fit (cyan solid line) for the 85:15 film experimental data. Panel (a) shows the corresponding tr-PL decay and (b) the steady-state photoluminescence. Both fits require one trap only.

From our analysis of the tr-PL in section **Error! Reference source not found.**, we noticed that good fits are possible with both a rather high and a rather low trap density. To explain this phenomenon, it is most instructive to study the analytical approximations that describe the differential decay time in situations, where the trapped electron density contributes to the charge-neutrality condition, and in situations, where it does not. For this more didactive purpose of understanding the difference in SRH lifetimes, it is advantageous to study just the recombination via one single defect and ignore both radiative recombination and the possibility of having a second defect. **Figure 10a** summarizes these findings by showing moderately good fits to the differential decay time of the 85:15 film, whereby the trap density is fixed to either the low value of $N_{t,low} = 10^{14}\text{cm}^{-3}$ or the rather high value of $N_{t,high} = 10^{17}\text{cm}^{-3}$ while keeping the trap energy level fixed to $E_{t,high} = E_{t,low} = 1.46$ eV. Both fits have a quite similar error ($\epsilon_{rms,low} \approx 0.14$ and $\epsilon_{rms,high} \approx 0.18$) and thus a quite similar likelihood if only the tr-PL is considered. However, we note that the SRH lifetimes $\tau_{SRH,n}$ and $\tau_{SRH,p}$ required to fit the experimental data are interestingly different (see Table 5). For the case of an extremely shallow acceptor-like defect close to the conduction band, the analytical approximation for the differential decay time in situations of a high defect density is given by[53] $\tau_{diff} \approx$ $\tau_{\text{p,SRH}}\frac{\sqrt{n_1 N_t}}{n_i}\exp\left(-\frac{\Delta E_{\text{F}}}{2kT}\right)$ if we neglect radiative recombination (see also Equation 10 and Figure 3 for the situation with radiative recombination). Here, the assumption is that $n_{\text{t}} = p$, i.e. the free electron density is small relative to the trapped electron density and therefore does not contribute to the charge-neutrality condition. As the free electron density is also small compared to the free hole density, this situation is also referred to as photodoping. If we instead assume that the charge-neutrality condition reads $n = p$, i.e. $n_t$ is small relative to $n$, then we find given the same assumptions as before that[53] $\tau_{\text{diff}} \approx \tau_{\text{SRH,p}}\frac{n_1}{n_{\text{i}}}\exp\left(-\frac{\Delta E_{\text{F}}}{2kT}\right)$ (see Equation 11 and Figure 3). These assumptions are verified for the two simulations in Figure 10 by checking the transient electron and hole carrier concentrations $n, p$ and the trapped electron concentration $n_{\text{t}}$ versus the Fermi-level splitting (see Figure S16).

**Table 5.** Trap parameters obtained from one trap fits to the tr-PL decay of the 85:15 film.

| | Parameter | Low $\boldsymbol{N_t}$ | High $\boldsymbol{N_t}$ |
|---|---|---|---|
| Basic | Radiative recombination coefficient $k_{\text{rad}}$ (cm$^3$/s) | $3\times10^{-10}$ | $6\times10^{-12}$ |
| Defect | Defect level $E_{\text{t}}$ (eV) | 1.463 | 1.463 |
| | $n_{\text{t}}$ (1/cm$^3$) | $4.2\times10^{15}$ | $4.2\times10^{15}$ |
| | Defect density $N_{\text{t}}$ (1/cm$^3$) | $1\times10^{14}$ | $1\times10^{17}$ |
| | Electron SRH lifetime $\tau_{\text{SRH,n}}$ | 1.1 µs | 4 ns |
| | Hole SRH lifetime $\tau_{\text{SRH,p}}$ | 0.4 µs | 86 ns |

Thus, to keep the same fit quality, we can equate the right-hand sides of the two analytical approximations for shallow traps close to the conduction band neglecting the influence of radiative recombination (see photodoping and no photodoping equations in Figure 3) and find the condition that

$$\frac{\tau_{\text{SRH,p}}(N_{\text{t,high}})}{\tau_{\text{SRH,p}}(N_{\text{t,low}})} \approx \sqrt{\frac{n_1}{N_{\text{t,high}}}}. \tag{17}$$

The value of $n_1$ that is the same between the two fits is $4.2\times10^{15}$ cm$^{-3}$. We find that the left-hand-side of **Equation 17** evaluates to $2.2 \times 10^{-1}$ while the right-hand-side yields $2.1 \times 10^{-1}$ verifying the validity to use our approximations. Note that the result does not depend on the exact value of the lower of the two trap densities (here $N_{\text{t,low}} = 10^{14}$ cm$^{-3}$). The lower of the two densities is only relevant insofar as the condition $n = p$ needs to hold during the experimentally observable part of the decay ($N_{\text{t,low}} \ll n_1$), but as the trapped electron density never actually contributes to the charge-neutrality condition, its exact density is irrelevant and only the values of $\tau_{\mathrm{SRH,p}}$ and $n_1$ matter.

At this stage, it is logical to attempt to include additional information about the sample to further restrict the parameter range and discriminate between situations where the trapped charge density matters or where it does not. Figure 10b shows the Fermi-level splitting (black symbols) calculated from the steady-state PL quantum efficiency of the samples at different excitation densities. The excitation is done with a 532 nm laser diode. In addition to the experimental data, we also show the Fermi-level splittings as a function of excitation intensity that follow from the steady-state version of the one trap SRH model for the same parameters used in Figure 10a. We note that now a difference emerges that was hidden in the tr-PL decay. We could now calculate the relationship between Fermi-level splitting and generation rate $G$ analytically. In the steady state, the recombination rate $R$ equals the generation rate $G$ whereby the recombination rate is the sum of the radiative and SRH recombination rates[27,28] $R = R_{\mathrm{rad}} + R_{\mathrm{SRH}} = k_{\mathrm{rad}}(np - n_{\mathrm{i}}^2) + \frac{np-n_{\mathrm{i}}^2}{\tau_{\mathrm{SRH,p}}(n+n_1)+\tau_{\mathrm{SRH,n}}(p+p_1)}$. We can neglect $n_{\mathrm{i}}^2$ relative to $np$. By using the relation $n_1 p_1 = n_{\mathrm{i}}^2$ and remembering that we have a shallow trap close to the conduction band with a large $n_1$, we find that $p_1$ is so small that it can be neglected. For both cases where the trap density influences or does not influence the charge-neutrality equation, we find that $n_1 \gg n, p$ for generation rates equivalent to 1 sun and below (see Figure S16) and the only term left in the denominator of $R_{\mathrm{SRH}}$ is $\tau_{\mathrm{SRH,p}} n_1$. Hence, for the simplified case of one shallow, acceptor-like defect that dominates recombination relative to radiative recombination, we find that

$$\Delta E_{\mathrm{F}} = k_{\mathrm{B}} T \ln\left(\frac{G \tau_{\mathrm{SRH,p}} n_1}{n_{\mathrm{i}}^2}\right), \tag{18}$$

whereby we have made use of the relation $np = n_{\mathrm{i}}^2 \exp\left(\frac{-\Delta E_{\mathrm{F}}}{k_{\mathrm{B}} T}\right)$. Thus, the differences in $\tau_{\mathrm{SRH,p}}$ needed to fit the data shown in Figure 10a lead to the differences in the relation between Fermi-level splitting and light intensity (which directly controls the parameter $G$). By making use of **Equation** 17 and **18**, we arrive at

$$\delta\Delta E_{\mathrm{F}} = \Delta E_{\mathrm{F}}^{\mathrm{low}} - \Delta E_{\mathrm{F}}^{\mathrm{high}} = \frac{1}{2} k_{\mathrm{B}} T \ln\left(\frac{N_{\mathrm{t,high}}}{n_1}\right). \tag{19}$$

**Equation 19** describes the difference in Fermi-level splittings across the two simulations with a high and low trap density and we see that this difference is only attributed to the ratio of the high trap density $N_{\mathrm{t,high}}$ to $n_1$ yielding $\delta\Delta E_{\mathrm{F}} \approx 41$ meV. We then subtracted the constant $\delta\Delta E_{\mathrm{F}}$ from the Fermi-level splitting for the low defect density $\Delta E_{\mathrm{F}}^{\mathrm{low}}$ and observe that at high excitation intensities ($\varphi > 0.2$ suns) this correction matches the Fermi-level splitting for the high defect density simulation $\Delta E_{\mathrm{F}}^{\mathrm{high}}$. At lower excitation intensities ($\varphi < 0.2$ suns) slight differences emerge between $\Delta E_{\mathrm{F}}^{\mathrm{low}} - \delta\Delta E_{\mathrm{F}}$ and $\Delta E_{\mathrm{F}}^{\mathrm{high}}$ whereby the correction seems to be more closely aligned to the experimentally observed Fermi-level splittings. This means that the discrepancy between the steady-state simulations originates from the difference in the influence of the trap density to the charge-neutrality condition viz. both the experimentally observed transient and steady-state PL are heavily influenced by the ratio $\frac{n_1}{N_{\mathrm{t}}}$ which makes sense for our shallow traps as $N_{\mathrm{t}}$ governs the trapping into the trap and $n_1$ the detrapping of the electrons back into the conduction band. This ratio therefore determines the trapped electron concentration $n_{\mathrm{t}} \approx \frac{N_{\mathrm{t}}}{n_1} n$ and heavily influences the recombination dynamics.

## 6. Conclusion

Transient photoluminescence is the standard probe for carrier dynamics in halide perovskites and it is asked to deliver parameters such as the defect energy level, the defect density, the SRH electron and hole lifetime and additionally the radiative recombination coefficient which allow to quantitatively compare the decay dynamics and device performance across samples. Here, we showed by simulating the coupled rate equations that there exist multiple fitting parameter combinations that equally well describe the same experimental decay even in the absence of noise. Therefore, least-square fitting selects only one arbitrary member of a broad set of data-compatible solutions and what becomes interpretable is the set of solutions itself weighted by how well each parameter combination reproduces the experiment and what is known about parameters before conducting the experiment which is referred to as the posterior from a Bayesian inference perspective. Practicing Bayesian inference so far has been time consuming due to a high computational cost. But the perovskite community has a need for fast Bayesian inference, especially in the case of high-throughput tr-PL. Here, we showed that Bayesian inference can be practiced within tens of seconds (see Figure S4 in the SI) instead of hours while still applying the full SRH and radiative recombination formalism without neglecting detrapping. In contrast to previous works[16,19,22,31], we carefully analyzed and compared the main uncertainty sources such as measurement noise or model error for both steady-state and transient photoluminescence and included the dominant uncertainties into our Bayesian inference scheme. The total computational cost of Bayesian inference and therefore the calculation time can be reduced either by lowering the computational cost of each simulation or by reducing the number of required simulations. First, we reduce the number of required simulations by classifying the decay into being either shallow or deep trap dominated with the help of analytical approximations to quickly find a high probability density region of data-compatible parameters. Then we employ a neural network acting as a surrogate model to the coupled rate equations to reduce the computational cost of the forward simulations for the parameter combinations that lie in this high probability region thereby unravelling parameter correlations and obtaining parameter posteriors. The obtained material parameter posteriors show how well the experiment can constrain parameters or in other words how certain we are about the obtained parameters.

## 7. Methods

### Sample Preparation for Photoluminescence Measurement

We prepared two films on glass viz. a $Cs_{0.05}(FA_{0.77}MA_{0.23})_{0.95}Pb(I_{0.85}Br_{0.15})_3$ triple-cation perovskite thin film post-treated with n-octylammonium iodide (OAI) referred to as a 85:15 film in the manuscript and a $Cs_{0.05}(FA_{0.77}MA_{0.23})_{0.95}Pb(I_{0.77}Br_{0.23})_3$ film with an additive of 3mol% $MAPbCl_3$ referred to as a 77:23 film. For the uncertainty quantification of the steady-state photoluminescence measurement, a GaAs wafer was used. The details about the material and the film fabrication for the 85:15 film can be found in Ye et al.[32]. For the 77:23 film, the substrate is Corning glass. Cesium iodide (CsI, 99.999%) was purchased from Thermo Fisher Scientific Inc. Formamidinium iodide (FAI, 99.99%) and methylammonium bromide (MABr, 99.99%) and methylammonium chloride (MACl, 99.99%) was purchased from GreatCell Solar. Lead iodide ($PbI_2$, 99.99%), lead bromide ($PbBr_2$, 98%) and lead chloride ($PbCl_2$, 99.0%) were obatined from TCI. N, N-dimethyl formamide (DMF, anhydrous), dimethyl sulfoxide (DMSO, anhydrous), isopropanol (IPA, anhydrous) and ethyl acetate (anhydrous) were purchased from Sigma Aldrich whereas piperazine iodide (PI, 99.5%) was ordered from YuriSolar. The glass substrates were ultrasonically cleaned with soap (Hellmanex III), deionized water, acetone, and IPA in turns for ten minutes. Subsequently, the glass substrates were treated with oxygen plasma (Diener Zepto, 50 W) for 12 min and immediately transferred to $N_2$-filled glovebox. Then, 100 µl perovskite precursor (1.7 M, $Cs_{0.05}(FA_{0.77}MA_{0.23})_{0.95}Pb(I_{0.77}Br_{0.23})_3$ with 3% $MAPbCl_3$ dissolved in DMF and DMSO with volume ratio of 4 to 1) was spin-coated onto glass substrates at 1000 rpm for 7 s and then 3000 rpm for 25 s, where 300 µl ethyl acetate was dropped in the center of the substrate at 10s before the end of spin-coating. After annealing at 100 °C for 15 minutes, 80 µl PI (0.2 mg/ml, dissolved in IPA) was spin-coated onto the perovskite surface, followed by annealing at 100 °C for 10 minutes.

**Time-Resolved Photoluminescence Measurement**

The transient and steady-state photoluminescence experiments were carried out as described by Yuan et al.[32]. More detailed instructions may also be found by Krückemeier et al.[13]. For the 77:23 film, the used laser wavelength was 514 nm and the measured fluence was 4.46 $\mu J/cm^2$. The film has a thickness of approximately 1 μm and the excited carrier concentration was 1.47 × $10^{17}$ $cm^{-3}$. Details about the specific measurement settings for the 85:15 film can be found in reference[32].

**Numerical Simulation**

Tr-PL and steady-state PL simulations were performed on the self-developed Python scripts based on the coupled rate equations. The scripts are available via the Jülich data repository and a GitHub repository.

For the transient photoluminescence experiments, first the thermal equilibrium carrier concentrations of the electrons in the conduction band, the holes in the valence band and the trapped electrons $(n_0, p_0, n_{t0})$ were calculated by making use of[61]

$$n_0 = N_C \exp\left(\frac{E_F - E_C}{k_B T}\right), \tag{20}$$

$$p_0 = N_V \exp\left(\frac{E_V - E_F}{k_B T}\right), \tag{21}$$

$$n_{t0} = \frac{N_t}{1 + \exp\left(\frac{E_t - E_F}{k_B T}\right)}, \tag{22}$$

$$\delta Q = p_0 - n_0 - n_{t0}. \tag{23}$$

Here, the valence band energy level was set to zero eV $E_V = 0$ and the conduction band energy level was set to the bandgap energy level $E_C = E_g$. For the calculation of the carrier concentrations, an initial guess for the Fermi-level $E_F$ of $E_g/2$ was used to estimate $(n_0, p_0, n_{t0})$ and then the Fermi-level was changed by employing Brent's method[62] to find the root $E_F$ that fulfils the charge-neutrality condition (Equation S4, $\delta Q \approx 0$). This in turn gives us the correct $(n_0, p_0, n_{t0})$ by Equations (20-23). Here, the effective densities of states at the conduction band and valence band were assumed to both equal $N_C = N_V = 2.2 \times 10^{18} \text{cm}^{-3}$ and the temperature was fixed to $T = 300$ K. The electron and hole effective masses $m^*_{e,h}$ have been reported to be similar[44–46] and by the relation

$$N_{C,V} = 2\left(\frac{2\pi m^*_{e,h} k_B T}{h^2}\right) \tag{24}$$

so are the effective density of states at the conduction and valence band $N_{\mathrm{C,V}}$ for perovskites. The value used for $N_{\mathrm{C,V}}$ our simulations was obtained from reference[63].

**Bayesian Inference Workflow**

During the development of data-analysis, neural-network training and the Bayesian inference workflow scripts accompanying this work, the authors used Claude Sonnet 4.6 (accessed August 2026) as an AI-assisted coding tool. The tool was employed to support script development, debugging, code documentation and review of the grid-based inference scheme. All AI-generated code and suggestions were reviewed, tested, and revised by the authors, who confirmed accuracy and reproducibility of results.

For a rapid parameter inference workflow, first a deep neural network was trained by creating a training dataset of 65.536 numerical simulations for the one trap model and 262.128 simulation for the two trap model. Here, only simulations where $E_{\mathrm{t1}} \leq E_{\mathrm{t2}}$ were used to reduce redundancy. Then the training set was split into a validation and test set such that the training, validation and test set ratio was 80/10/10. The validation set was used to check the quality and progress of the neural network during training and the test set was used to evaluate how well the network extends to unseen data within the given parameter bounds (see Table S1for bounds and TableS2 for training metrics). The networks were trained on an Nvidia RTX A600 with 48 GB VRAM (limited to 24 GB during training) for 250 epochs with early stopping implemented when the training loss (mean square error) did not change by at least $10^{-7}$ in 30 epochs, thus returning the model's weights from the epoch with the lowest validation loss. The initial learning rate was set to $10^{-3}$ which was halved whenever the validation loss plateaus for 8 consecutive epochs. For the weight adjustment, the Adam optimizer[64] was utilized. The batch size was 300. Each epoch the training set was reshuffled. A visualization of the neural network architectures can be found in Section 2 Figure S1. The input parameters of the networks consist of the bandgap energy $E_{\mathrm{g}}$, the excited carrier concentration from the laser $n_{\mathrm{pulse}}$, the radiative recombination coefficient $k_{\mathrm{rad}}$ and the 4 (8) trap specific material parameters $\left(E_{\mathrm{t}}, N_{\mathrm{t}}, \tau_{\mathrm{SRH,n}}, \tau_{\mathrm{SRH,p}}\right)$. The parameter specific bounds used to create the synthetic training datasets can be found in Table S1. For the networks used in this paper, $E_{\mathrm{g}}$ and $n_{\mathrm{pulse}}$ were fixed but could be variables for networks that should not only describe one absorber type or transient photoluminescence experiment. For better training results, the logarithm base 10 was taken for all input parameters θ, which are unitless, except the defect energy level which was expressed as a fraction relative to the bandgap energy level. Then standard scaling was applied

$$z = \frac{(\theta - \mu)}{s}, \tag{25}$$

where $\mu$ is the mean value for a specific material parameter taken over the training set and $s$ is its standard deviation.

The network was trained to output the logarithm base 10 of 256 differential decay times $\log_{10}(\tau_{\mathrm{diff}}/1s)$ scaled between 0 and 1 at equally spaced Fermi-level splittings ranging from $\Delta E_F = 2k_{\mathrm{B}}T\ln(n_{\mathrm{pulse}}/n_{\mathrm{i}})$ to a Fermi-level splitting that corresponds to a photoluminescence decay of 12 orders of magnitude. The neural network gained a speed up of about 100 times compared to the numerical simulations (see Figure S3) and acts as a high-fidelity surrogate model for both the one trap and two trap case (see neural network's performance metrics in TableS2).

Having trained the neural network, we use the CMA-ES optimizer together with the trained network to conduct the tr-PL and ss-PL fits. Here, the CMA-ES optimizer converged either when the root mean square error reached $10^{-3}$ or as defined by the original implementation of the algorithm in ref.[65,66].

To assess parameter uncertainty and correlations beyond the point estimate obtained from the CMA-ES optimizer, we performed a grid-based Bayesian inference scheme around the optimizer best fit parameter vector $\boldsymbol{\theta}_{\mathrm{best}}$. For each pair $(\theta_i, \theta_j)$ of the 5 (9) material parameters, we independently varied $\theta_i$ and $\theta_j$ over their respective ranges (linearly and uniformly spaced for the trap energy levels $E_{\mathrm{t}}$ and logarithmically and uniformly spaced otherwise) while keeping the remaining 3 (7) parameters fixed at $\boldsymbol{\theta}_{\mathrm{best}}$. We therefore obtained 10 (36) slices through the best fit rather than a full 5-or-9-dimensional grid. The benefit of taking cross-sections is that the computational cost scales quadratically instead of exponentially with the number of free material parameters. For each grid point, we calculated the mean-squared error between simulation and experiment independently for the transient and the steady-state by using Equation 16 and assigned a Gaussian log-likelihood by taking the logarithm of Equation 13. The uncertainty was determined from a systematic model error for the transient photoluminescence and from measurement noise for the steady-state photoluminescence (see Section 2, Figure S4-S6). For the joint inference, of steady-state and transient photoluminescence, the datasets are viewed as conditionally independent given a parameter combination $\boldsymbol{\theta}$. Therefore, their log-likelihoods

$$\ln L(\boldsymbol{\theta}) \equiv \ln P(D \mid \boldsymbol{\theta}) = -\frac{\epsilon_{\mathrm{rmse}}^2(\boldsymbol{\theta})}{2\sigma^2} - \ln(\sigma\sqrt{2\pi}) \tag{26}$$

were summed

$$\ln L_{\text{joint}}(\boldsymbol{\theta}) = \ln L_{\text{tr-PL}}(\boldsymbol{\theta}) + \ln L_{\text{ss-PL}}(\boldsymbol{\theta}), \tag{27}$$

where the uncertainties in the log-likelihoods for the steady-state and transient photoluminescence were distinct (see discussion in Section S2). Within each two-dimensional sliced corner plot panel $p = (i, j)$, the posterior $P_p(\boldsymbol{\theta}^{(g)} | D)$ was approximated by normalizing the likelihood over the panel's grid $G_p$ via the log-sum-exp (softmax) operation

$$P_p(\boldsymbol{\theta}^{(g)} | D) = \exp\left[\ln L\left(\boldsymbol{\theta}^{(g)}\right) - \ln Z_p\right], \qquad Z_p = \sum_g \exp\left[\ln L\left(\boldsymbol{\theta}^{(g)}\right)\right], \tag{28}$$

where $\boldsymbol{\theta}^{(g)}$ denotes the full parameter vector whose $i$-th and $j$-th components take their grid values $\theta_i, \theta_j$ and whose remaining components are set to their best fit value. Here, the prior was assumed to be log-uniform for all parameters except the trap energies and uniform for the trap energies, so that it could be crossed out from Bayes' theorem (Equation 12), leading to Equation S4. Note, Equation (28) is a probability mass to each grid cell rather than a probability density, so that the masses within each panel sum to unity, $\sum_{g \in G_p} P_p(\boldsymbol{\theta}^{(g)}|D) = 1$. The one-dimensional marginal posteriors for a given parameter $\theta_i$, referred to as $P(\theta_i|D)$, were obtained from the 4 (8) panels containing $\theta_i$, for the one or two traps, respectively. We marginalized each two-dimensional panel over its companion parameter by summing the posterior mass at a fixed $\theta_i$

$$\widetilde{P}_p(\theta_i) = \sum_{\substack{g \in G_p \\ \theta_i^{(g)} = \theta_i}} P_p(\boldsymbol{\theta}^{(g)}|D), \tag{29}$$

which by definition (see Equation 28) results in a profile of unit mass $\sum_{\theta_i} \widetilde{P}_p(\theta_i) = 1$.

The per-panel marginal profiles from all panels sharing $\theta_i$ were then combined with equal weight,

$$P(\theta_i|D) = \frac{1}{d-1} \sum_{p \ni \theta_i} \widetilde{P}_p(\theta_i), \tag{30}$$

where $d$ denotes the number of dimensions (5 for one trap 9 for two traps).

For the visualization in the diagonal panels of the corner plots, **Equation 30** was converted to a probability density $p(\theta_i|D)$ by dividing the grid spacing $\Delta_i$ of the sampled coordinate. The resulting density is defined with respect to $\log_{10}\theta_i$ for all parameters except the trap energies. The credible 1 $\sigma$ intervals for each parameter $\theta_i$ were computed as the smallest, symmetric interval about its best fit value $\boldsymbol{\theta}_{\text{best}}$ from the one-dimensional posteriors $p(\theta_i|D)$ enclosing 68 % probability in $\log_{10}\theta_i$ for all parameters and in $\theta_i$ for the trap energy level. Therefore, when transforming back to the correct physical units the interval becomes asymmetric for the logarithmic parameters. For the off-diagonal panels, the log-likelihoods (see Equation 26) are referenced to the log-likelihood at the optimizer best fit $\ln L\,(\boldsymbol{\theta}) - \ln L\,(\boldsymbol{\theta}_{\text{best}})$ using the same reference value for all panels. Since the panel normalization $Z_p$cancels in this difference and the prior is constant across grid points, it is identical to the corresponding difference of log posterior densities $\ln p\left(\boldsymbol{\theta}^{(g)} \mid D\right) - \ln p\,(\boldsymbol{\theta}_{\text{best}} \mid D)$; the uniform grid spacing within each panel then makes it equal to the difference of the log posterior masses, $\ln P_p\left(\boldsymbol{\theta}^{(g)} \mid D\right) - \ln P_p\,(\boldsymbol{\theta}_{\text{best}} \mid D)$, of that panel. Therefore, only the log-likelihoods had to be calculated. We emphasize, that while the individual log densities differ between panels, their difference with respect to the best-fit point does not. Hence, the depicted quantity is one and the same function of $\boldsymbol{\theta}$ throughout the corner plot figure.

Note, this procedure is necessitated by our pairwise-sliced grid design yielding a conditional posterior distribution and should only be interpreted as a locally valid estimate around the optimizer best-fit parameter combination rather than an exact marginal posterior from a full 5- or 9-dimensional posterior.

**Acknowledgements**

We acknowledge funding by the Helmholtz Association via the POF IV funding, via the project "Beschleunigter Transfer der nächsten Generation von Solarzellen in die Massenfertigung - Zukunftstechnologie Tandem-Solarzellen", via the Helmholtz.AI project "AISPA - AI-driven instantaneous solar cell property analysis" as well as by the Deutsche Forschungsgemeinschaft (German Research Foundation) via the project "Correlating Defect Densities with Recombination Losses in Halide-Perovskite Solar Cells" (SPP 2196). Open-access publication was funded by the German Research Foundation (DFG) 491111487. CD acknowledges that part of this project was funded by the Deutsche Forschungsgemeinschaft (DFG, German Research Foundation) – 576639778.

We acknowledge the usage of Claude Opus 5 (accessed August 2026) for generating a first draft of the Table of Contents text and as a language and writing aid in preparing the paragraphs about Bayesian statistics in the subsection about the Bayesian inference workflow of the methods section. All AI-assisted text was reviewed and revised by the authors to ensure accuracy and clarity of meaning.

**Data Availability Statement**

The scripts for the simulations, training of the neural networks and fitting of the experimental data as well as the trained neural network can be found in the repository called Heumann, Robin, 2026, "Scripts for Rapid Parameter Estimation from Photoluminescence Decays of Halide Perovskite Thin Films", https://doi.org/10.26165/JUELICH-DATA/2IEHBF, Jülich DATA. Future updates of these scripts can be found in the related GitHub repository https://github.com/roheumann/rapid-trpl-bayes.

**Supporting Information**

Supporting Information is available from the Wiley Online Library or from the author.

# Supporting Information

**Rapid Parameter Estimation from Photoluminescence Decays of Halide Perovskite Thin Films**

*Robin Heumann, Toby Rudolph, Gaosheng Huang, Thomas Kirchartz**, *Chris Dreessen**

R. Heumann, T. Rudolph, G. Huang, T. Kirchartz, C. Dreessen
IMD-3 Photovoltaics, Forschungszentrum Jülich, 52425 Jülich, Germany

T. Kirchartz
Faculty of Electrical Engineering and Information Technology, RWTH Aachen, 52062 Aachen, Germany

E-mail: t.kirchartz@fz-juelich.de, c.dreessen@fz-juelich.de

The outline of the supplementary information is as follows. The Figures and Tables related to the neural network training are to be found in Section S1. The uncertainty estimation for the photoluminescence experiments is discussed in more detail in Section S2. Section S3 includes all the remaining corner plots and neural network versus rate equation simulations for the synthetic one trap tr-PL inference and the remaining tr-PL corner plot panels for the experimental film data. Additionally, the experimental ss-PL fit and the inference results for the joint steady-state and transient photoluminescence are also included. In Section S4, the carrier concentration versus Fermi-level splitting plots that are used in Section 5.2 of the main paper can be found.

## S1. Neural Network Training

**Table S1.** Trap parameters boundaries used for the scope of the numerical simulations. These simulations are used for the training of the neural network. The bandgap $E_g$ is fixed to 1.625 eV, the initially excited carrier concentration by the laser pulse $n_{pulse}$ is $1.6\times10^{17}$and the intrinsic carrier concentration $n_i$ is $4.96\times10^4$.

| Parameter | Lower Simulation Bound | Upper Simulation Bound |
|---|---|---|

| | | | |
|---|---|---|---|
| Defects | Radiative recombination coefficient $k_{\text{rad}}$ (cm$^3$/s) | 1×10$^{-12}$ | 1×10$^{-10}$ |
| | Relative Defect level $E_{\text{t}}/E_{\text{g}}$ | 0.5 | 0.975 |
| | Defect density $N_{\text{t}}$ (1/cm$^3$) | 1×10$^{12}$ | 2.2×10$^{18}$ |
| | Electron SRH lifetime $\tau_{\text{SRH,n}}$ (s) | 1×10$^{-12}$ | 1×10$^{-4}$ |
| | Hole SRH lifetime $\tau_{\text{SRH,p}}$ | 1×10$^{-12}$ | 1×10$^{-4}$ |

**Table S2** Metrics of the trained neural networks. The mean squared error is calculated via equation 16 and the $R^2$ value via equation 17.

| Network Type | Dataset Type | Mean Squared Error | $R^2$ Value |
|---|---|---|---|
| 1 Defect | Training | 4×10$^{-5}$ | 0.999 |
| | Validation | 5×10$^{-5}$ | 0.998 |
| | Test | 5×10$^{-5}$ | 0.998 |
| 2 Defects | Training | 2×10$^{-5}$ | 0.998 |
| | Validation | 5×10$^{-5}$ | 0.995 |
| | Test | 5×10$^{-5}$ | 0.995 |

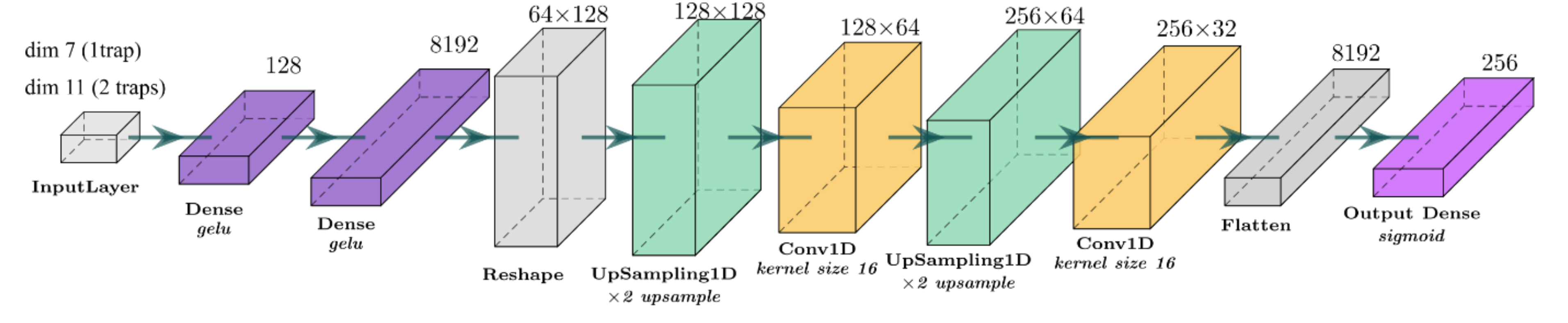

**Figure S1** Neural network architecture acting as a surrogate model to the Shockley-Read-Hall model with radiative recombination. Only the input dimensions differ between the one trap and two trap model for this architecture.

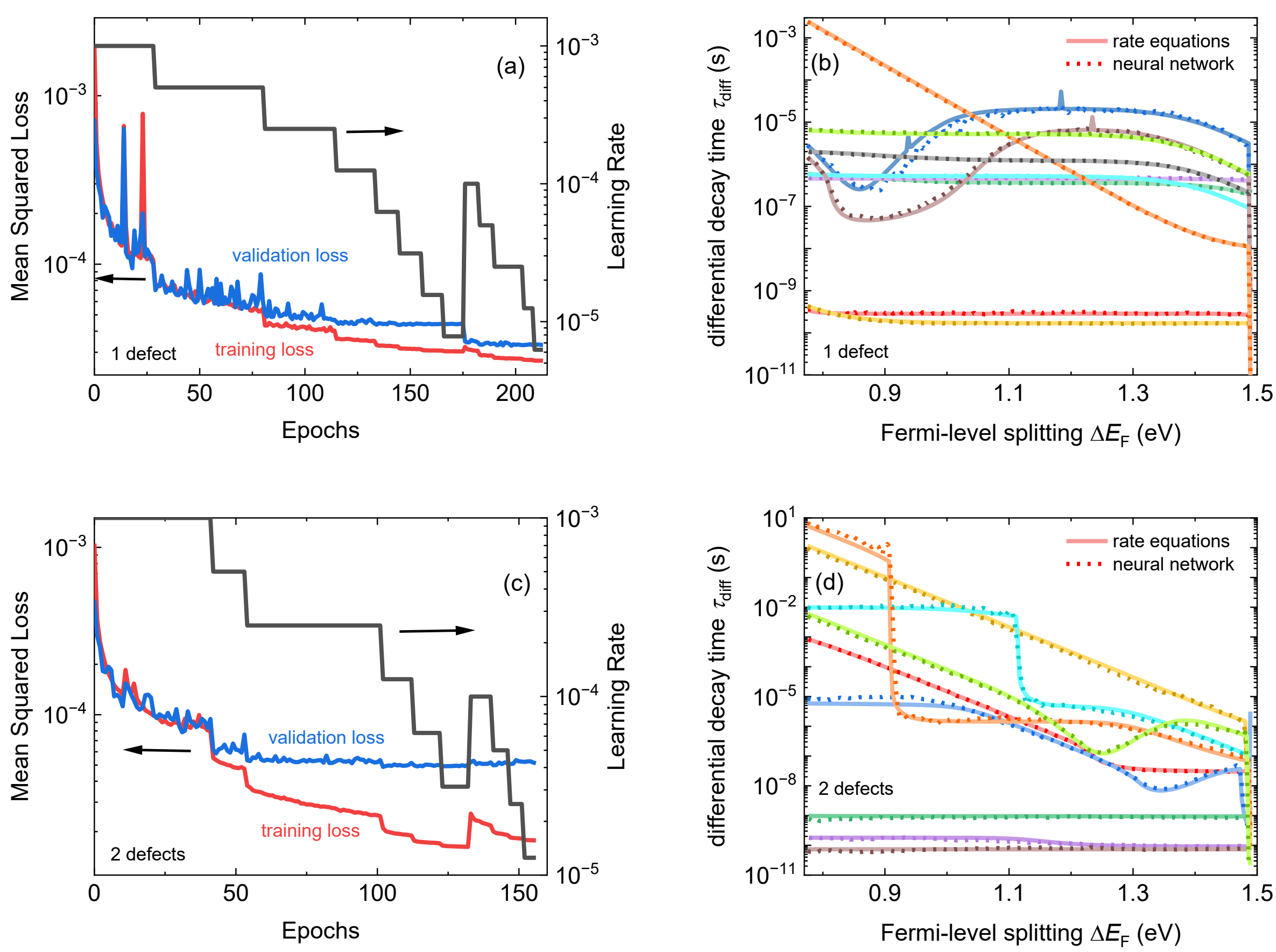

**Figure S2.** Overview of the networks predictive performance. Panel a (c) show the network training and validation loss as well as the learning rate versus the epochs whereas panel b (d) illustrates the neural network output compared to the rate equations for ten randomly selected curves from the test set for one (two) traps. In b and d the dotted lines are the output from the neural network and the solid lines the solutions to the rate equations from the SRH+radiative recombination model.

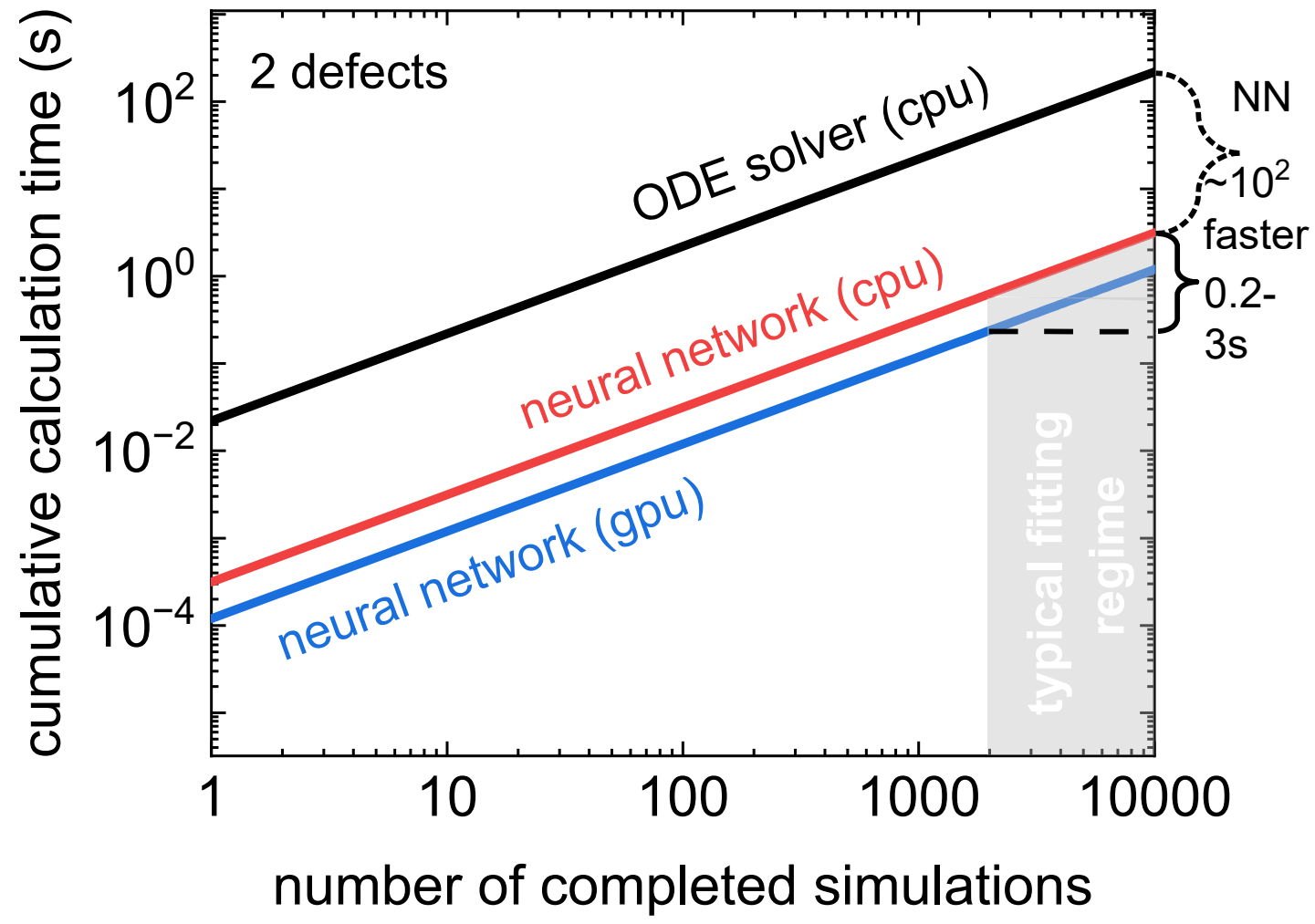

**Figure S3.** Comparison of the performance of the trained neural networks to solving the ordinary differential equations of the SRH+rad. recombination model for 2 defects.

## S2. Uncertainty Quantification for Photoluminescence Experiments

### S2.1. Estimation of 0D Tr-PL Model Error

The uncertainty that stems from the transient photoluminescence measurement noise $\sigma_{\mathrm{noise}}$ was calculated via the logarithmic residuals $r$ in Figure S4 by applying

$$\sigma_{\mathrm{noise}} = \langle r \rangle = \left\langle \left| \log_{10} \tau_{\mathrm{diff}}^{\mathrm{exp}} - \log_{10} \tau_{\mathrm{diff}}^{\mathrm{spline}} \right| \right\rangle, \tag{S1}$$

where, the $\langle \quad \rangle$ operator denotes the mean operator. The uncertainty results for the spline fit to the experimental data of the 85:15 film and the neural network to the rate equation output at the best fit can be found in **Figure S4**.

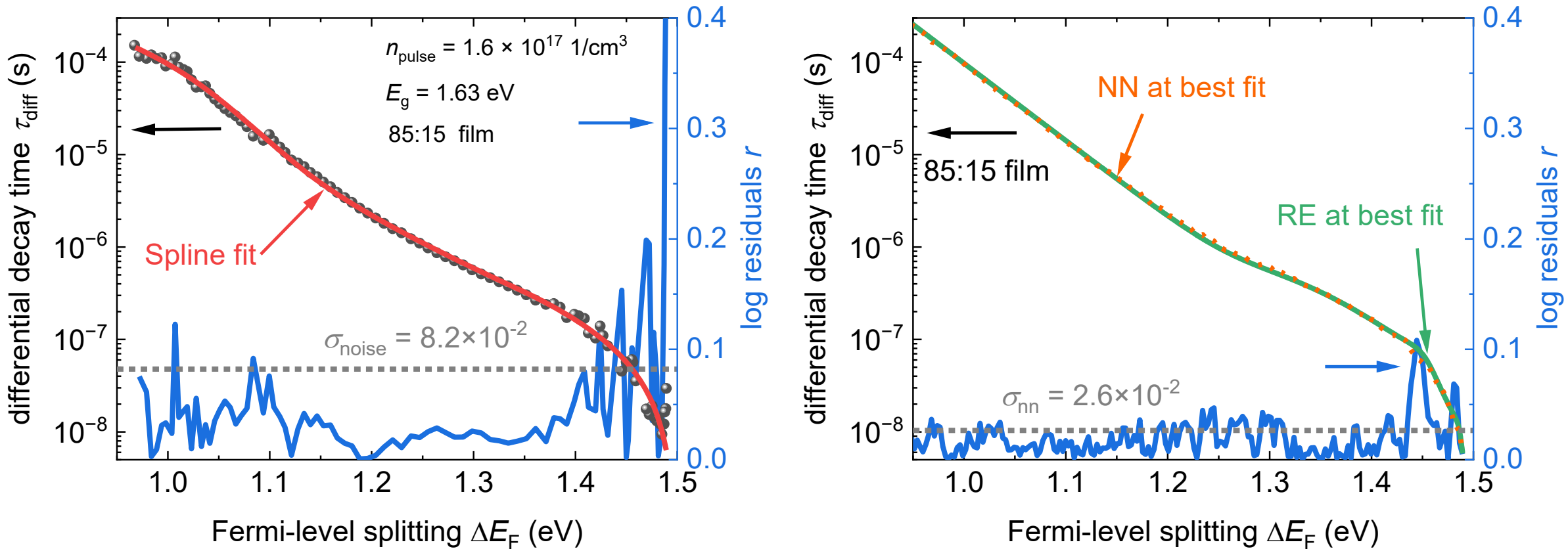

**Figure S4** Comparison of differential decay times versus Fermi-level splittings of (a) the raw experimental 85:15 film data to its spline fit and (b) the rate equation (RE) output to the neural network (NN) output at the best fit. The right axis shows the resulting logarithmic residuals $r$. The measurement uncertainty $\sigma_{\mathrm{noise}}$ and the neural network uncertainty $\sigma_{\mathrm{nn}}$ are also shown for comparison.

We neglected the influence of diffusion in our proposed model as this would introduce at least two more variables (mobility $\mu$ and absorption coefficient $\alpha$) to the material parameter vector $\theta$ making Bayesian inference exponentially more computationally costly. Nevertheless, we approximate this introduced error as a model uncertainty. According to the Beer-Lamber law[1], the absorption coefficient directly affects the initial generation profile. At initial times when the carriers have not diffused fully yet there exist regions with a greater carrier concentration in the perovskite absorber layer compared to a uniform carrier concentration. Due to the non-linearity of the PL$\sim np$, the carriers recombine faster leading to a smaller differential decay time (tail at higher Fermi-level splittings and a shift in Fermi-level splittings (see Figure S5) compared to the zero-dimensional case. We computed the differential decay time over an ensemble of simulations spanning the plausible range of α and took the resulting spread at each time point as a heteroscedastic uncertainty estimate at every Fermi-level splitting via

$$\sigma_{\mathrm{hc}}(\Delta E_{\mathrm{F}}) = \left\langle \left| \frac{\mathrm{d}\log_{10}\tau_{\mathrm{diff}}}{\mathrm{d}\log_{10}\alpha} \right| \right\rangle. \quad \text{(S2)}$$

Here, we set the mobility constant to $\mu = 1\ \mathrm{cm^2V^{-1}s^{-1}}$as we find that for our simulations the absorption coefficient has a strong influence on the systematic error introduced by chosen the wrong model.  The heteroscedastic uncertainty was then transformed into a homoscedastic uncertainty that does not vary with the Fermi-level splitting, referred to as the model uncertainty $\sigma_{\mathrm{model}}$, by taking the root-mean-square

$$\sigma_{\mathrm{model}} = \sqrt{\langle \sigma_{\mathrm{hc}}^2(\Delta E_{\mathrm{F}}) \rangle} \quad \text{(S3)}$$

stopping the posterior probability density to contract to a shape which is not permitted by the uncertainty in the absorption coefficient. Here, correlations between channels induced by the coherent response to α are neglected. We find that the measurement noise $\sigma_{\mathrm{noise}} = 8.2 \times 10^{-2}$ and the neural network uncertainty $\sigma_{\mathrm{nn}} = 2.6 \times 10^{-2}$ are considerably smaller than the uncertainty introduced by neglecting diffusion $\sigma_{\mathrm{model}} = 4.9 \times 10^{-1}$ and can therefore be neglected.

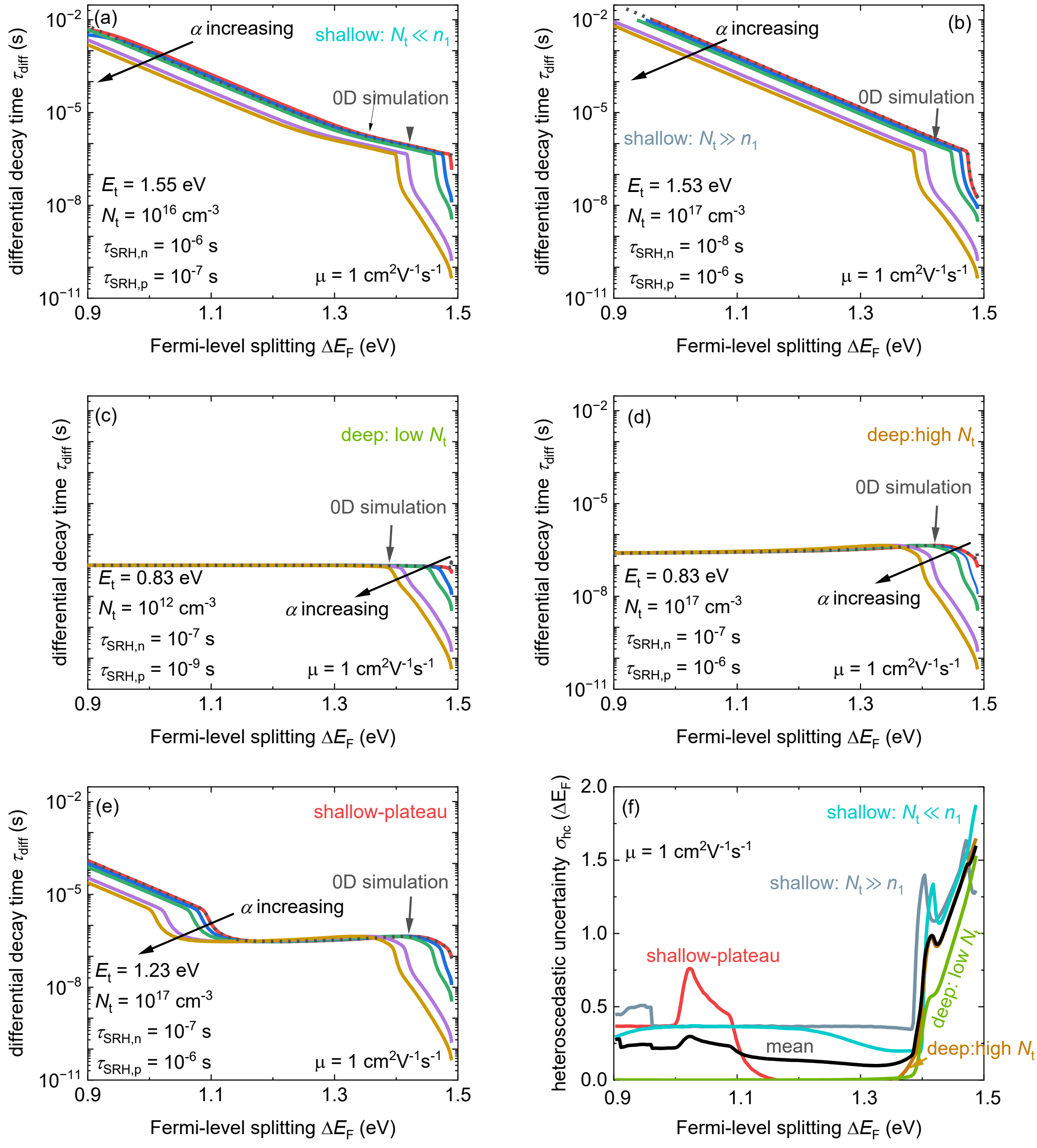

**Figure S5.** Simulations for five different cases including diffusion to the SRH + radiative recombination model (panels a to e). Panel f shows the derived heteroscedastic uncertainty $\sigma_{\text{hc}}(\Delta E_{\text{F}})$ for these five cases and their mean. The electron and hole mobilities were assumed to be equal and were set to $\mu = 1\ \text{cm}^2\text{V}^{-1}\text{s}^{-1}$ for all the simulations. The absorption coefficient $\alpha$ was varied from $10^4$ cm$^{-1}$ to $10^6$ cm$^{-1}$. The radiative recombination coefficient was constant across cases $k_{\text{rad}} = 10^{-11}\text{cm}^3\text{s}^{-1}$ as was the bandgap $E_{\text{g}} = 1.625$ eV and the excited carrier concentration $n_{\text{pulse}} = 1.6 \times 10^{17}\text{cm}^{-3}$. The perovskite absorber layer used in the simulation had a thickness of $d = 450$ nm.

### S2.2. Estimation of Steady-State Photoluminescence Uncertainty

The steady-state photoluminescence measurement's uncertainty is estimated from the time-dependent variation of the external quantum yield and the statistical error introduced by repeating an intensity-dependent steady-state measurement five times (see Figure S6) as in the steady-state. For this, we chose a highly stable, doped GaAs wafer for which the external photoluminescence quantum yield $Q_e^{\text{lum}}$ should not change for either the time- or the intensity-dependent measurement. The combined steady-state uncertainty is calculated as follows

$$\sigma_{\text{ss-pl}} = \sqrt{\sigma_{\text{time}}^2 + \sigma_{\text{intensity}}^2}, \tag{S4}$$

where $\sigma_{\text{time}}$ and $\sigma_{\text{intensity}}$ were extracted as the standard deviations from the time-dependent and intensity-dependent measurements and using the $\log_{10} Q_e^{\text{lum}}$, respectively. We find that $\sigma_{\text{ss-pl}} = 3.4 \times 10^{-2}$. For the steady-state diffusion already occurred and we do not have to take this into account for our model. Note, the measurement noise here is a bit smaller than the transient photoluminescence experiment.

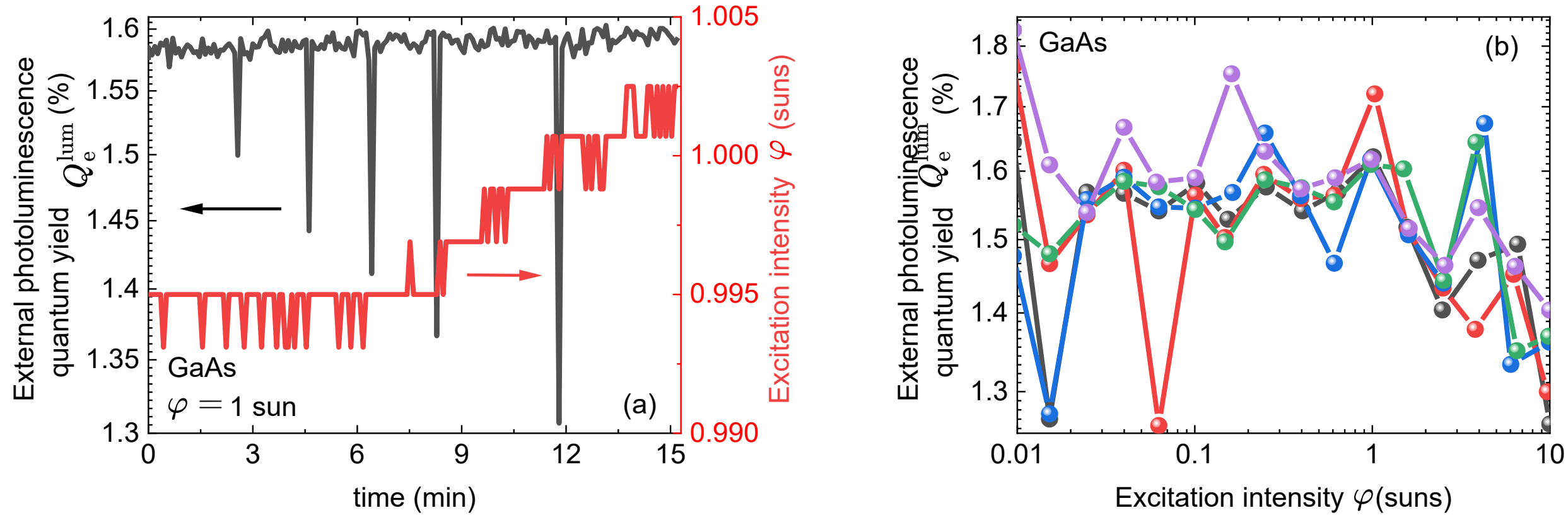

**Figure S6** Measured steady-state photoluminescence of a GaAs wafer. Panel a shows the time-dependent variation in external photoluminescence quantum yield and photon excitation intensity for a measurement where the excitation intensity was set to 1 sun. Panel b shows the intensity-dependent external photoluminescence quantum yield for several repeated experiments (different colors).

### S3. Remaining Corner Plot Panels and Synthetic 1 Trap Simulations

The remaining panels for the synthetic tr-PL corner plot in Figure S7 show that once a suitable $k_{\text{rad}}$ has been found the trap density $N_t$ can vary over orders of magnitude. The SRH lifetime of the electrons or holes ($\tau_{\text{SRH,n}}$ ; $\tau_{\text{SRH,p}}$) do not constrain the radiative recombination coefficient and $k_{\text{rad}}$ can vary over orders of magnitude without changing the probability density.

This is also reflected in the low posterior probability density. This trend is also reflected in the corner plot panels containing $k_{\mathrm{rad}}$ for the experimental data (see Figure S9 and S10). At the optimizer best fit, $k_{\mathrm{rad}}$ can vary over orders of magnitude for all trap parameter combinations. From Figure S9 we see that the probability landscapes regarding defect densities for defect 1 and 2, $N_{\mathrm{t1,2}}$, reveal the possibility of a second minimum at low trap densities $N_{\mathrm{t1,2}}, \approx 10^{-12}\mathrm{cm}^{-3}$ (see $N_{\mathrm{t2}}$ vs $E_{\mathrm{t2}}$ and $N_{\mathrm{t1}}$ vs $N_{\mathrm{t2}}$ subpanels). Interestingly, $\tau_{\mathrm{SRH,n1}}$ remains the most unconfined parameter (see Figure S8).

The simulations between the rate equations of our model and the neural network surrogate for the synthetic dataset of one shallow trap close to the conduction band show that there is no visible difference between the transients at the ground truth and best fit in the experimentally accessible region (see Figure S8).

We now look at the Bayesian inference results when we include also the steady-state photoluminescence to the transient photoluminescence (see Figures S11-S14). Firstly, we see that our fitting approach considering both steady-state and transient photoluminescence also leads to a good steady-state fit (see Figure S11). When looking at the joint ss-pl and tr-pl corner plots (Figures S12-S14), we see that the parameter relationships and posterior shapes are very similar to the tr-PL only corner plots (see Figure 9 and S9-10). The posterior of the defect energy-levels seems to have become slightly narrower, but the posterior of the SRH electron lifetime seems to have become significantly flatter whereas the posteriors of the remaining parameters did not change much (see Figure S12 and Table S3). Taking the results from the main paper into account, we can therefore conclude that the steady-state photoluminescence in addition to the transient photoluminescence can be used to rule out initially promising tr-PL fits (see one trap fits of Section 5.2). However, from the joint ss-PL and tr-PL inference we conclude that the steady-state results do not provide enough information to uniquely identify material parameters to help solve the structural identifiability problem. Other, complementary experiments in addition to steady-state and transient photoluminescence are needed to resolve the structural identifiability problem.

### S3.1. Synthetic Data

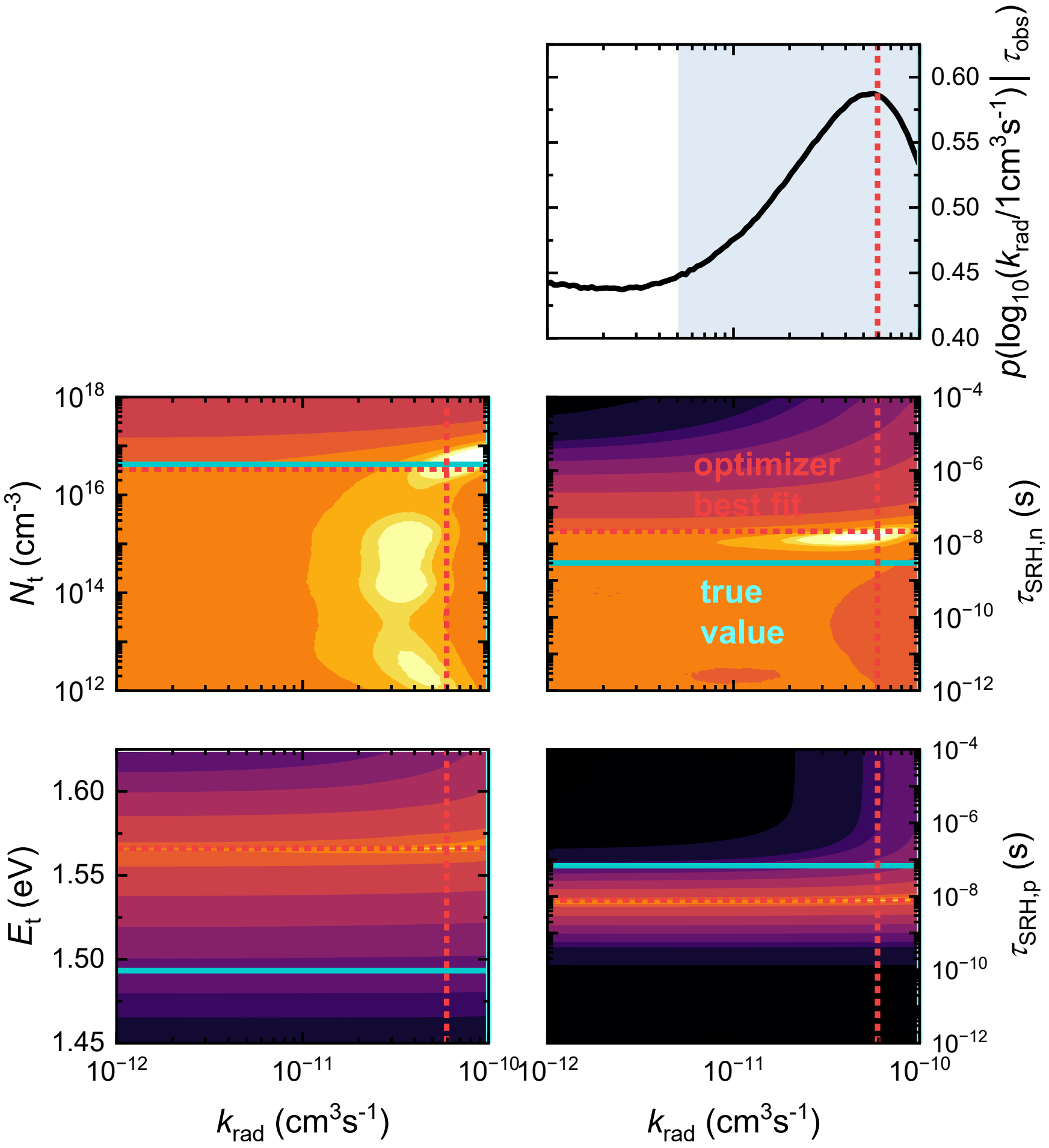

**Figure S7.** Remaining panels including the radiative recombination coefficient $k_{\mathrm{rad}}$ of the sliced corner plot around the optimizer best-fit for the synthetic shallow trap fit from Figure 5. The probability density function of $k_{\mathrm{rad}}$ is shown in the top right and the blue shaded regime around the optimizer best fit is the approximate $1\sigma$ (68 %) conditional credible interval. The joint probability densities are depicted in the lower four panels. For each axis 100 points were used yielding $10^4$ points per panel. The dotted red lines indicate the optimizer best-fit value and the solid cyan lines show the true parameter values of the artificial dataset. The color scheme was constructed as stated in the section about the Bayesian workflow or Figure 5. The uncertainty for the likelihood is $\sigma = 3.3\times10^{-2}$.

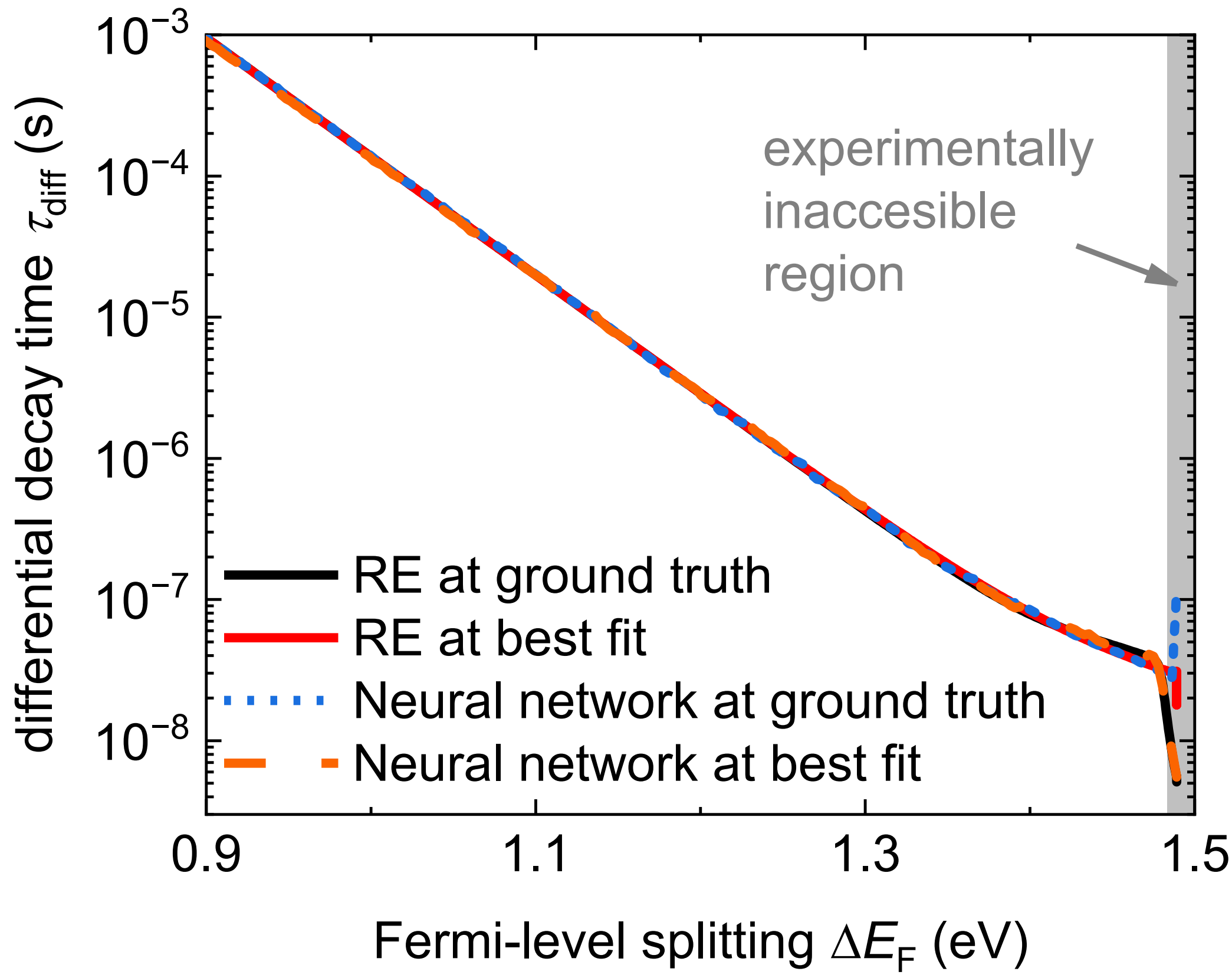

**Figure S8.** Comparison between transient simulations of one shallow trap close to the conduction band. The rate equations (RE) and the neural network outputs for the ground truth parameters and the best fit parameters are shown. The experimentally inaccessible region is highlighted in grey.

### S3.2. Shallow Trap Experimental Data

#### *S3.2.1. Transient Photoluminescence Inference*

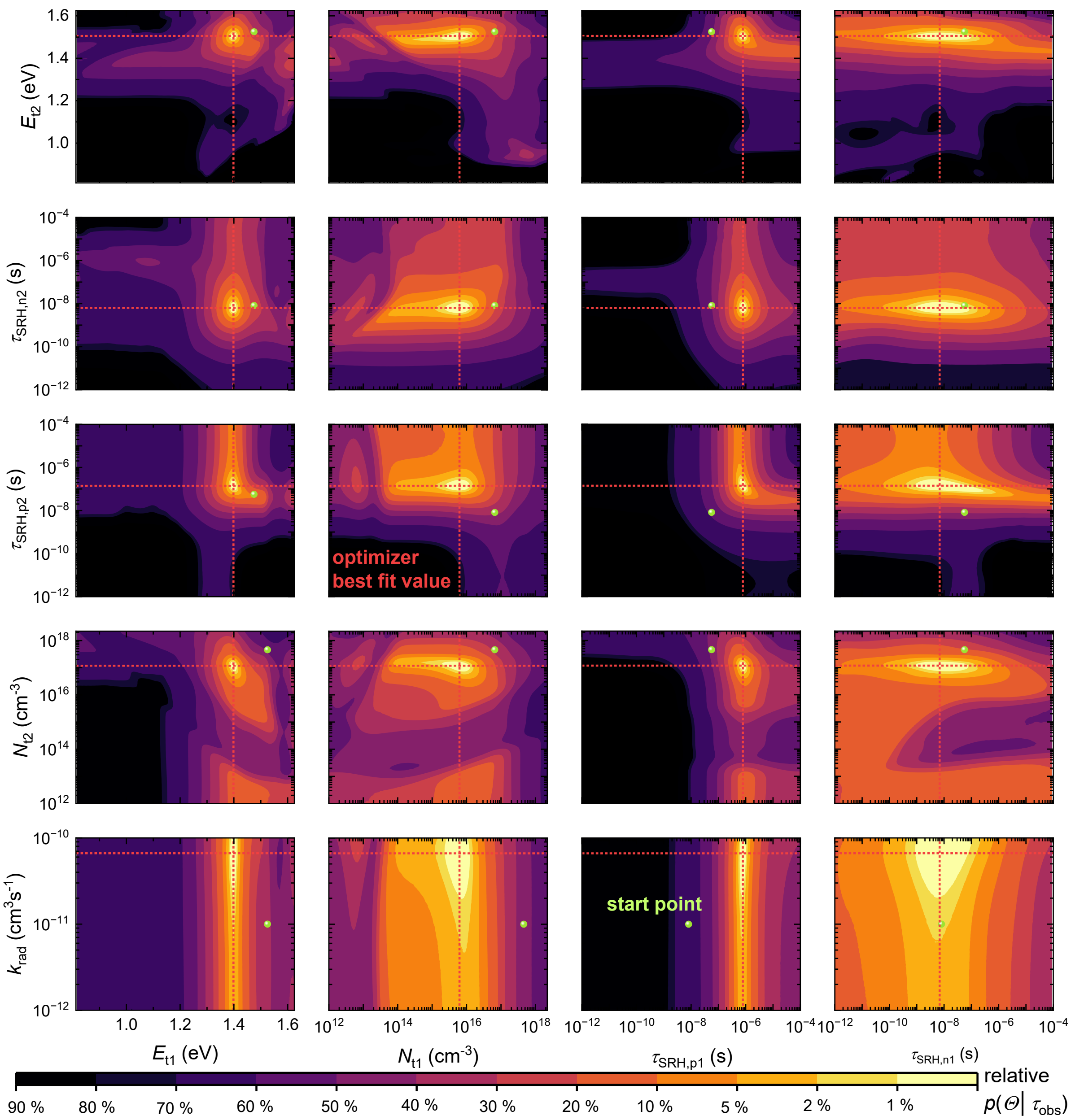

**Figure S9.** Remaining sliced corner plot panels of defect 1 parameters to defect 2 parameters around the optimizer best-fit for the experimental shallow trap fit from Figure 9. The probability density function of $k_{\mathrm{rad}}$ is shown in the top right and the blue shaded regime around the optimizer best fit is the approximate $1\sigma$ (68 %) conditional credible interval. The joint probability densities are depicted in the lower four panels. For each axis 100 points were used yielding $10^4$ points per panel. The dotted red lines indicate the optimizer best-fit value and the green dots the start point value for the fitting procedure. The color scheme was constructed as stated in the section about the Bayesian workflow or Figure 5. The uncertainty for the likelihood is $\sigma = \sigma_{\mathrm{model}} = 4.9\times10^{-1}$.

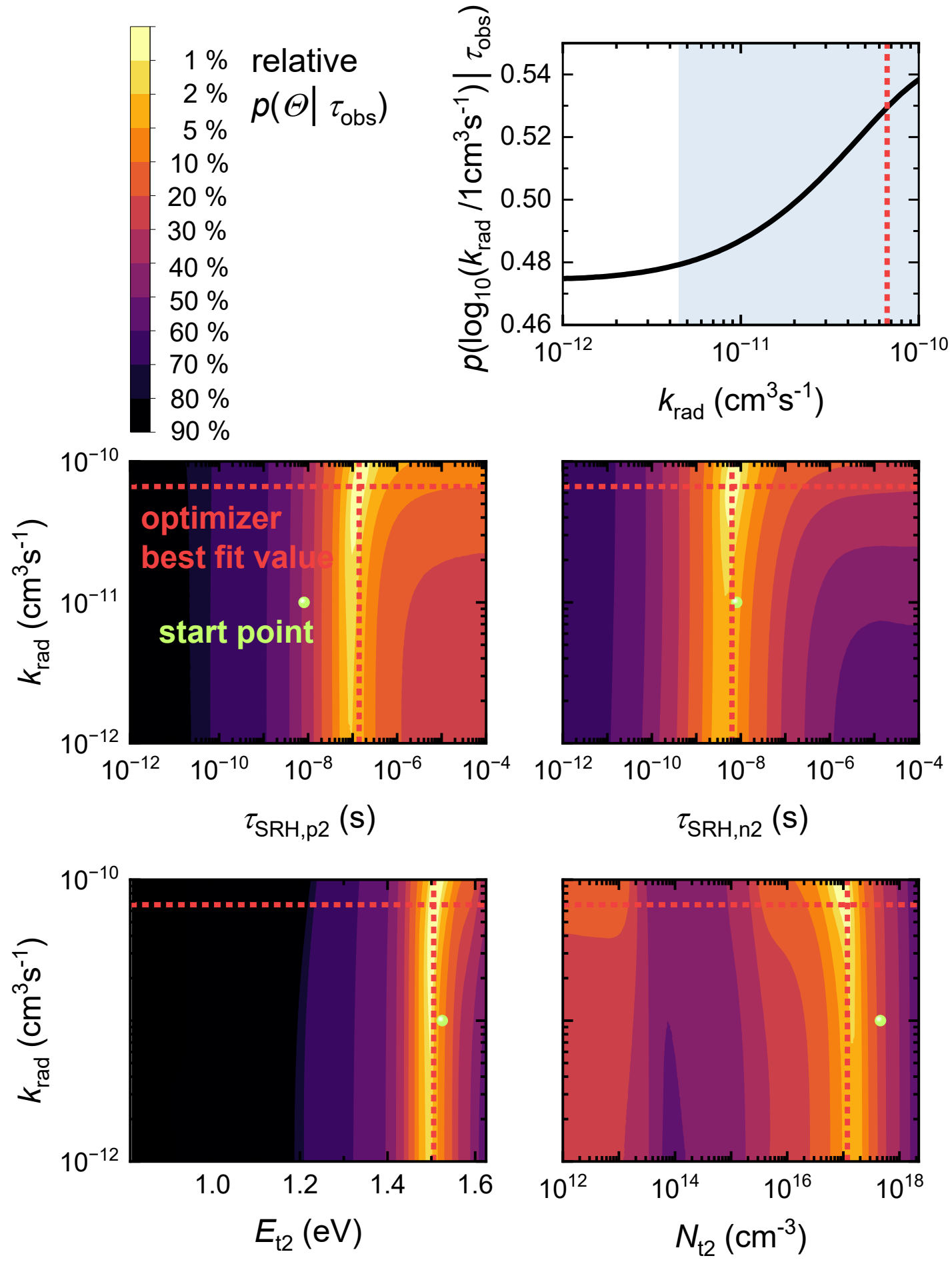

**Figure S10.** Remaining radiative recombination coefficient $k_{\text{rad}}$ panels for defect 2 of the sliced corner plot around the optimizer best-fit for the experimental shallow trap fit from Figure 9. The probability density function of $k_{\text{rad}}$ is shown in the top right and the blue shaded regime around the optimizer best fit is the approximate $1\sigma$ (68 %) conditional credible interval. The joint probability densities are depicted in the lower four panels. For each axis 100 points were used yielding $10^4$ points per panel. The dotted red lines indicate the optimizer best-fit value and the green dots the start point value for the fitting procedure. The color scheme was constructed as stated in the section about the Bayesian workflow or Figure 5. The uncertainty for the likelihood is $\sigma = \sigma_{\text{model}} = 4.9\times10^{-1}$.

*S3.2.2. Steady-State and Transient Photoluminescence Inference*

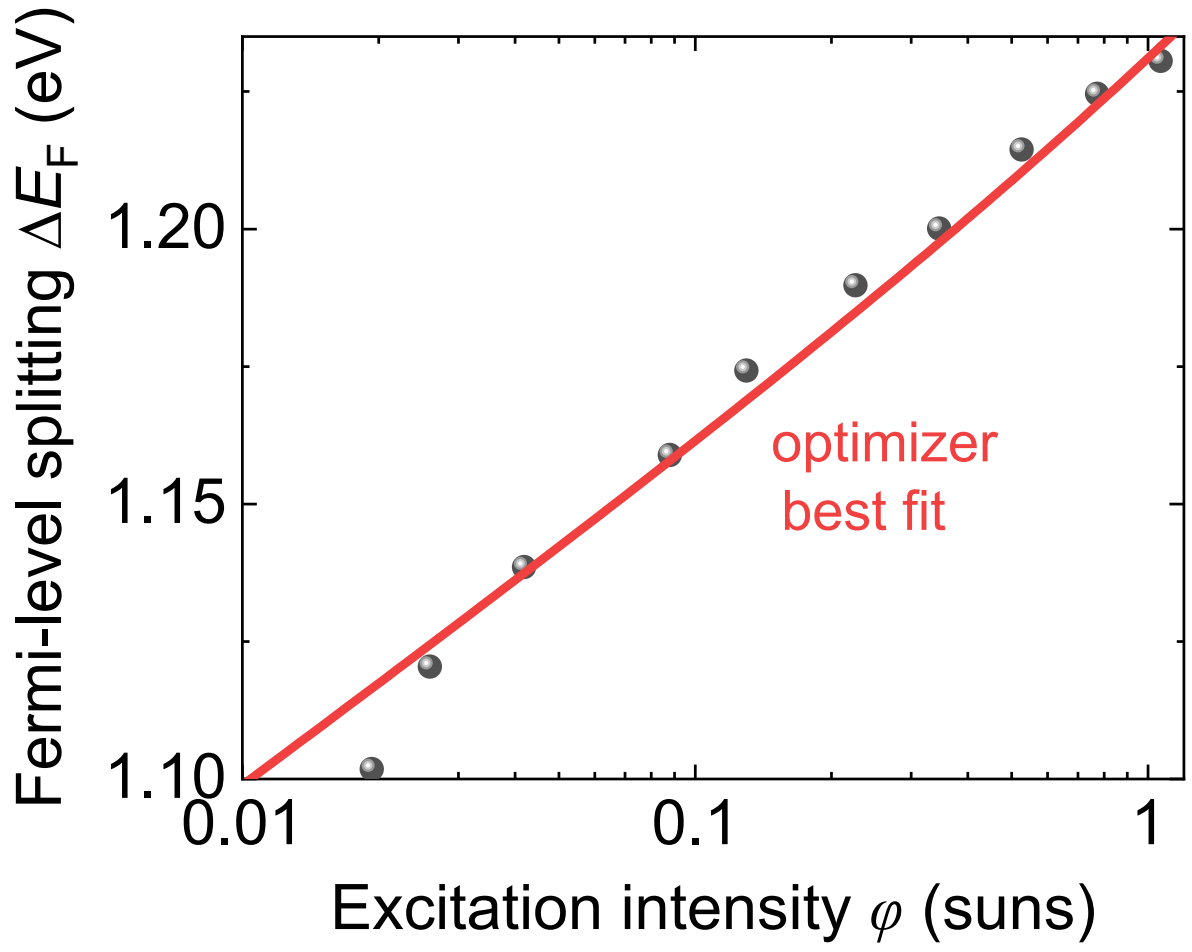

**Figure S11.** Steady-state photoluminescence experimental data and fit (red line) for the 85:15 film. The fit used two traps and the steady-state was fitted together with the transient photoluminescence data.

**Table S3.** Trap parameters obtained from the machine learning assisted parameter estimation workflow for the 85:15 film considering transient and steady-state photoluminescence data. The bandgap $E_g$ is 1.625 eV, the initially excited carrier concentration by the laser pulse $n_{pulse}$ is $1.6\times10^{17}$and the intrinsic carrier concentration $n_i$ is $4.96\times10^4$. The lower and upper credibility bounds are the conditional smallest symmetric contiguous interval around the optimizer best fit and are only approximations of the $1\sigma\,\chi^2$ credibility interval within which the true parameter value lies with 68% probability.

| | Parameter | 85:15 Film Best Fit | Lower Credibility Bound | Upper Credibility Bound |
|---|---|---|---|---|
| Defect 1 | Radiative recombination coefficient $k_{rad}$ (cm$^3$/s) | $6.6\times10^{-11}$ | $4.7\times10^{-12}$ | $9.4\times10^{-10}$ |
| | Defect level $E_{t1}$ (eV) | 1.39 | 1.31 | 1.49 |
| | Defect density $N_{t1}$ (1/cm$^3$) | $6.2\times10^{15}$ | $8.6\times10^{13}$ | $4.5\times10^{17}$ |
| | Electron SRH lifetime $\tau_{SRH,n1}$ | 6.9 ns | 26 ps | 1.8 µs |
| | Hole SRH lifetime $\tau_{SRH,p1}$ | 0.8 µs | 0.1 µs | 4.2 µs |
| Defect 2 | Defect level $E_{t2}$ (eV) | 1.51 | 1.42 | 1.59 |
| | Defect density $N_{t2}$ (1/cm$^3$) | $1.2\times10^{17}$ | $1.2\times10^{14}$ | $1.2\times10^{20}$ |

| | | | |
|---|---|---|---|
| Electron SRH lifetime $\tau_{SRH,n2}$ | 6.3 ns | 29 ps | 1.4 μs |
| Hole SRH lifetime $\tau_{SRH,p2}$ | 0.1μs | 3.4 ns | 5.8 μs |

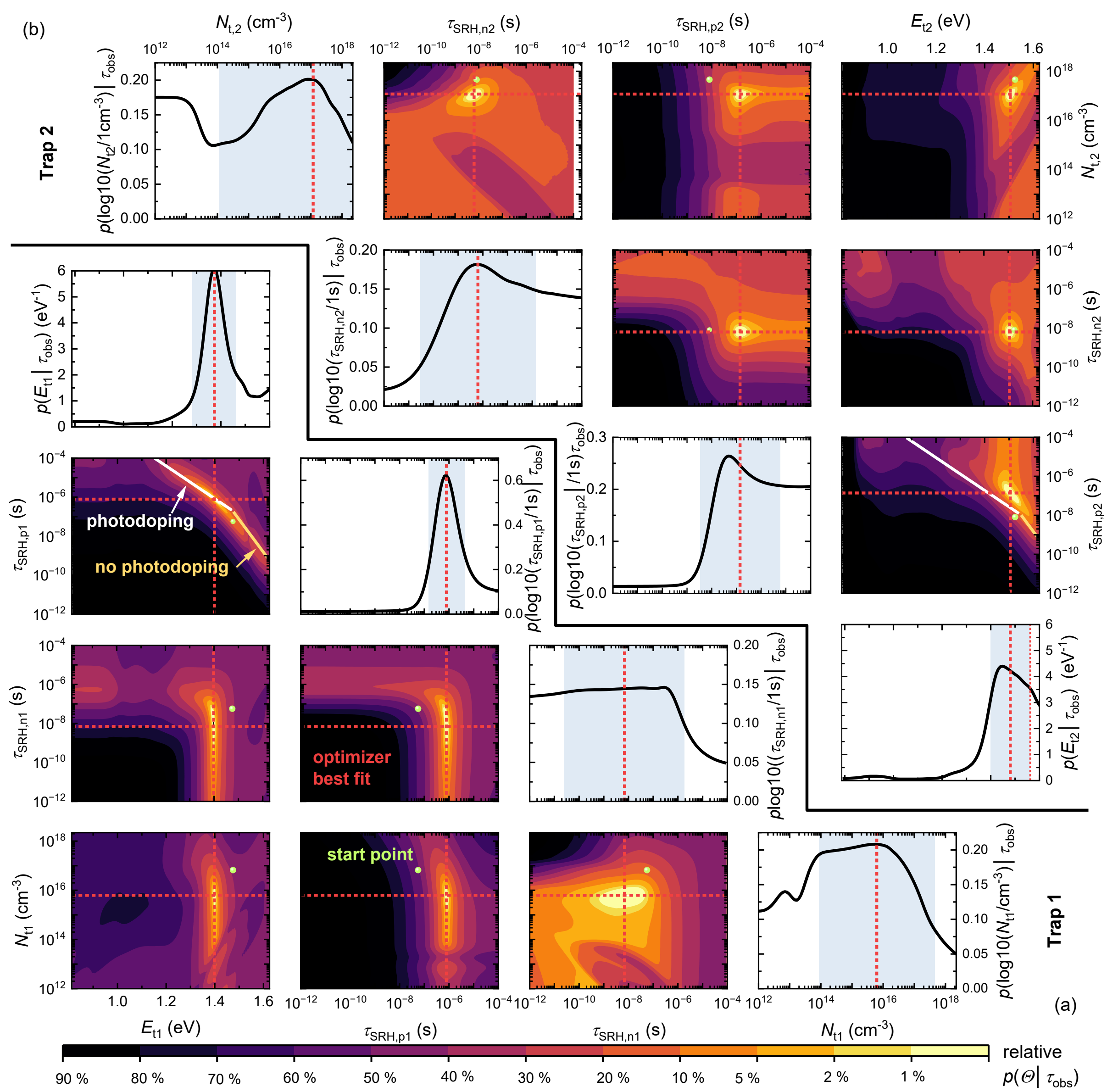

**Figure S12,** Sliced corner plot around the optimizer best fit values for the transient and the steady-state PL decay of the experimental 85:15 film data (see figure 6). Panel (a) focuses on the defect parameters associated with defect 1 and panel (b) with defect 2. The diagonal panels show the one-dimensional conditional marginal probability density functions for each free parameter, while the off-diagonal panels display the conditional joint probability density functions, revealing correlations and degeneracies in multidimensional parameter space. For each axis 100 points were used yielding $10^4$ points per panel. The dotted red lines denoted the optimizer best fit values. The black point in the off-diagonal panels marks the physics informed start point used for the fitting routine. In panels ($E_t$, $\tau_{SRH,p}$ the solid white line originates from Equation 10 for high trap densities (photodoping) and the orange solid line for low trap densities from Equation 11 (no photodoping, see ref. [26]). The color scheme was constructed as stated in the section about the Bayesian workflow or Figure 5. The diagonal panels show the approximate $1\sigma$ (68 %) conditional credibility intervals around the best fit (blue shaded regime). The uncertainty for the likelihood of the tr-PL is $\sigma_{\text{tr-PL}} = \sigma_{\text{model}} = 4.9\times10^{-1}$ and for the ss-PL $\sigma_{\text{ss-PL}} = \sigma_{\text{noise}} = 3.4\times10^{-2}$.

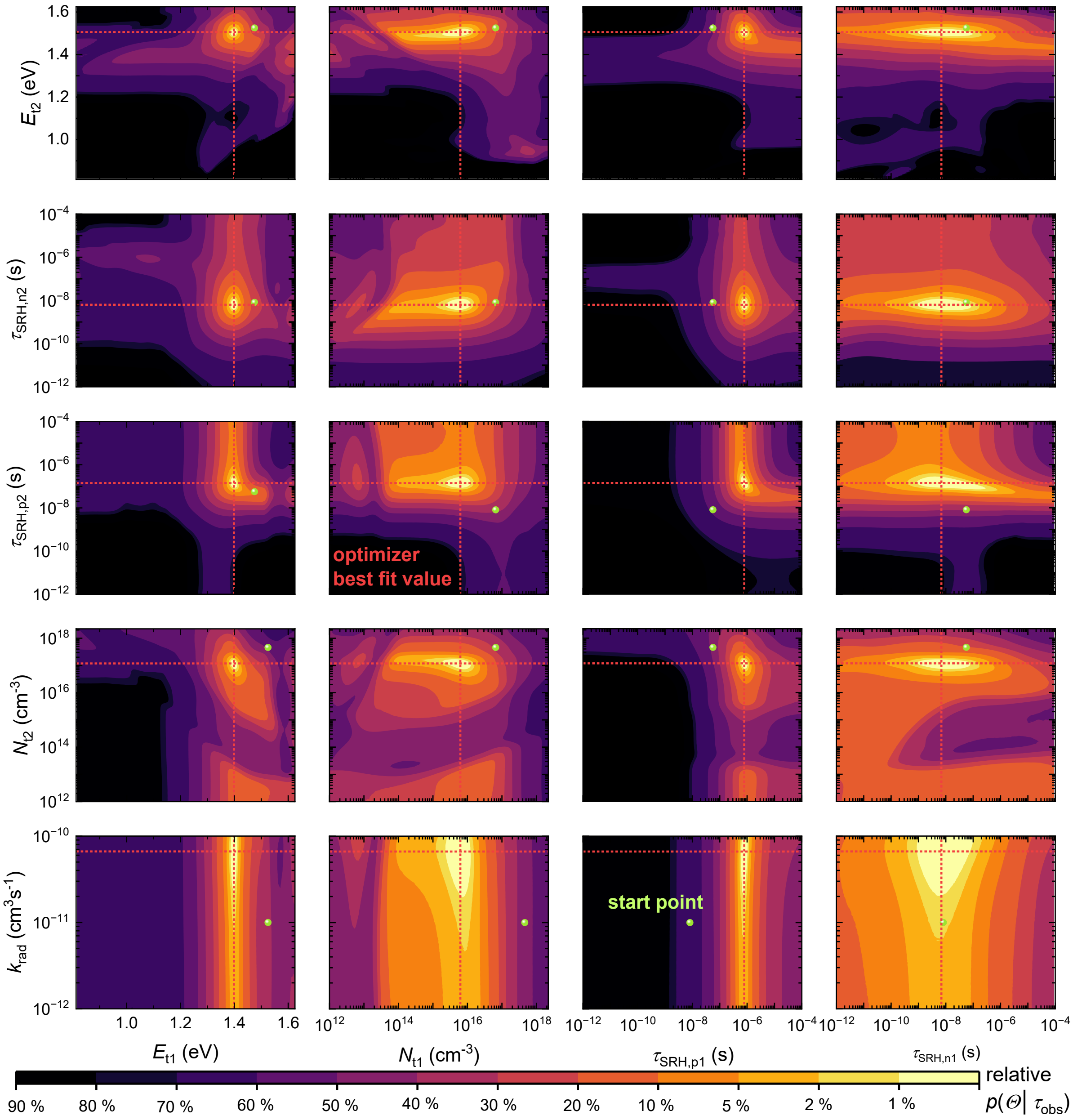

**Figure S13**. Remaining sliced corner plot panels of defect 1 parameters to defect 2 parameters around the optimizer best fit values for the transient and the steady-state PL decay of the experimental 85:15 film data (see figure 6). The conditional joint probability density functions reveal correlations and degeneracies in multidimensional parameter space. For each axis 100 points were used yielding $10^4$ points per panel. The dotted red lines denoted the optimizer best

fit values. The green point marks the physics informed start point used for the fitting routine. The color scheme was constructed as stated in the section about the Bayesian workflow or Figure 5. The uncertainty for the likelihood of the tr-PL is $\sigma_{\text{tr-PL}} = \sigma_{\text{model}} = 4.9\times10^{-1}$ and for the ss-PL $\sigma_{\text{ss-PL}} = \sigma_{\text{noise}} = 3.4\times10^{-2}$.

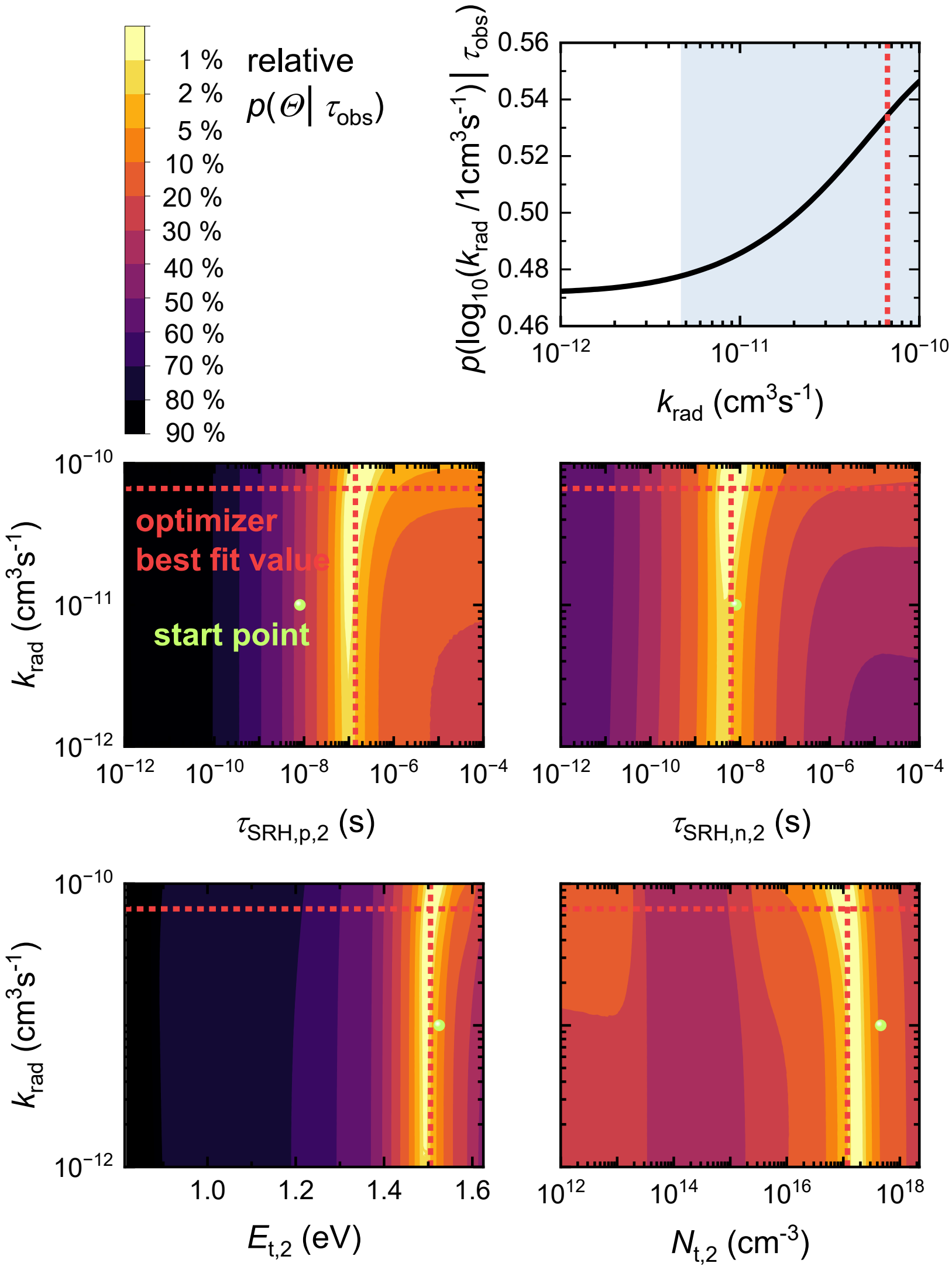

**Figure S14.** Remaining sliced corner plot panels of defect 2 parameters to the radiative recombination coefficient around the optimizer best fit values for the transient and the steady-state PL decay of the experimental 85:15 film data (see figure 6). The top right panel shows the one-dimensional conditional marginal probability density functions for the radiative recombination coefficient. Here, the approximate $1\sigma$ (68 %) conditional credibility intervals around the best fit are shaded in blue. The lower four panels display the conditional joint probability density functions, revealing correlations and degeneracies in multidimensional parameter space. For each axis 100 points were used yielding $10^4$ points per panel. The dotted red lines denoted the optimizer best fit values. The color scheme was constructed as stated in the section about the Bayesian workflow or Figure 5. The uncertainty for the likelihood of the tr-PL is $\sigma_{\text{tr-PL}} = \sigma_{\text{model}} = 4.9\times10^{-1}$ and for the ss-PL $\sigma_{\text{ss-PL}} = \sigma_{\text{noise}} = 3.4\times10^{-2}$.

## S4. Carrier concentrations

### S4.1. Synthetic Data

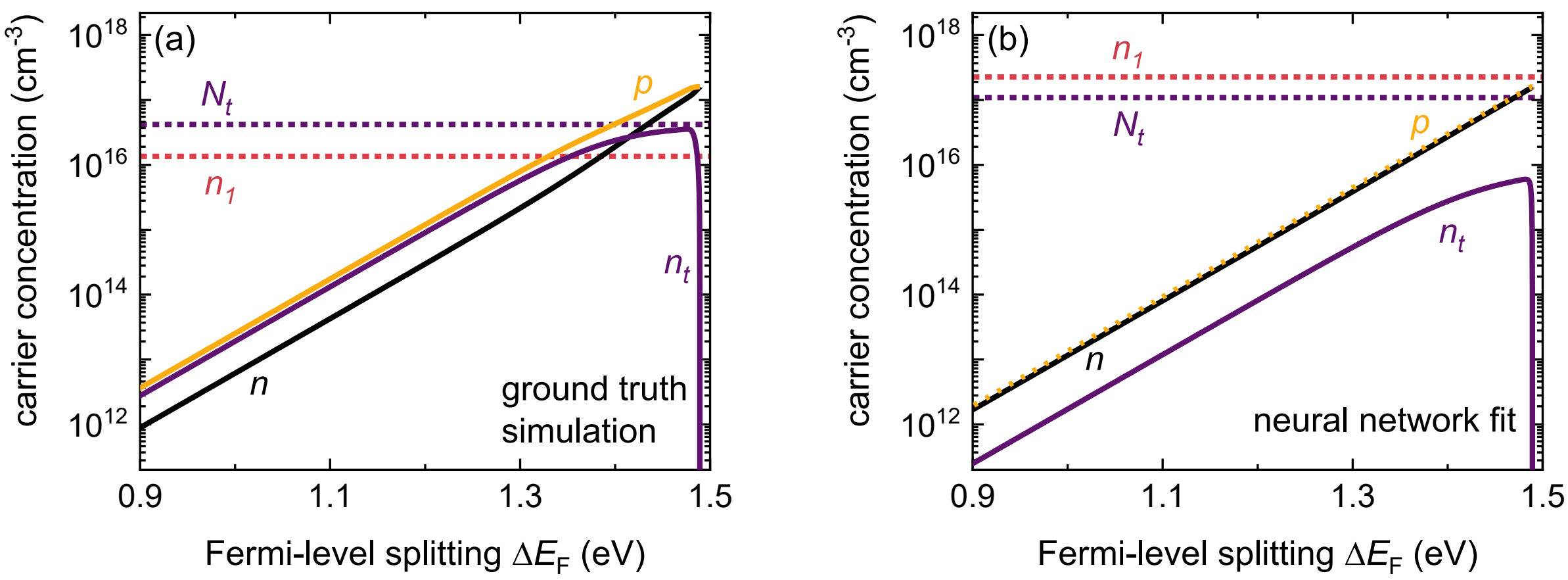

**Figure S15.** Overview of the simulated charge carrier concentrations versus Fermi-level splittings for the transient photoluminescence of the 1 defect synthetic dataset presented in the main paper. Panel a shows the ground truth simulation and panel b the neural network best fit simulation.

### S4.2. Experimental Transient and Steady-State Photoluminescence

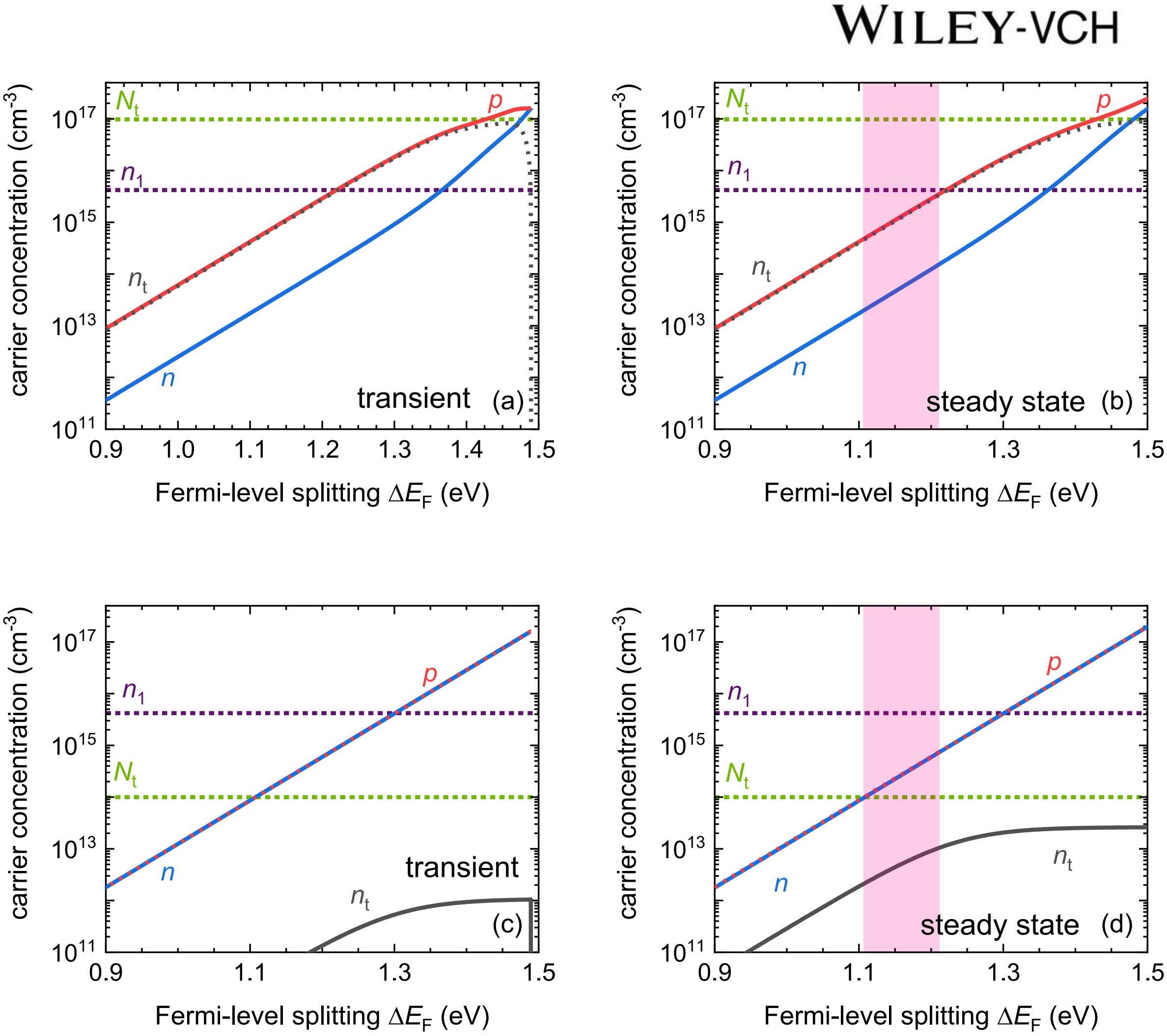

**Figure S16.** Overview of the simulated charge carrier concentrations versus Fermi-level splittings of the transient and steady-state photoluminescence for a shallow trap close to the conduction band $E_t = 1.463$ eV with a high trap density $N_t = 10^{17}\mathrm{cm}^{-3}$ (low trap density $N_t = 10^{14}\mathrm{cm}^{-3}$) in panel a (c) and b (d), respectively. The bandgap energy was set to $E_g = 1.63$ eV and the excited charge carrier concentration set to $n_{pulse} = 1.6 \times 10^{17}\mathrm{cm}^{-3}$. More details about the model parameters can be found in Table 4.